\documentclass[fleqn,iop,numberedappendix]{emulateapj}
\usepackage{morefloats}
\usepackage{amsmath}
\usepackage{paralist}
\usepackage{adjustbox}
\usepackage{gensymb}
\usepackage[breaklinks,colorlinks,citecolor=blue]{hyperref}
\usepackage[all]{hypcap}
\usepackage{multirow}
\begin{document}
\title{The radial variation of H\,{\scriptsize{i}} velocity dispersions in dwarfs and spirals}
\author{R. Ianjamasimanana\altaffilmark{1}, W.J.G. de
  Blok\altaffilmark{2,3,4}, Fabian Walter\altaffilmark{5}, George
  H. Heald\altaffilmark{3,4}, Anahi Cald\'{u}-Primo\altaffilmark{5}, Thomas H. Jarrett\altaffilmark{2}} 
  \altaffiltext{1}{College of Graduate Studies, University of South Africa, P.O. Box 392, UNISA 003, South Africa; 
  ianja@ast.uct.ac.za}  
  \altaffiltext{2}{Astrophysics, Cosmology
  and Gravity Centre, Department of Astronomy, University of Cape
  Town, Private Bag X3, Rondebosch 7701, South Africa} 
  \altaffiltext{3}{ASTRON, the Netherlands Institute for Radio Astronomy, Postbus 2, 7990 AA, Dwingeloo, The Netherlands} 
    \altaffiltext{4}{Kapteyn Astronomical Institute, University of Groningen, PO Box 800, 9700 AV, Groningen, The Netherlands}
  \altaffiltext{5}{Max-Planck Institut
  f$\rm{\ddot{u}}$r Astronomie, K\"onigstuhl 17, 69117,
  Heidelberg, Germany}
  \slugcomment{Accepted for publication in The Astronomical Journal}
  \begin{abstract}
Gas velocity dispersions provide important diagnostics of the forces counteracting gravity to prevent collapse of 
the gas. We use the 21-cm line of neutral atomic hydrogen (H\,{\sc i}) to study H\,{\sc i} velocity dispersion ($\sigma_{\rm{H\,{\textsc i}}}$) 
and H\,{\sc i} phases as a function of galaxy morphology in 22 galaxies from The H\,{\sc i} Nearby Galaxy Survey (THINGS). 
We stack individual H\,{\sc i} velocity profiles and decompose them into broad and narrow Gaussian components.
We study the H\,{\sc i} velocity dispersion and the H\,{\sc i} surface density, $\Sigma_{\rm{H\,{\textsc i}}}$, as a function of radius. For spirals, the velocity dispersions of the narrow and broad components decline with radius and their radial profiles are well described by an exponential function. For dwarfs, however, the profiles are much flatter. 
The single Gaussian dispersion profiles are, in general, flatter than those of the narrow and broad components. In most cases, the dispersion profiles in the outer disks do not drop as fast as the star formation profiles, derived in the literature. This indicates the importance of other energy sources in driving $\sigma_{\rm{H\,{\textsc i}}}$ in the outer disks. The radial surface density profiles of spirals and dwarfs are similar. The surface density profiles of the narrow component decline more steeply than those of the broad component, but not as steep as what was found previously for the molecular component. As a consequence, the surface density ratio between the narrow and broad components, an estimate of the mass ratio between cold H\,{\sc i} and warm H\,{\sc i}, tends to decrease with radius. On average, this ratio is lower in dwarfs than in spirals. This lack of a narrow, cold H\,{\sc i} component in dwarfs may explain their low star formation activity. 
\end{abstract}
\keywords{galaxies: dwarf - galaxies: fundamental parameters - galaxies: ISM - galaxies: spiral - ISM: structure - radio lines: galaxies}
\section{introduction}
Studies of the shapes of the velocity profiles of neutral atomic hydrogen (H\,{\sc i}) offer valuable 
information about the kinematical properties and phase structure of the interstellar medium (ISM) of galaxies
\citep{vanderkruit82,vanderkruit84,shostak84,younglo96,younglo97a,younglo97b,braun97,
deblokwalter06,petricrupen07}. However, the interpretation of the shapes of H\,{\sc i} velocity profiles requires careful analysis to avoid confusion between the intrinsic shapes of the profiles, data artefacts and effects related 
to beam smearing or major bulk motions. The knowledge of the spatial distribution, variation and radial properties of the shapes 
of the profiles are important  to understand the inter-relationship between gas properties and energetic processes such 
as star formation. In a previous paper \citep{ianjamasimananaetal12}, we reported on the shapes of 
stacked H\,{\sc i} velocity profiles, averaged over the disks of galaxies from The H\,{\sc i} Nearby Galaxy Survey \citep[THINGS,][]{walter08}. 

This paper investigates the shapes of the stacked profiles as a function of radius. Earlier work has also reported on this. 
For example, \citet{kamphuissancisi93} and \citet{petricrupen07} found radially decreasing H\,{\sc i} velocity dispersions in the face-on galaxies NGC 6946 and  NGC 1058, respectively. \citet{braun97} analysed the H\,{\sc i} 
emission spectra of 11 nearby spiral galaxies 
at a resolution of $\sim$ 150 pc and found filamentary structures, forming a so-called \textit{High Brightness Network} (HBN). 
This HBN has a narrow line core with a velocity full width at half maximum (FWHM) of less than 
$\sim$ 6 $\rm{km~s^{-1}}$, superposed 
on broad Lorentzian wings having a velocity FWHM extending up to $\sim$ 30 $\rm{km~s^{-1}}$. 
The HBN contains about 60\% - 90\% of the total H\,{\sc i} flux within the optical radius, $\rm{r_{25}}$, 
but becomes less dominant or disappears in the outer disk. A simple radiative transfer 
modelling of the HBN indicates that its kinetic temperature increases with radius. \citet{braun97}
attributed the HBN to the Cold Neutral Medium (CNM) phase of the ISM.

In a recent study, \citet{tamburroetal09} analysed the H\,{\sc i} profiles of 11 face-on and 
intermediately inclined galaxies from THINGS, with the aim of understanding the 
physical mechanisms that determine the linewidth of the profiles. \citet{tamburroetal09} used 
second-moment  values (i.e. intensity-weighted standard deviation of line-of-sight 
velocities) as tracers of the random motion of 
gas in the galaxies. These values decrease with radius from 
$\gtrsim$ 20 $\rm{km~s^{-1}}$ to $\sim$ 5  $\rm{km~s^{-1}}$. \citet{tamburroetal09} also reported that their sample 
exhibits a characteristic velocity dispersion value of 10$\pm$2 $\rm{km~s^{-1}}$ 
at $\rm{r_{25}}$. By comparing the predicted energy input from different physical mechanisms 
with the gas kinetic energy inferred from the observed linewidth of the profiles, they concluded 
that star formation is the dominant mechanism that determines the width of the profiles inside 
$\rm{r_{25}}$. Outside this radius, ultra-violet (UV) heating from extragalactic sources or 
magneto-rotational instability (MRI) were proposed to be likely candidates that set the widths of the profiles.

\citet{stilpetal13} also analysed the shapes of H\,{\sc i} velocity profiles in a sample of 23 galaxies from the 
THINGS and the VLA-ANGST \citep{ott12} surveys. They applied a similar stacking method as we did in our previous analysis 
but they used a different interpretation of the shapes of the profiles. By fitting single Gaussians to the stacked profiles, 
they concluded that the profiles can be described by a central narrow peak with a velocity width of $\sim6-10~\rm{km~s^{-1}}$ 
and wings to either side of the Gaussian core containing $\sim10-15$ \% of the total flux. The wings were attributed to 
turbulence driven by star formation feedback, mainly supernova explosions.  

In our previous paper, we constructed 
high signal-to-noise (S/N) profiles, called super profiles, by stacking individual profiles. We identified a
sample of 22 galaxies from the THINGS sample where systematic effects from beam smearing, 
major bulk motions or individual asymmetric profiles play no large role. 
We refer the reader to \citet{ianjamasimananaetal12} for a detailed discussion of this. The main results are 
summarized below. The shapes of the super profiles are well described by the sum of a narrow and broad 
Gaussian components, with an average velocity dispersion of 7$\pm$2 $\rm{km~s^{-1}}$ and 17$\pm$4 $\rm{km~s^{-1}}$, respectively.
The super profile parameters (velocity dispersions and mass ratios between the broad and narrow components) 
correlate with metallicity and FUV--NUV colors.  

Here we extend this work by analysing the shapes of the super 
profiles as a function of radius in the 22 galaxies from \citet{ianjamasimananaetal12}. In Section \ref{sec:sup_radius}, we 
present  the method and motivation for the analysis.    
In Section \ref{sub:rad_disp_prof}, we present measurements of the velocity dispersion 
as a function of radius. In Section \ref{sub:comp}, we compare our 
measured velocity dispersion values with those from the literature. 
In Section \ref{sub:flux}, we analyse the surface density of the single Gaussian, the narrow and broad components 
as a function of radius. We give conclusions and summarize our results in Section \ref{sec:conc}. 

\section{Motivation and methodology}\label{sec:sup_radius}
The shapes of H\,{\sc i} velocity profiles are largely influenced by the local conditions within galaxies. Energy from stellar winds, 
supernova explosions, and magnetic fields (through MRI), is constantly injected into the ISM and its effect on 
the H\,{\sc i} can be 
investigated by analysing the shapes of the H\,{\sc i} velocity profiles. As the amount of energy from the ISM can be different 
from one location to another within galaxies, the shapes of the profiles are expected to vary accordingly. 
Here we will analyse the key parameters that characterize the shapes of the H\,{\sc i} velocity profiles as a function of radius. 
These are the velocity dispersion, the mass ratio between the narrow and broad components, as well as their surface densities.

Gas velocity dispersion is an important parameter as it quantifies the amount of random motion of gas induced by different energy sources \citep{tamburroetal09}. This random motion provides pressure support and prevents gravitational collapse of the gas. Thus, knowing the appropriate value of gas velocity dispersion is crucial for disk stability analysis. In an ideal case (i.e. purely thermal energy injection, small beam size), the velocity dispersion is a direct tracer of the gas kinetic temperature. In practice though, other sources of turbulence can increase the linewidth from its purely thermal value. 
Due to the lack of concrete knowledge about the appropriate value of the H\,{\sc i} velocity dispersion, it has usually been 
assumed to be independent of radius. In this paper, we derive the H\,{\sc i} velocity dispersion as a function of radial position 
for our sample galaxies. Note that our velocity dispersions include both the contribution of the thermal and the turbulent velocity dispersion of the 
H\,{\sc i} gas. In addition to the velocity dispersion analysis, if the narrow and broad Gaussian 
components can be associated with the CNM and the WNM, then the super profile shapes 
can be used to study the relative amount of the WNM and the CNM as a function of radius. 

We fit the super profiles with a double Gaussian function as described in detail by 
\citet{ianjamasimananaetal12}. 
We also include results from fitting with a single Gaussian. This will facilitate a direct comparison with previous estimates 
of H\,{\sc i} velocity dispersion in the literature, which did not take into account the 
presence of multiple components in H\,{\sc i}. We refer the reader to \citet{ianjamasimananaetal12} for a 
detailed description of the error analysis. In summary, we define the uncertainties in the data points of each super profiles as
 \begin{equation}\label{eq:equation1}
 \sigma = \sigma_{ch.map} \times \sqrt{N_{prof}/N_{prof,beam}}
 \end{equation}
where $\sigma_{ch.map}$ is the rms noise level in one channel map,
$N_{prof}$ is the number of stacked profiles at a certain velocity
$V$, and $N_{prof,beam}$ is the number of profiles per beam. 

Here, we investigate the shapes of the super profiles as a function of radius. We derive 
and fit super profiles within annular ellipses of 0.2 $\rm{r_{25}}$ width. We use the optical radius, $\rm{r_{25}}$, 
as indicative of the radius within which star formation is efficient. We also include scaling by the radial scale length of the stellar light, $\rm{r_{D}}$, taken from \citet{HunterElmegreen04} and \citet{leroy08}.  
In the following, we separate our sample into spirals and dwarfs following the definition by \citet{leroy08}. 
Therefore, galaxies with rotation
velocities $V_{rot}\leq125~\rm{km~s^{-1}}$, stellar masses
$M_\ast\lesssim10^{10}~M_\odot$, and absolute $B$ magnitude
$M_B\gtrsim-20~\rm{mag}$ are classified as dwarfs, whereas the more massive ones are classified as spirals. 
With this definition, our sample includes 12 dwarf and 10 spiral galaxies. 
\begin{figure*} 
\centering
    \begin{tabular}{l l l}
        \includegraphics[scale=.24]{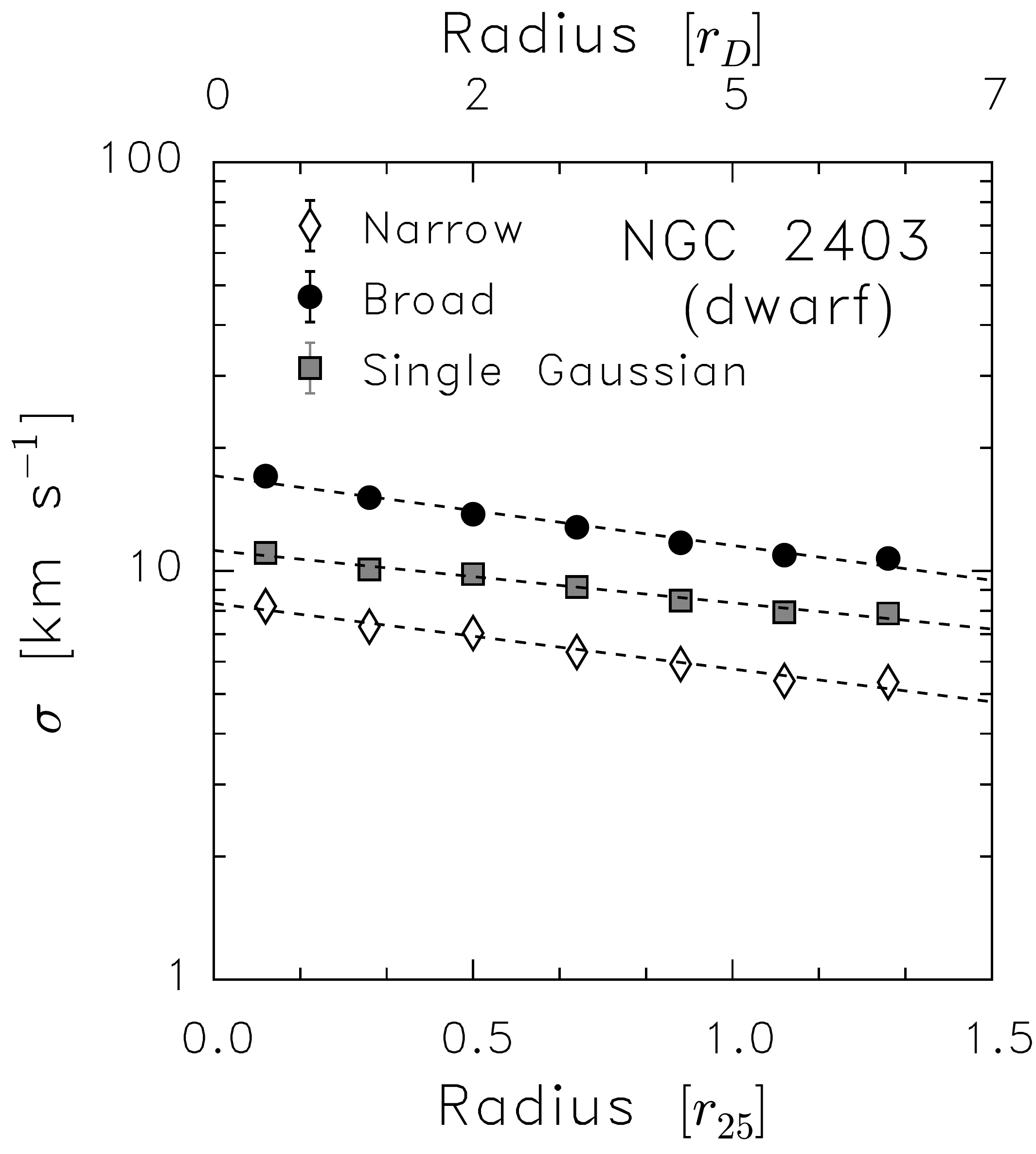}&
        \includegraphics[scale=.24]{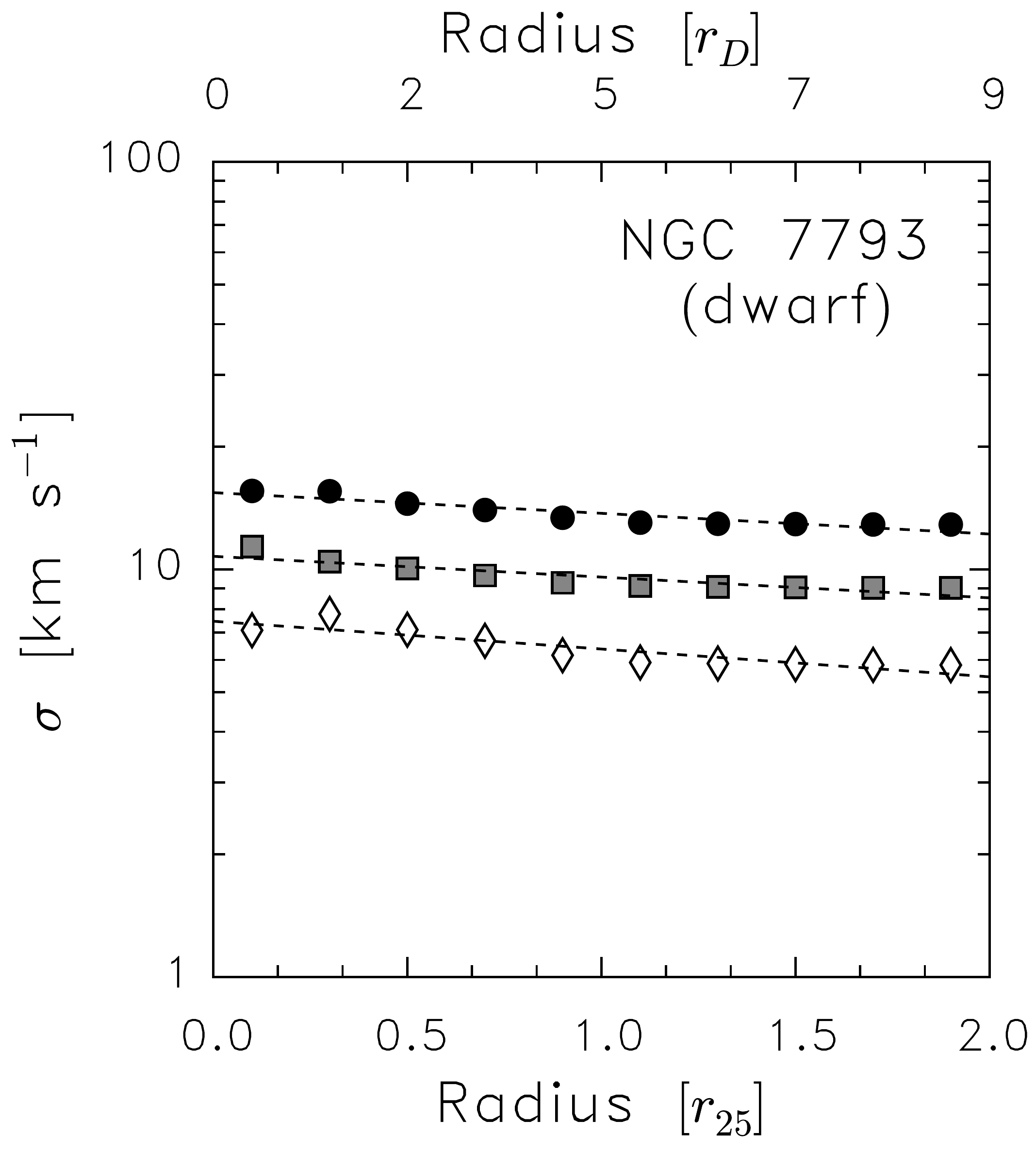}&
        \includegraphics[scale=.24]{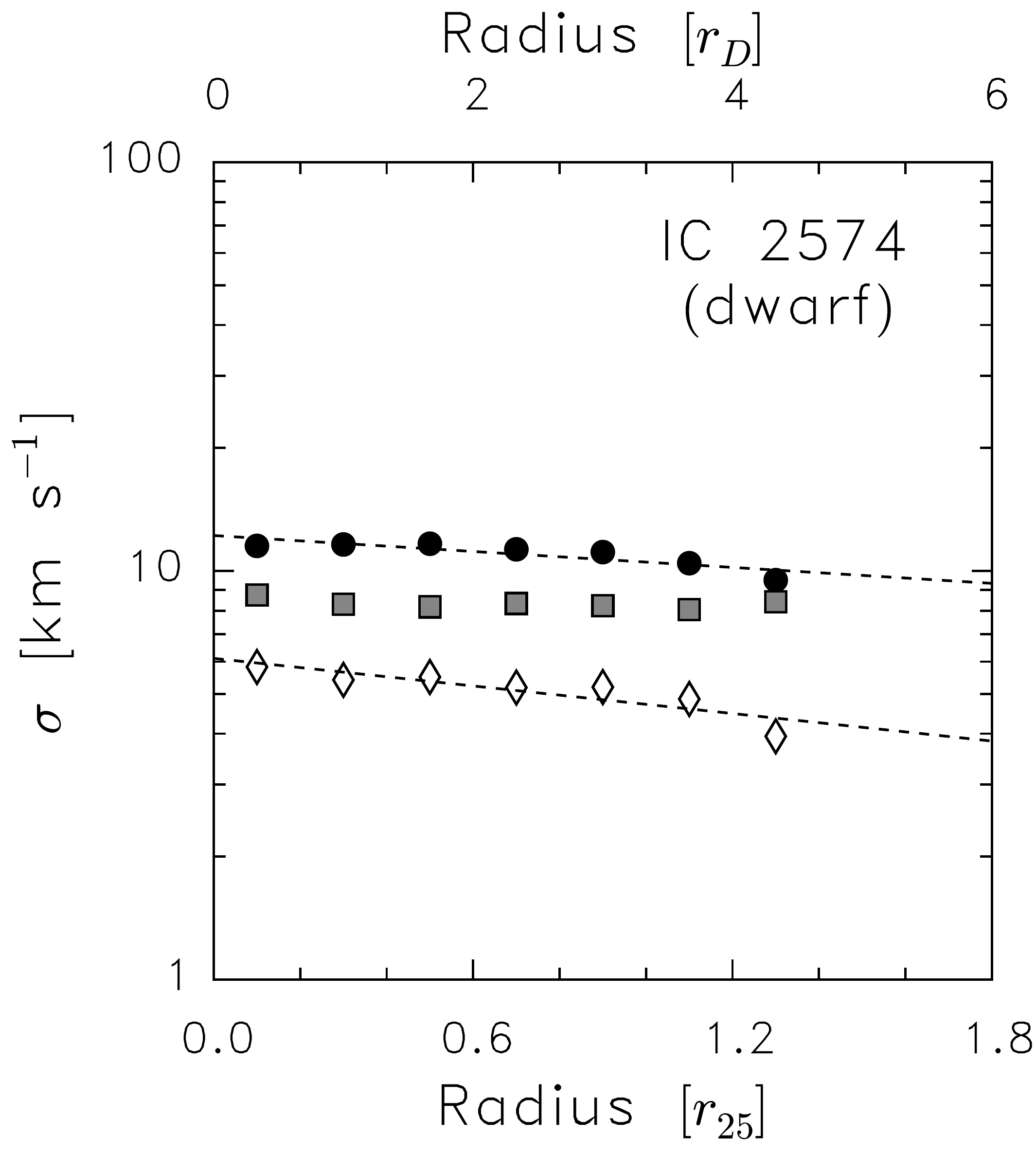}\\
        \includegraphics[scale=.24]{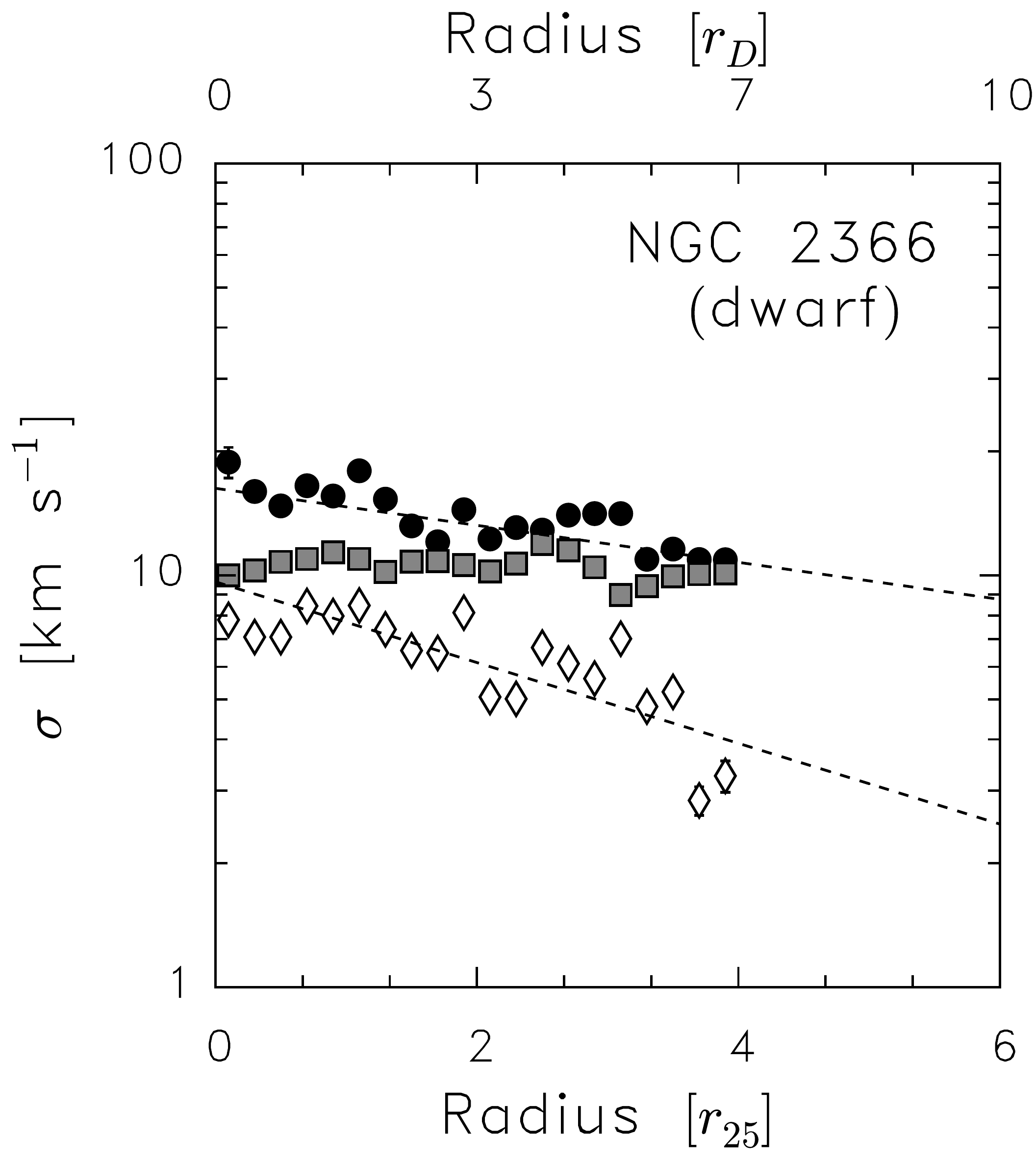}&
        \includegraphics[scale=.24]{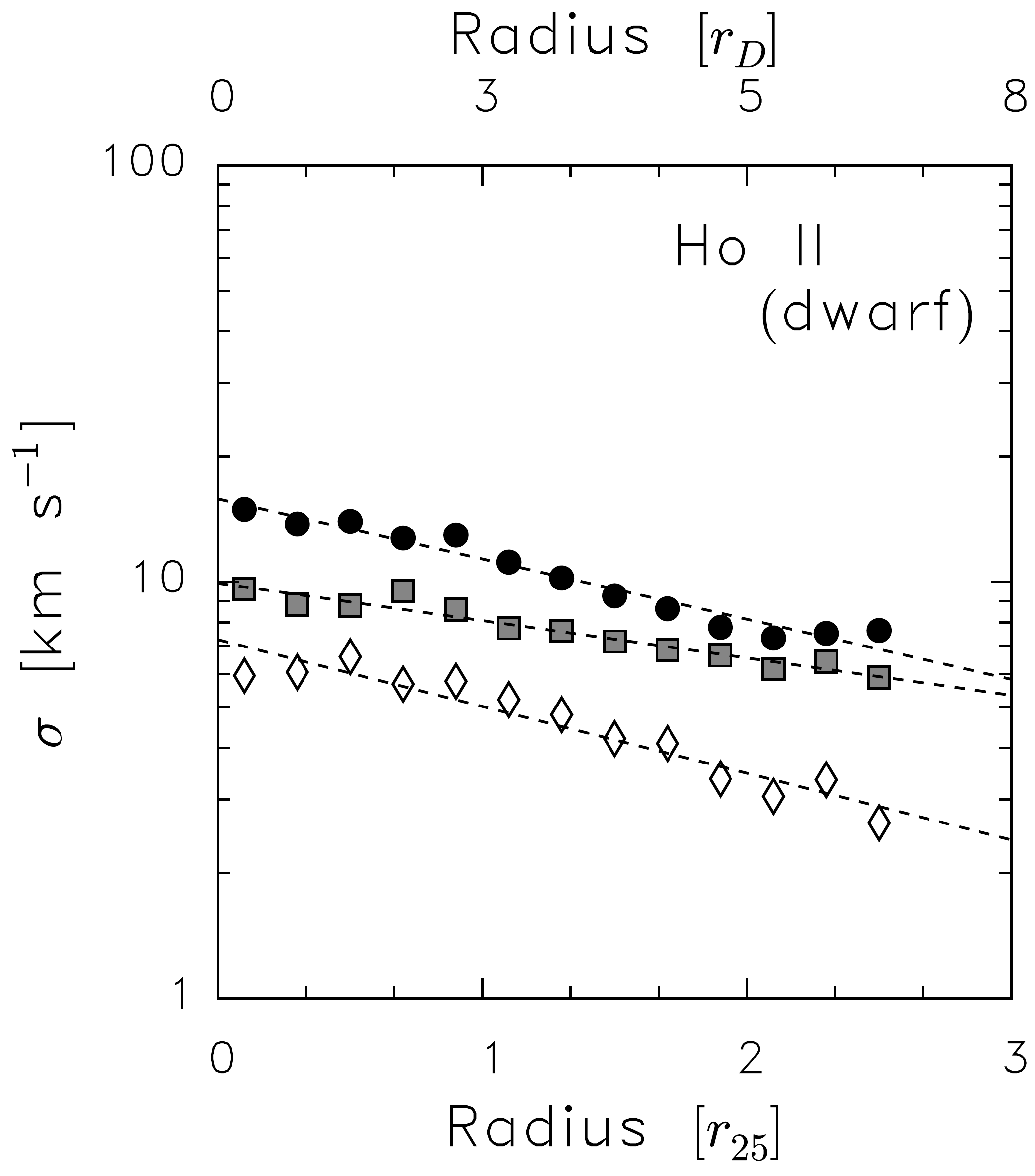}&
        \includegraphics[scale=.24]{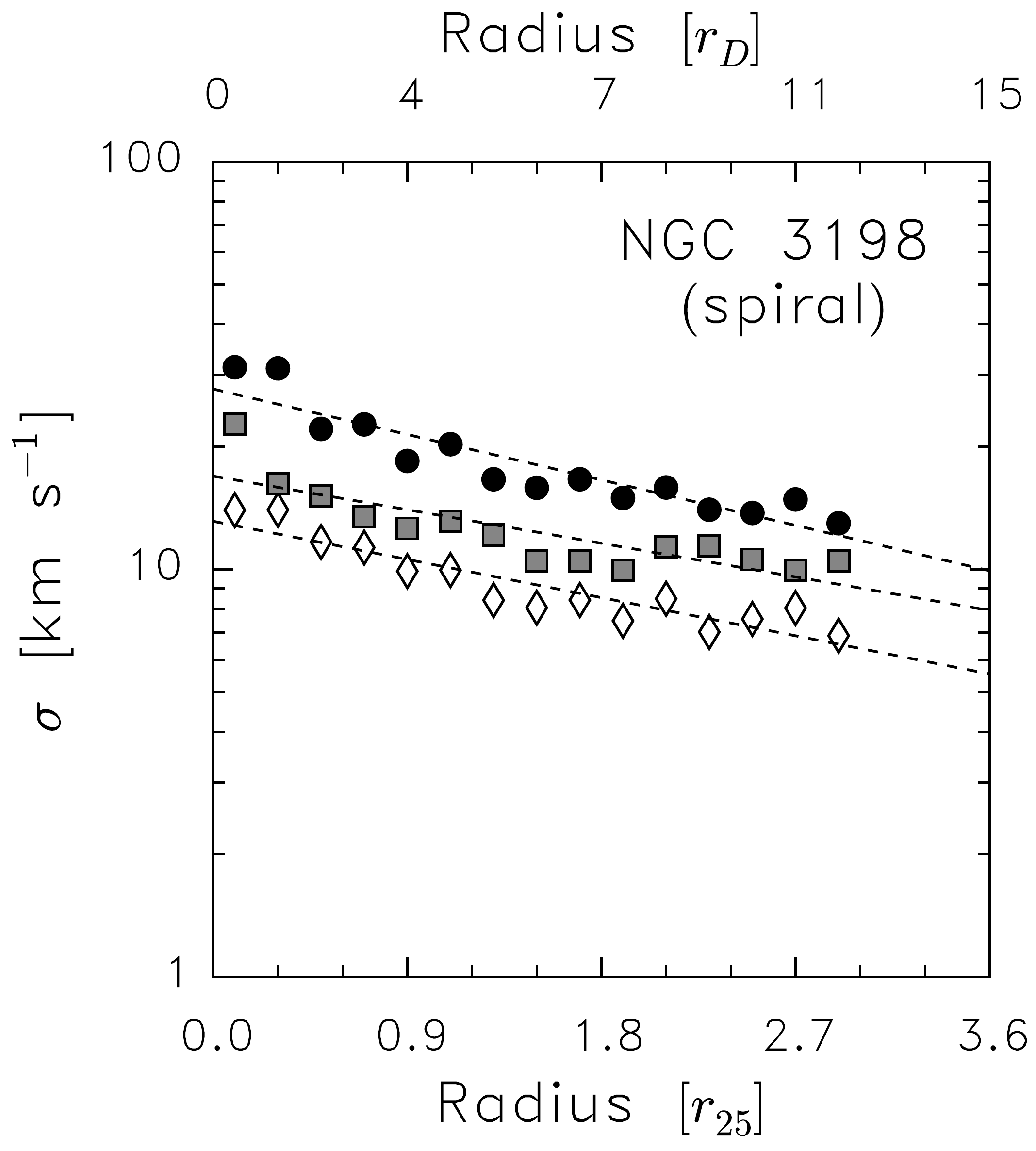}\\
        \includegraphics[scale=.24]{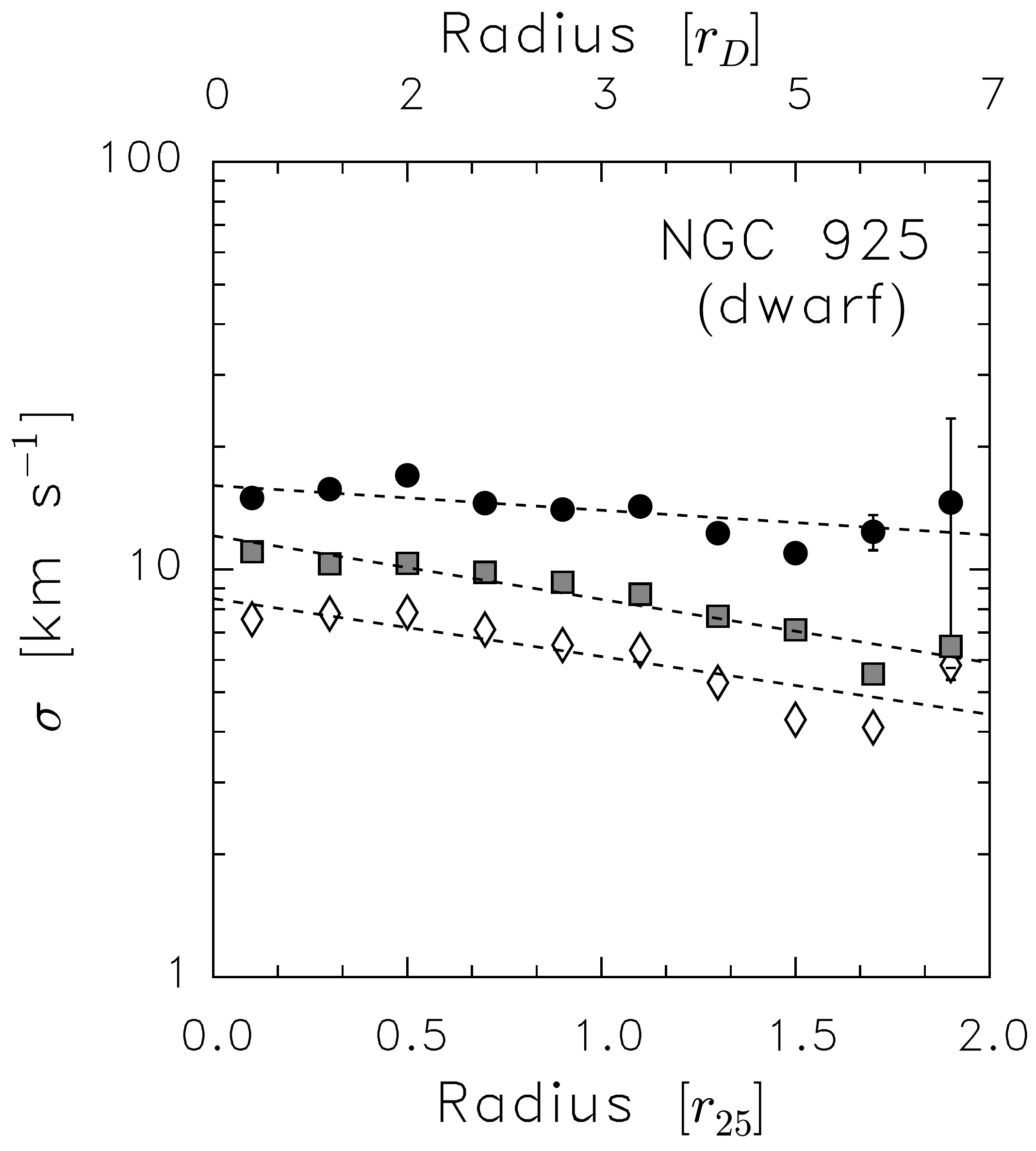}&
        \includegraphics[scale=.24]{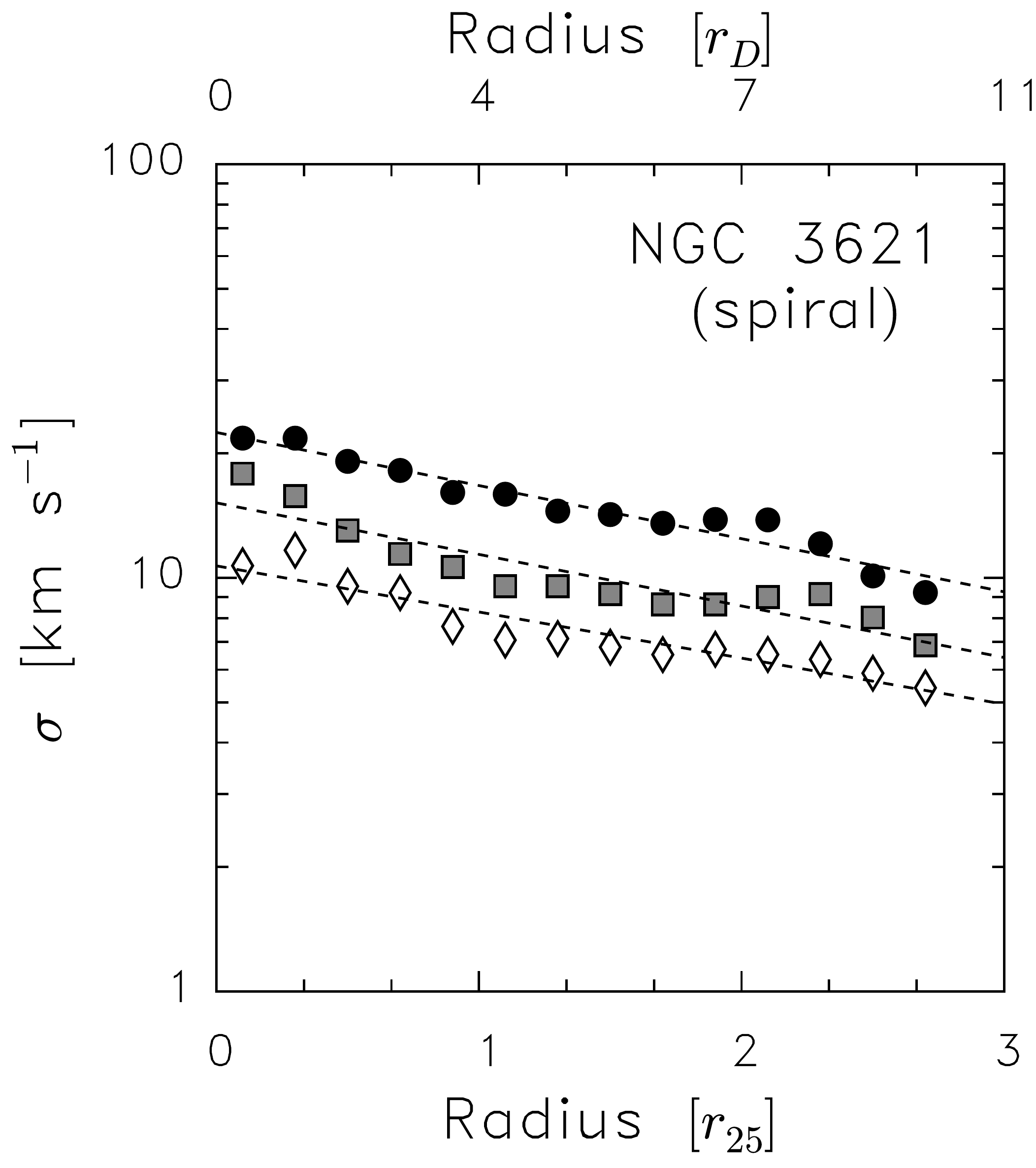}&
        \includegraphics[scale=.24]{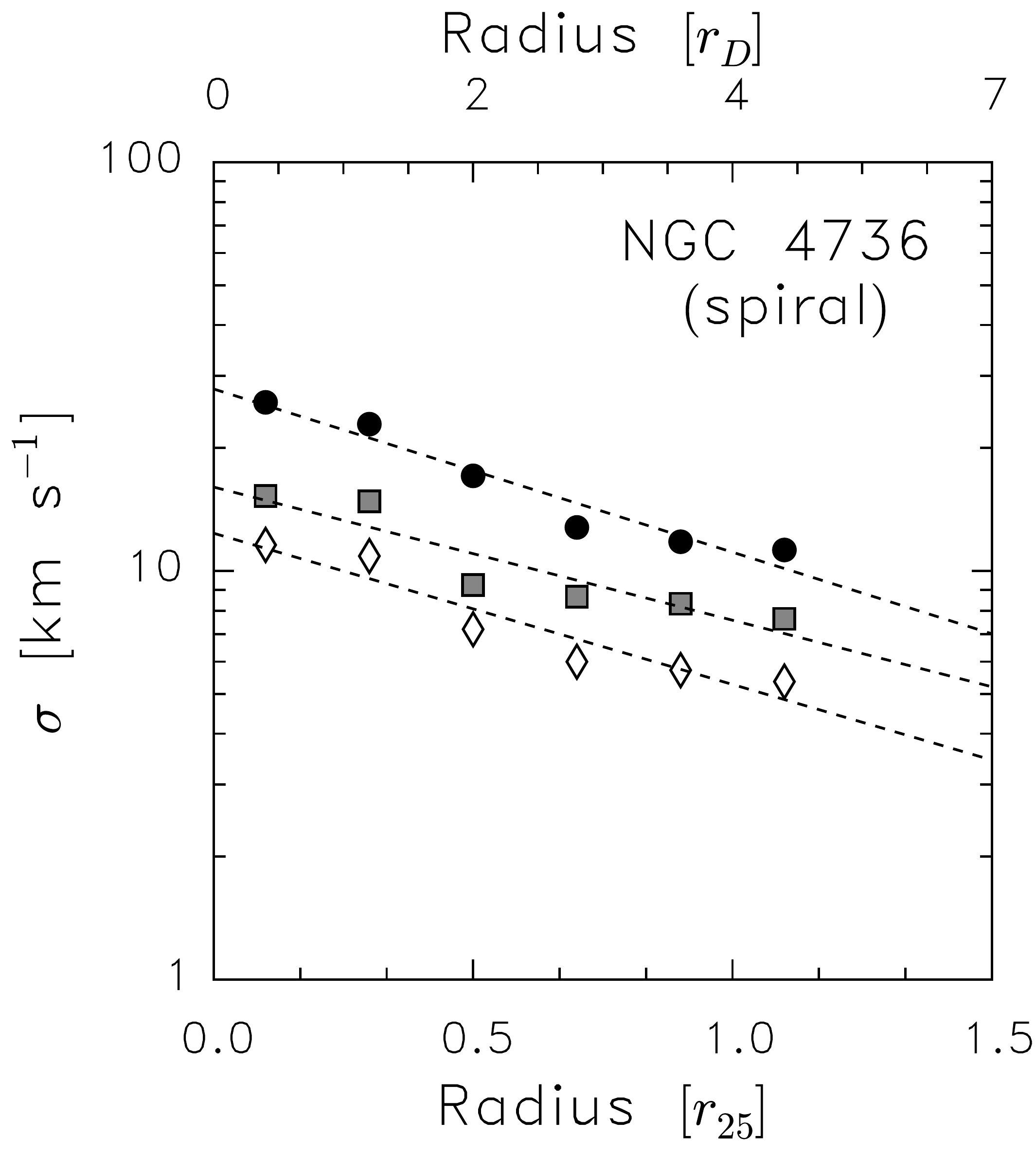}\\
        \includegraphics[scale=.24]{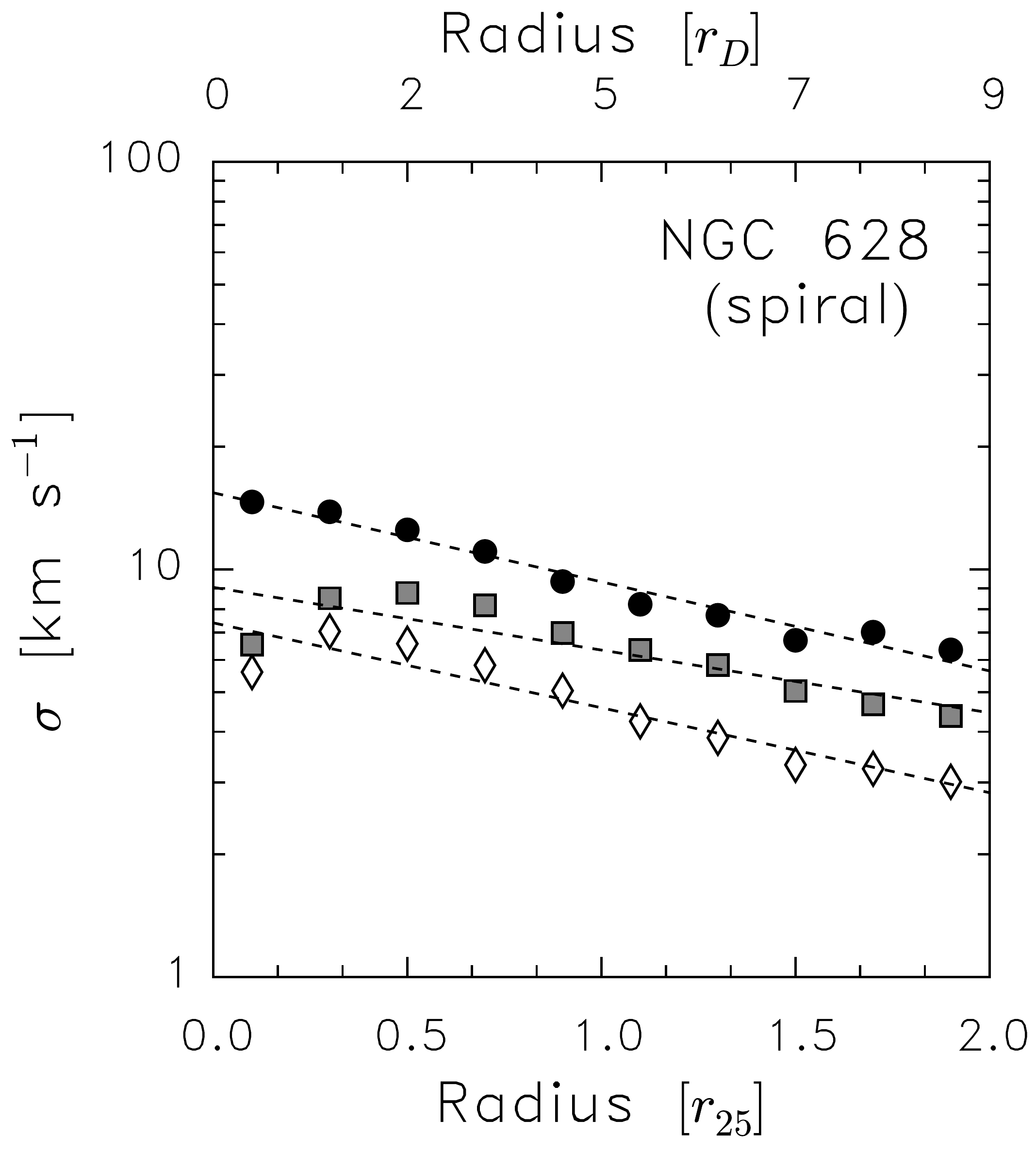}&
        \includegraphics[scale=.24]{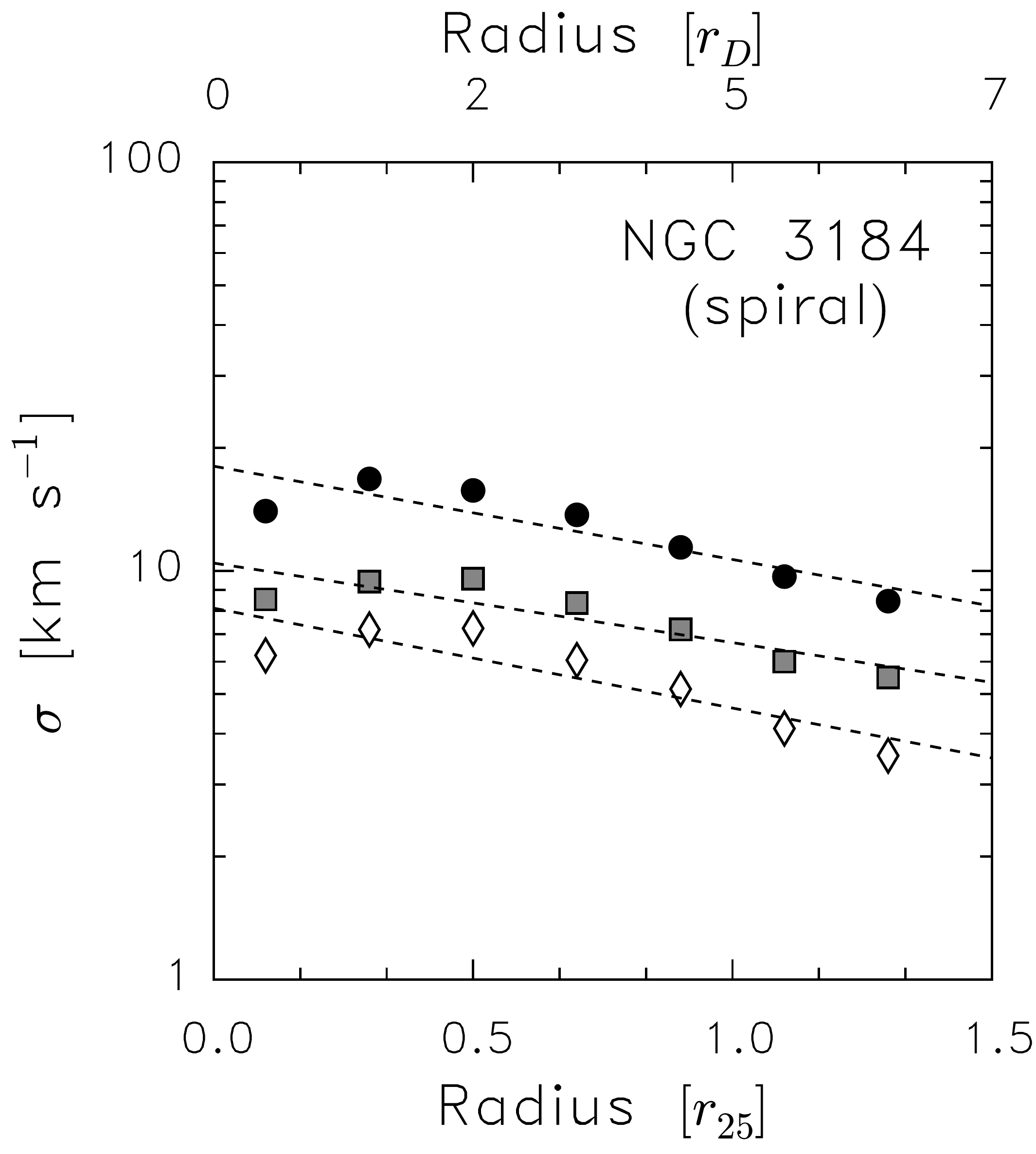}&
        \includegraphics[scale=.24]{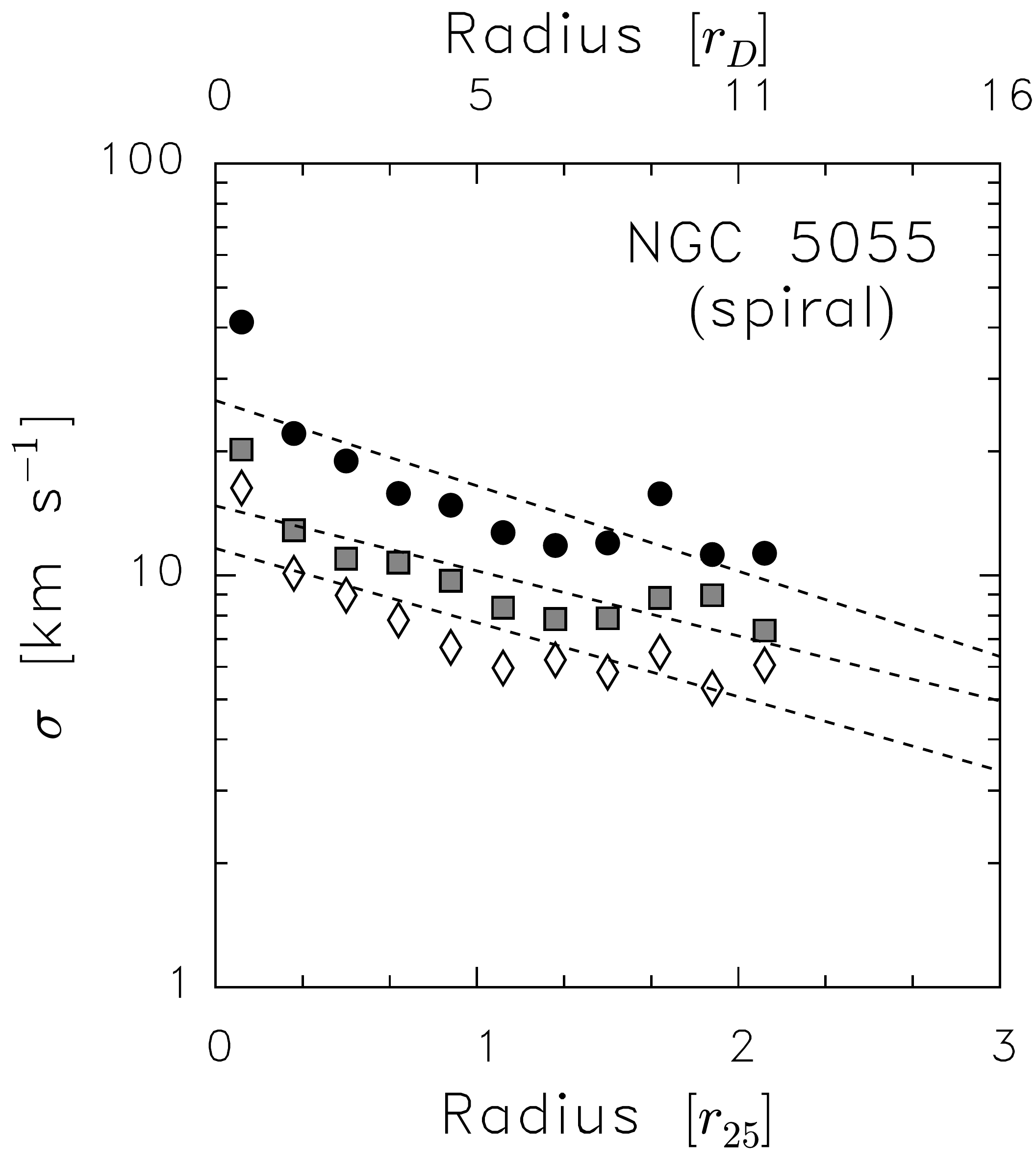} 
   \end{tabular}
\caption{Velocity dispersions as a function of radius normalised by 
$\rm{r_{25}}$ and $\rm{r_{D}}$. Black solid circle symbols represent the broad 
component velocity dispersions. The square gray symbols represent the velocity dispersions from the 
single Gaussian fits and the open diamond symbols show the velocity 
dispersions of the narrow component. The dashed lines are exponential fits. 
The uncertainties on the fitted velocity dispersion values 
are shown as error bars, though they are usually smaller than the size of the symbols.}
\label{fig:rad_disp_fit}
 \end{figure*}  
\capstartfalse
\begin{figure*}
\centering
\begin{tabular}{l l l}
    \includegraphics[scale=.24]{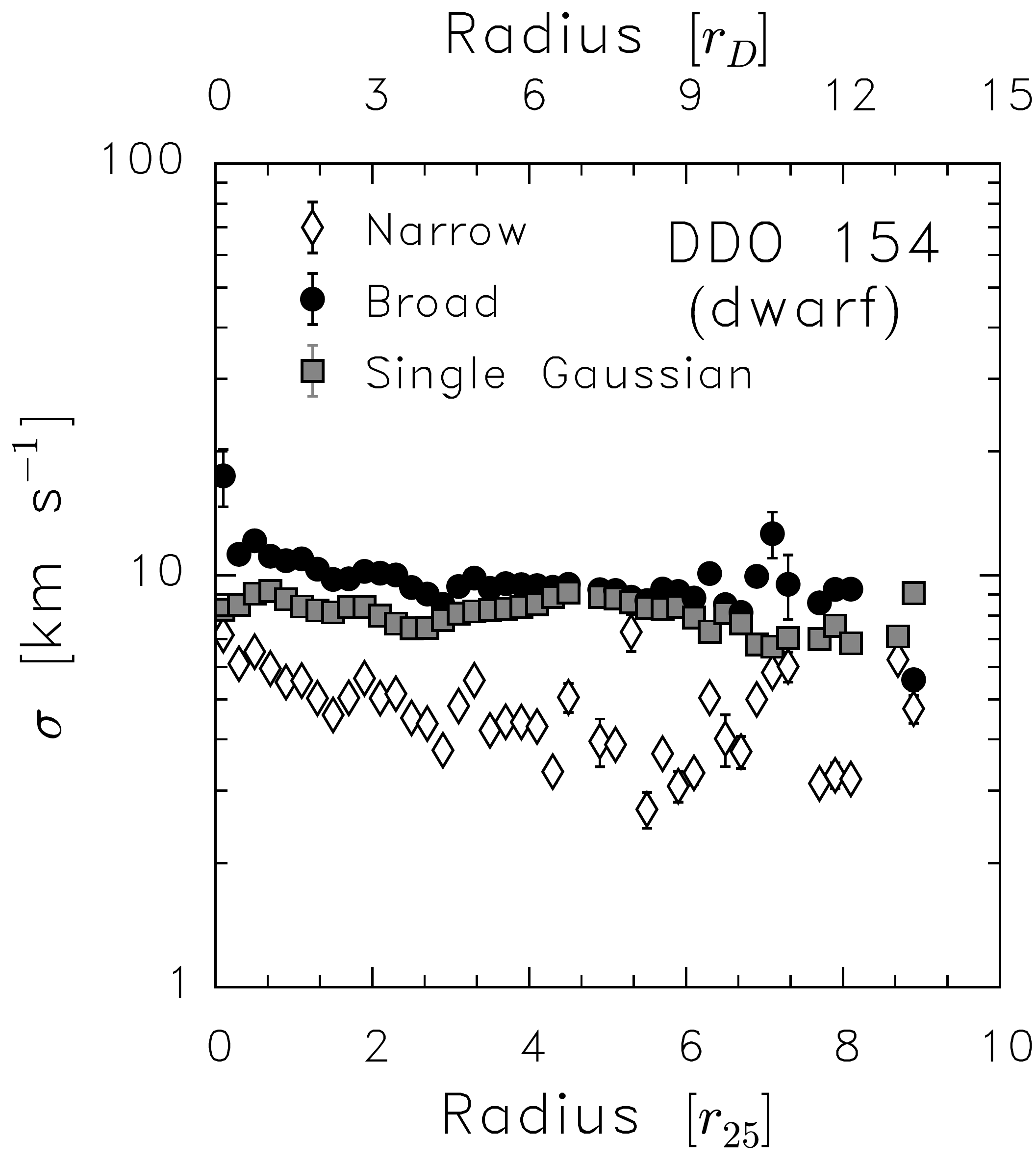}&
    \includegraphics[scale=.24]{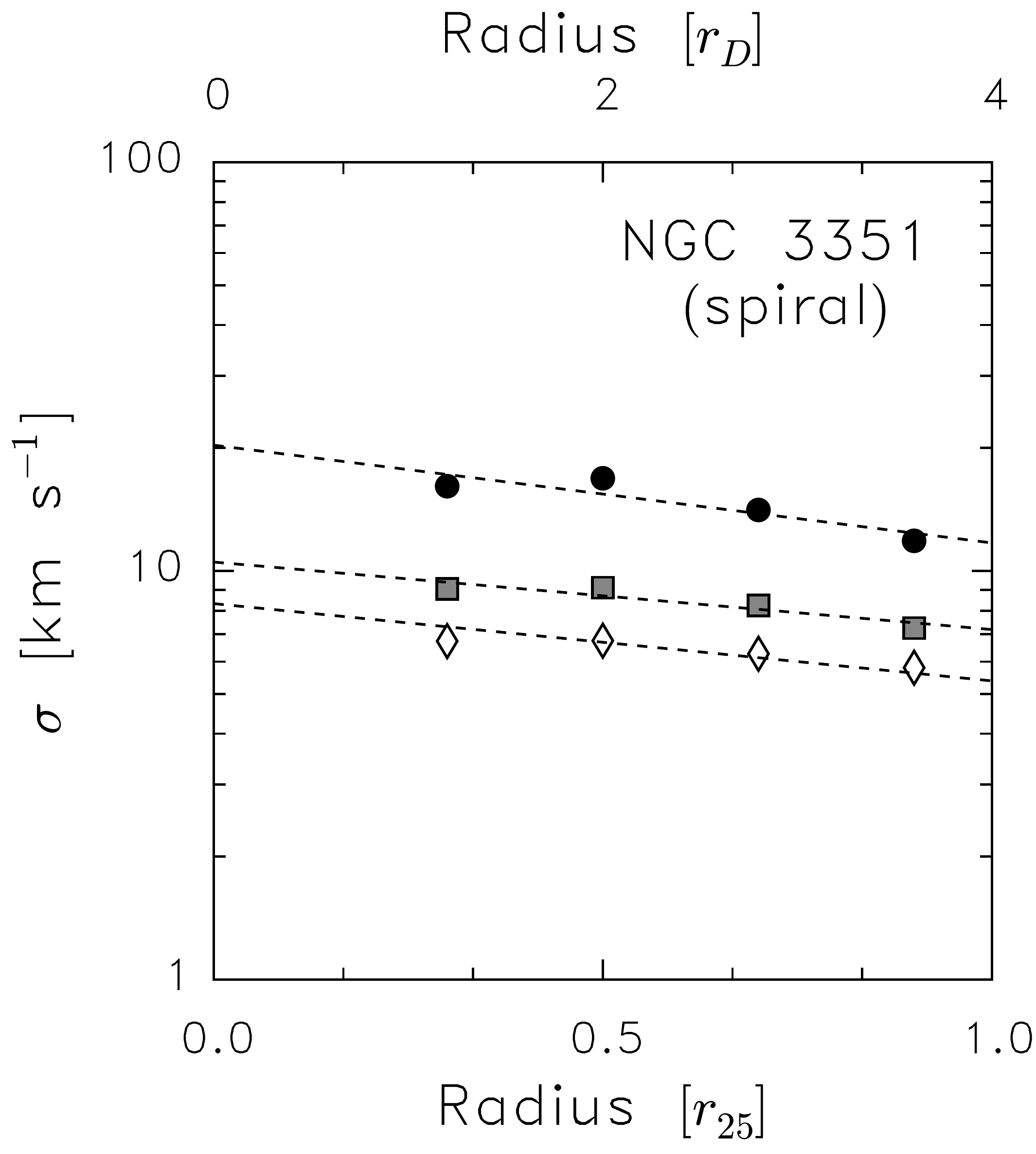}&
    \includegraphics[scale=.24]{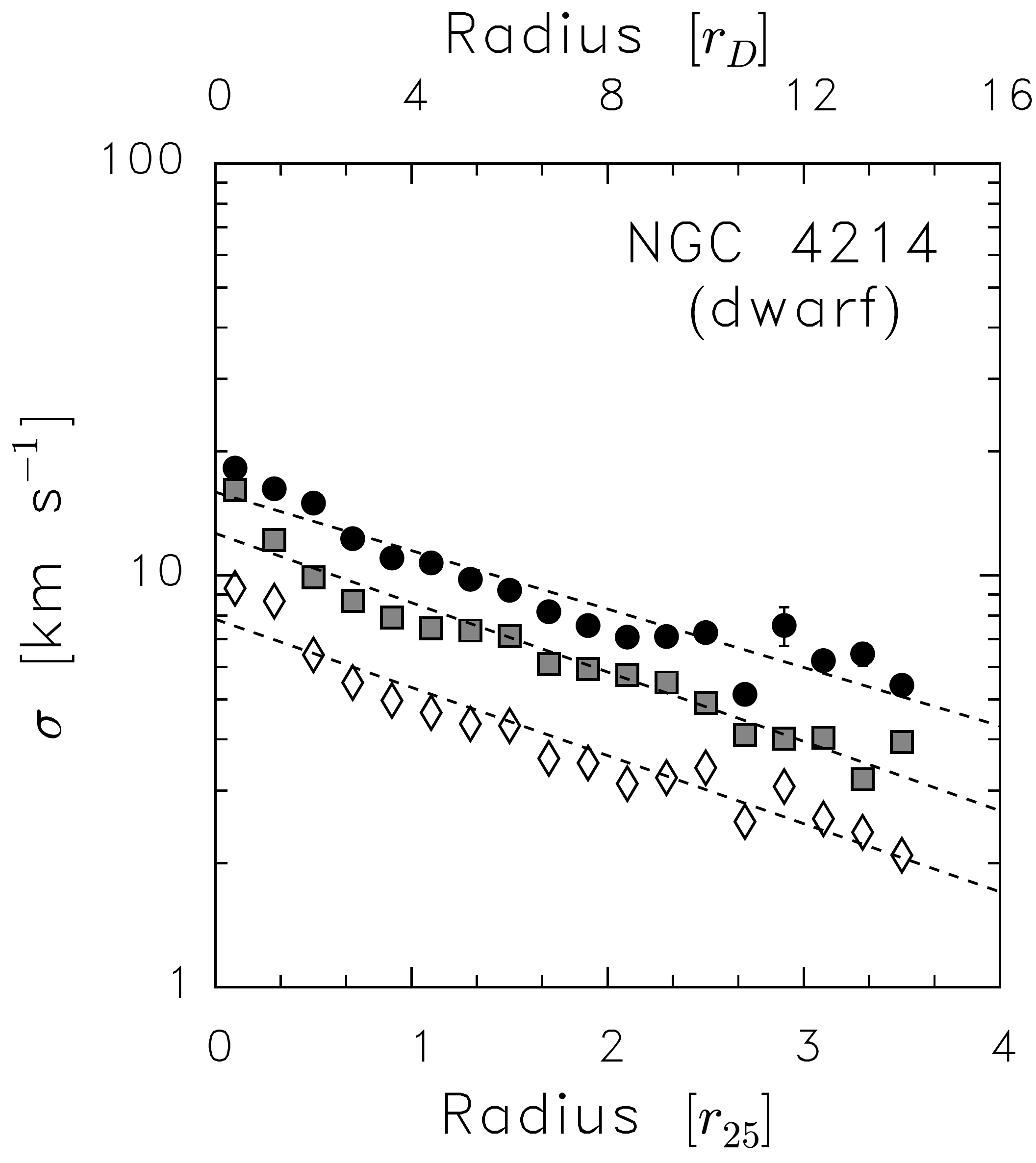}\\
    \includegraphics[scale=.24]{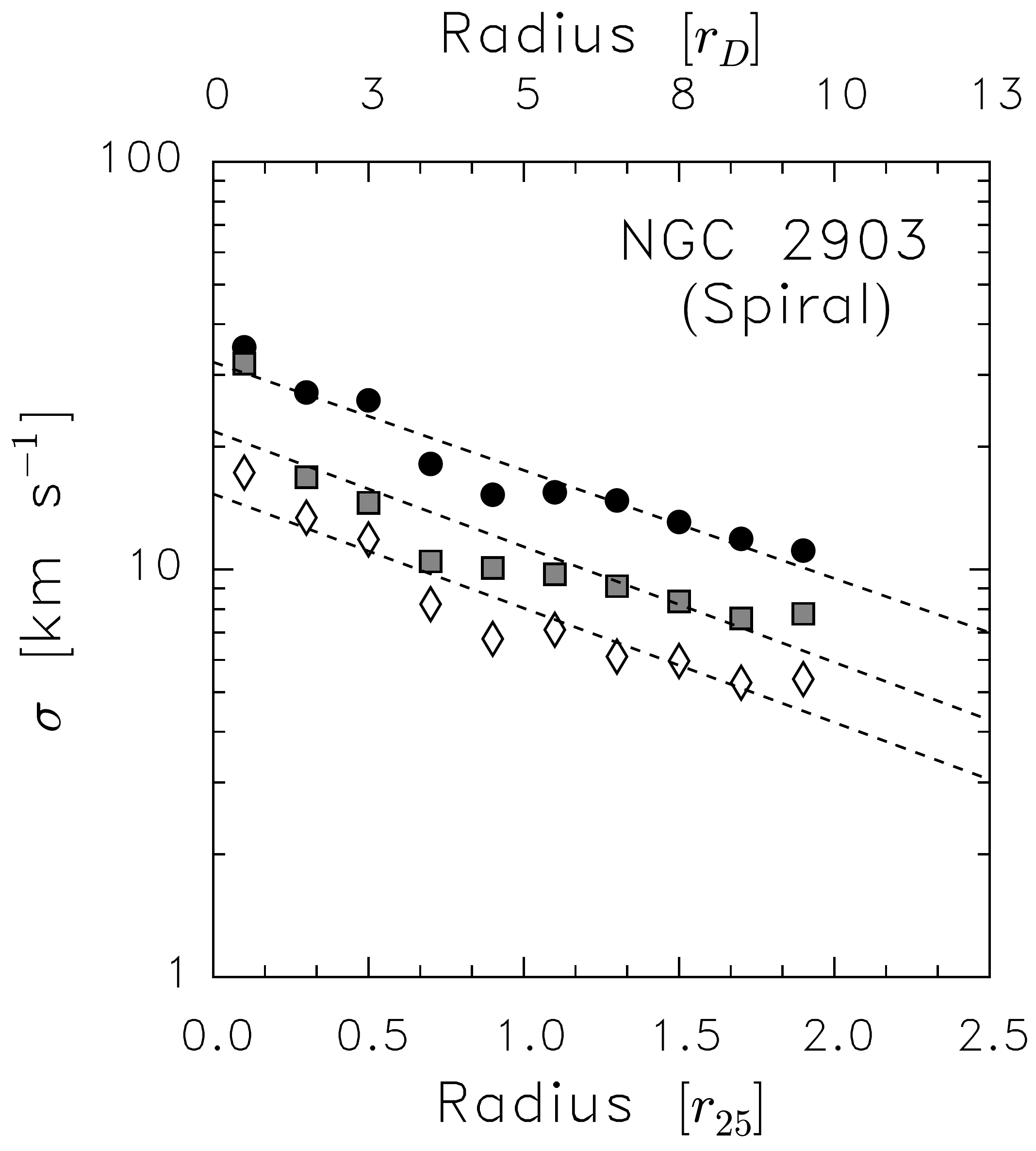}&
    \includegraphics[scale=.24]{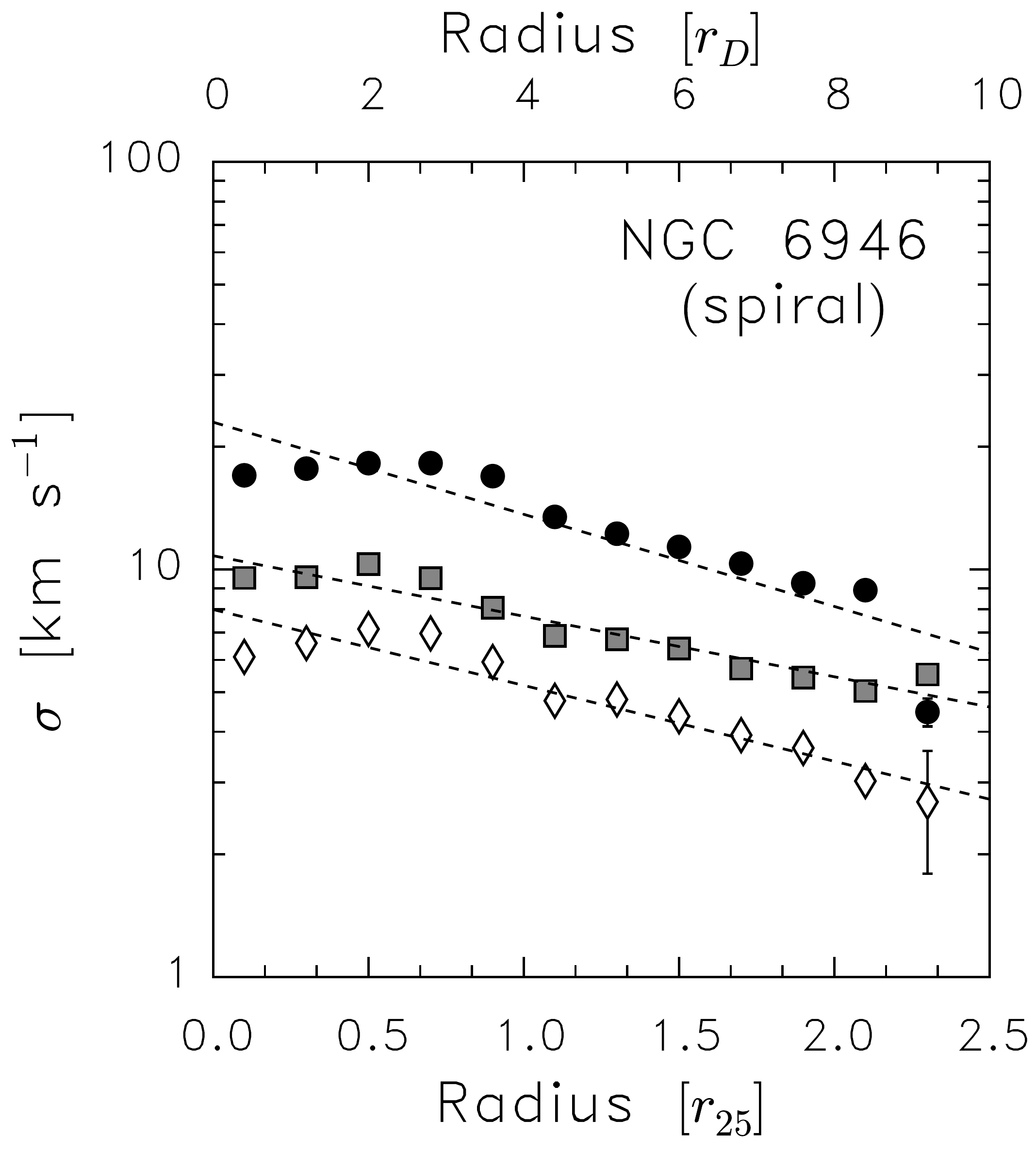}&
    \includegraphics[scale=.24]{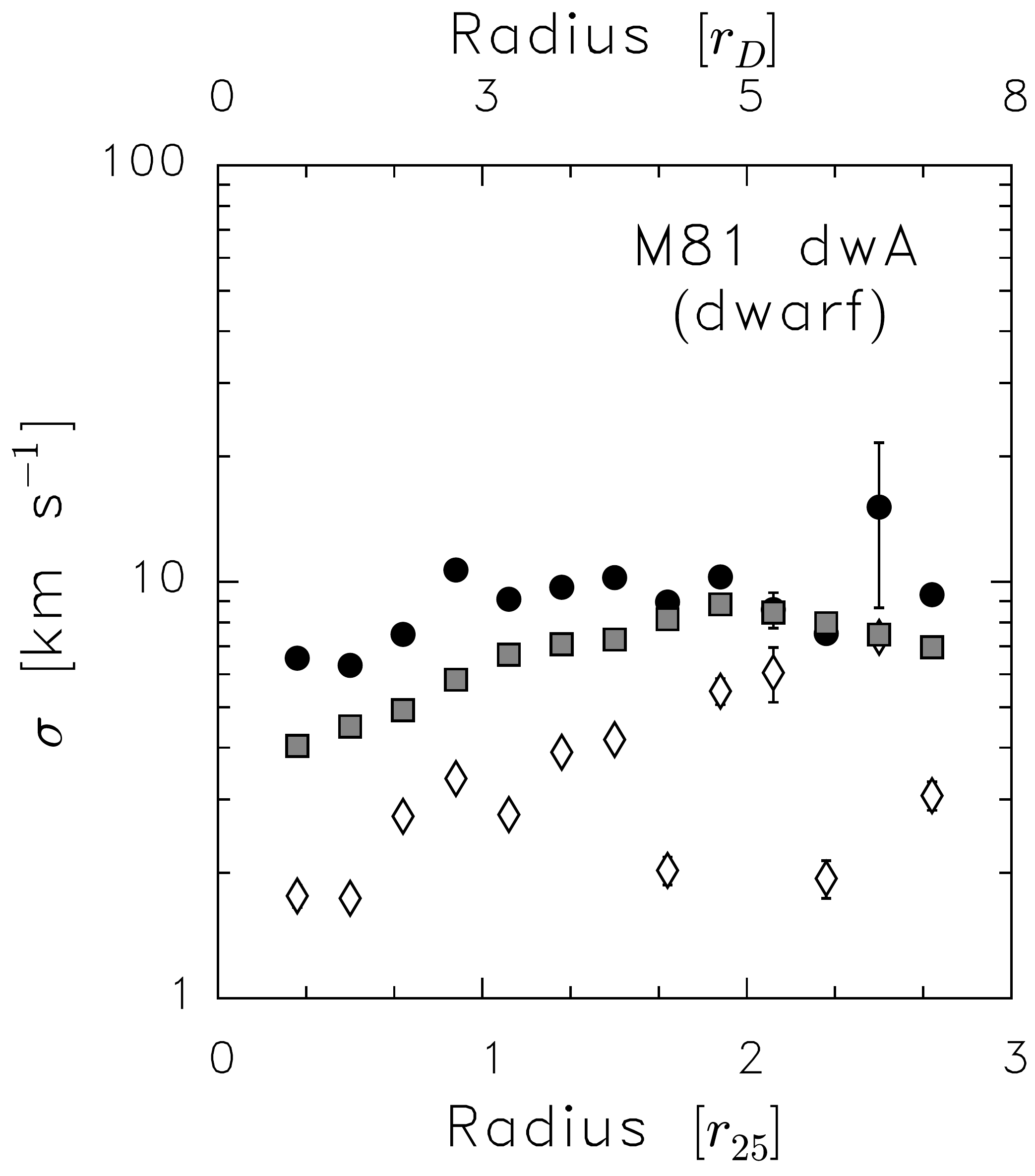}\\
    \includegraphics[scale=.24]{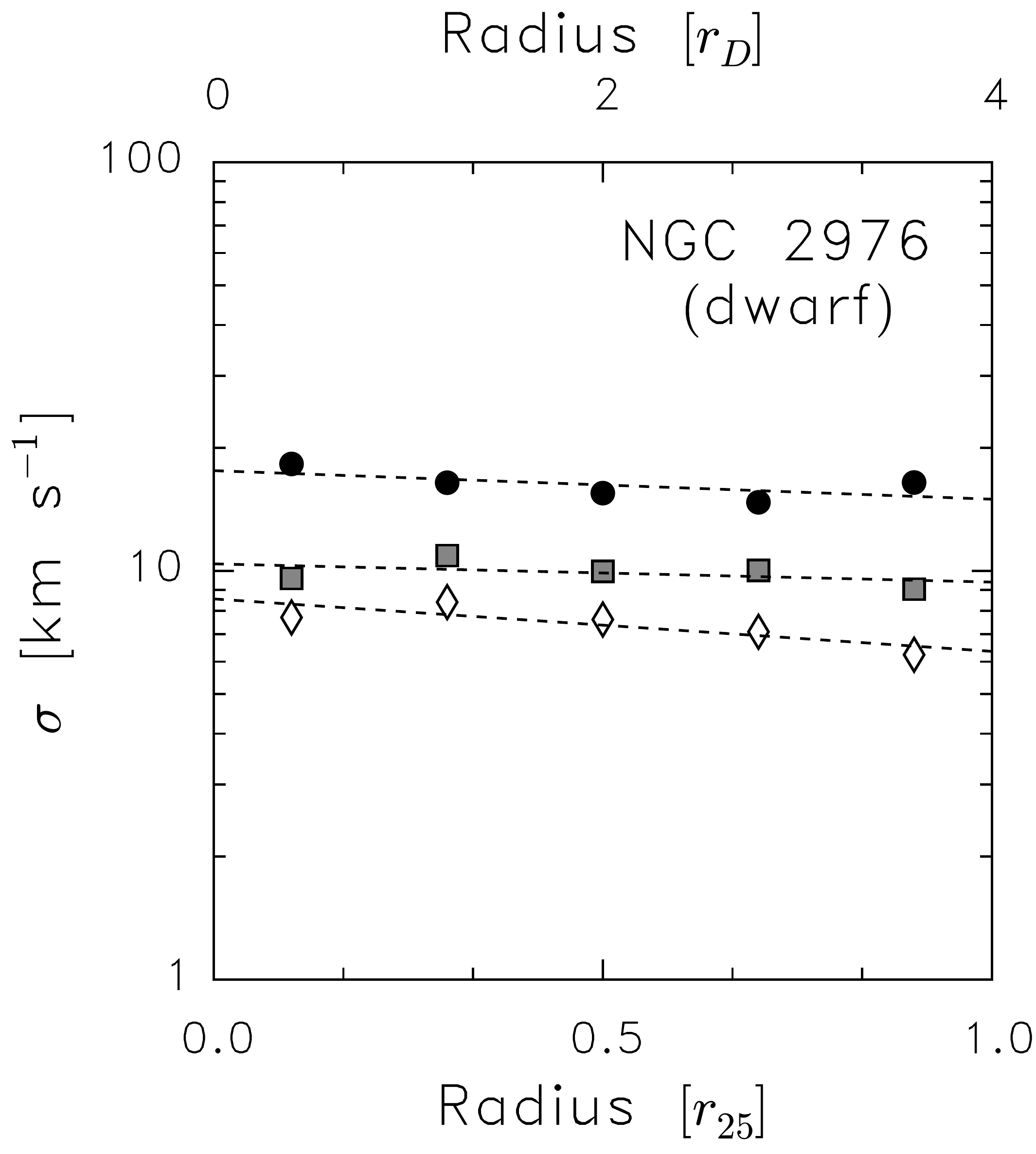}&
    \includegraphics[scale=.24]{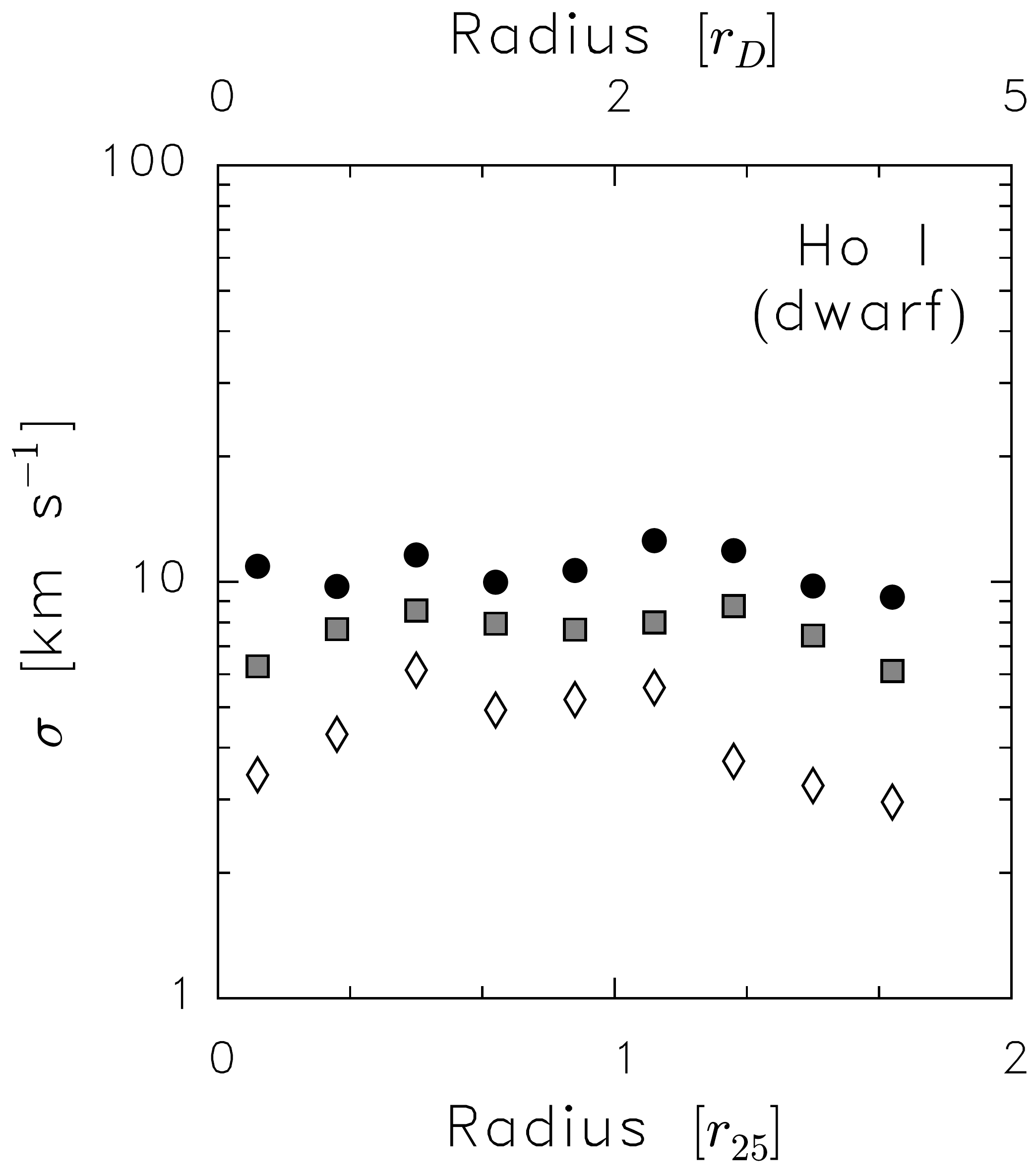}&
    \includegraphics[scale=.24]{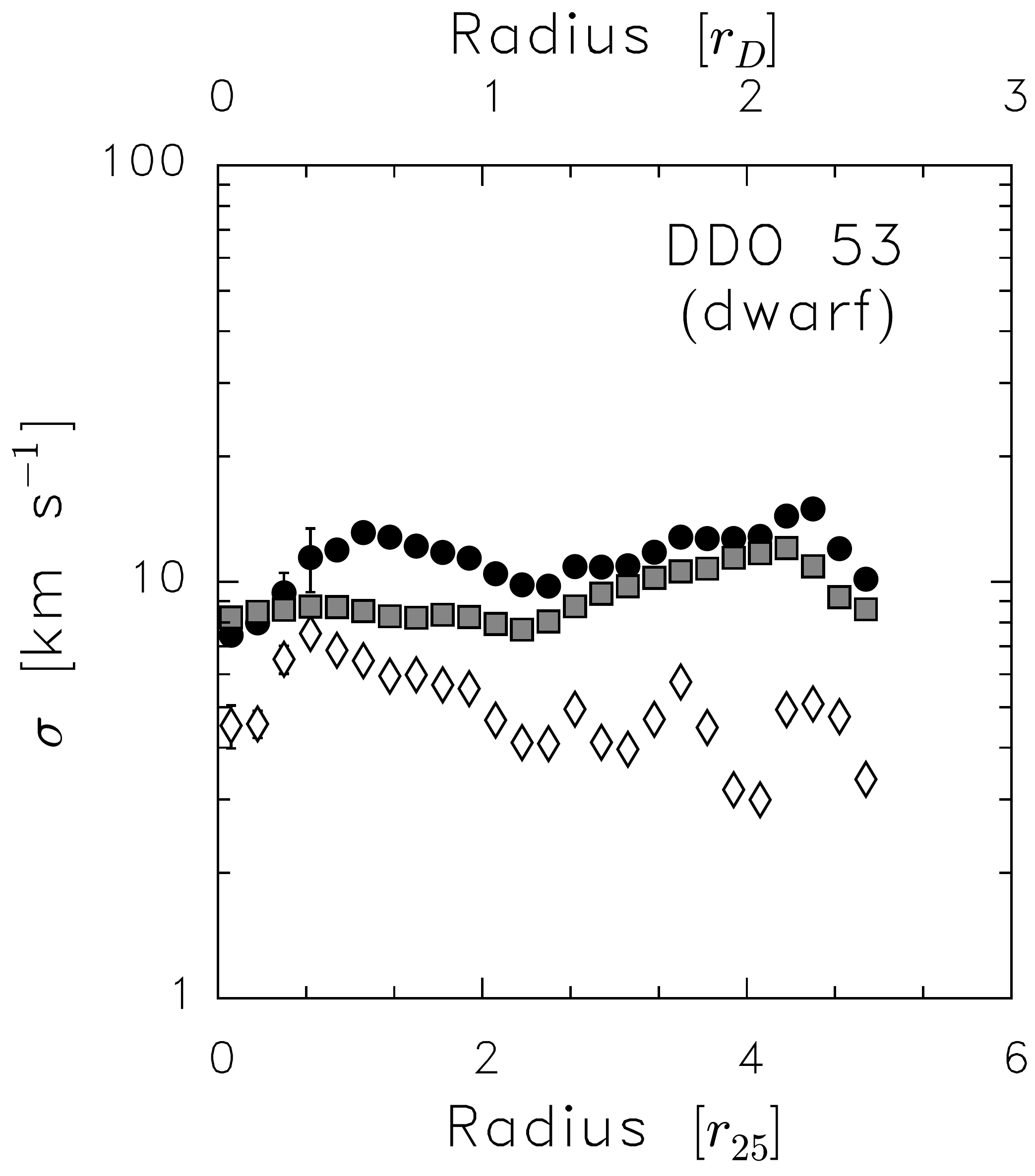}\\
    \includegraphics[scale=.24]{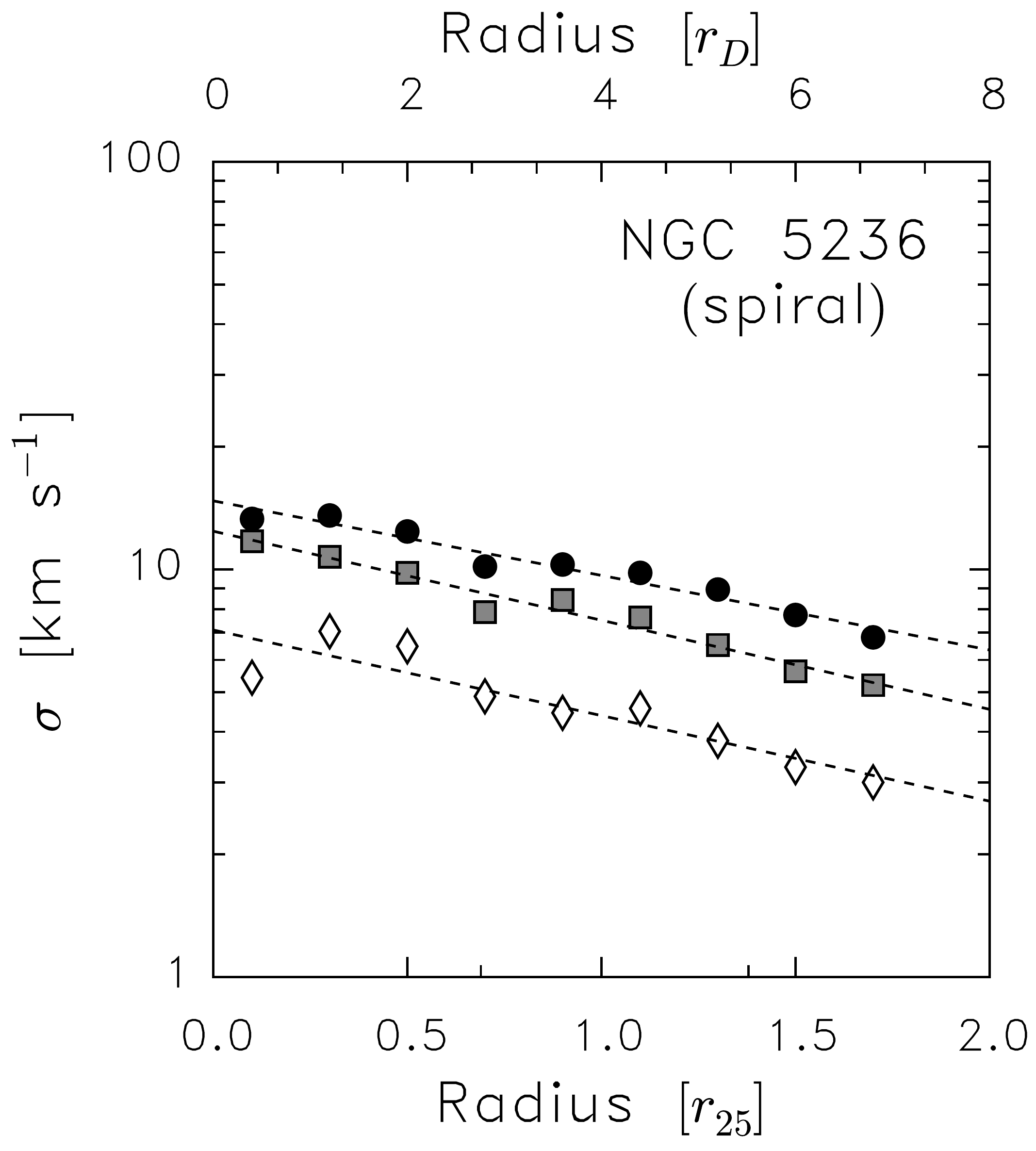}
\end{tabular}
\\\textbf{Figure \ref{fig:rad_disp_fit}}: (Continued).
\end{figure*} 

\section{Results}
\subsection{Velocity dispersion as a function of radius}\label{sub:rad_disp_prof}
Figure~\ref{fig:rad_disp_fit}  shows the single Gaussian, narrow, and broad components' velocity dispersions 
as a function of radius, normalised by the optical radius, $\rm{r_{25}}$, and the radial scale length, $\rm{r_{D}}$. 
In most cases, the velocity dispersions decrease exponentially with radius. 
The single Gaussian velocity dispersion 
profiles tend to be flatter than those of the narrow and broad components. 
This behaviour can be understood as follows. 
In the inner disks, both narrow and broad individual line profiles
make equally important contributions to the shapes of the derived super profiles, 
resulting in super profiles with a narrow peak and broad wings. 
Single Gaussian fits to these super profiles give velocity dispersion values between those of the narrow and broad components. 
As  we go further out in the disk, the contribution of the narrow component to the overall 
shape of the super profiles becomes less important and the shapes of the super profiles become more Gaussian 
(see Section \ref{sub:flux}). In this case, single Gaussian fits to the super profiles result in 
velocity dispersion values close to those of the broad components. 
In conclusion, the change in the relative importance of the narrow and broad components seems to be the reason why here 
and in previous work \citep[e.g.,][]{leroy08}, the H\,{\sc i} velocity dispersions derived from single Gaussian fitting of H\,{\sc i} profiles 
hardly decline with radius.

We fit the radial velocity dispersion profiles with an exponential function of 
the form $\rm{\sigma_{0}\exp(-\frac{r}{\textit{l}})}$, where $\sigma_{0}$ is the fitted 
value of the central velocity dispersion, and \textit{l} is the scale length. 
The results of the fits are shown as dashed lines 
in Figure~\ref{fig:rad_disp_fit}. Spiral galaxies exhibit a clear radial decline in velocity dispersion, which can be 
well fitted by the exponential function. However, dwarf galaxies show flatter radial velocity dispersion profiles. 
Ho I, M81 dwA, DDO 53, and DDO 154 have flat profiles, giving unrealistically large scale lengths ($l~\gg~\rm{r_{25}}$). 
For this reason, the fit for these galaxies are not shown in Figure~\ref{fig:rad_disp_fit}. We list the scale length values for 
each galaxies in Table~\ref{tab:scale_length}.  Note that, for some galaxies, 
the results of the exponential fits depend on the radial range being fitted. For example, for NGC 3621, NGC 5055, 
NGC 2903, and NGC 3198 the radial decline starts to flatten at about the optical radius. Thus fitting only those 
points inside $\rm{r_{25}}$ will give a steeper slope. Here we fit the entire radial range.  

\capstartfalse

\begin{deluxetable}{l c c c c c}
\tabletypesize{\scriptsize}
\tablewidth{0pt}
\tablecaption{Fitted velocity dispersion scale lengths \label{tab:scale_length}}
\tablehead{
	\multicolumn{1}{c}{Galaxy}& $\rm{r_{25}}$ &$\rm{r_{D}}$ &\multicolumn{1}{c}{$l_{1g}$}&
	\multicolumn{1}{c}{$l_{n}$}&\multicolumn{1}{c}{$l_{b}$}\\
	&(kpc)&(kpc)&\multicolumn{1}{c}{$(\rm{r_{25}})$}&$(\rm{r_{25}})$&$(\rm{r_{25}})$\\
	\multicolumn{1}{c}{1} & \multicolumn{1}{c}{2} &\multicolumn{1}{c}{3} &\multicolumn{1}{c}{4}&\multicolumn{1}{c}{5}
}
\startdata 
NGC 2976 & 3.8 & 0.9 & 9.7 $\pm$ 10.6 & 3.4 $\pm$ 1.3 & 6.2 $\pm$ 4.4 \\ 
NGC 2366 & 2.2 & 1.3 & 70.4 $\pm$ 84.1 & 4.4 $\pm$ 0.9 & 9.7 $\pm$ 2.0 \\ 
HOII & 3.3 & 1.2& 4.8 $\pm$ 0.4 & 2.7 $\pm$ 0.2 & 3.0 $\pm$ 0.2 \\ 
NGC 4214 & 2.9 & 0.7& 2.6 $\pm$ 0.2 & 2.6 $\pm$ 0.2 & 3.1 $\pm$ 0.3 \\ 
\textbf{NGC 3184} & 12.0 & 2.4 & 2.2 $\pm$ 0.5 & 1.8 $\pm$ 0.4 & 1.9 $\pm$ 0.4 \\ 
NGC 2403 & 7.4 & 1.6 & 3.4 $\pm$ 0.2 & 2.7 $\pm$ 0.2 & 2.5 $\pm$ 0.2 \\ 
NGC 7793 & 5.9 & 1.3& 8.5 $\pm$ 1.5 & 6.4 $\pm$ 1.1 & 8.5 $\pm$ 1.3 \\ 
IC2574 & 7.5 & 2.1 & 32.6 $\pm$ 25.0 & 3.9 $\pm$ 0.9 & 6.7 $\pm$ 1.7 \\ 
\textbf{NGC 628} & 10.4 & 2.3 & 2.8 $\pm$ 0.6 & 2.1 $\pm$ 0.3 & 2.0 $\pm$ 0.1 \\ 
NGC 925 & 14.3 & 4.1&2.8 $\pm$ 0.3 & 3.1 $\pm$ 0.7 & 7.2 $\pm$ 3.0 \\ 
\textbf{NGC 2903} & 15.2 & 2.9 & 1.5 $\pm$ 0.3 & 1.5 $\pm$ 0.2 & 1.6 $\pm$ 0.2 \\ 
\textbf{NGC 3351} & 10.6 & 2.5 &2.6 $\pm$ 0.8 & 3.8 $\pm$ 1.0 & 1.8 $\pm$ 0.6 \\ 
\textbf{NGC 5055} & 17.3 & 3.2& 2.7 $\pm$ 0.6 & 2.4 $\pm$ 0.5 & 2.1 $\pm$ 0.5 \\ 
\textbf{NGC 4736} & 5.3 & 1.2& 1.3 $\pm$ 0.3 & 1.2 $\pm$ 0.2 & 1.1 $\pm$ 0.1 \\ 
\textbf{NGC 3198} & 13.0 & 3.2 & 4.7 $\pm$ 0.9 & 4.2 $\pm$ 0.5 & 3.5 $\pm$ 0.4 \\ 
\textbf{NGC 3621} & 9.4 & 2.6 & 3.5 $\pm$ 0.4 & 3.9 $\pm$ 0.4 & 3.4 $\pm$ 0.2 \\ 
\textbf{NGC 6946} & 9.9 & 2.5& 2.9 $\pm$ 0.3 & 2.3 $\pm$ 0.3 & 1.9 $\pm$ 0.3 \\
\textbf{NGC 5236} & 10.1 & 2.5& 2.0 $\pm$ 0.1 & 2.1 $\pm$ 0.3 & 2.4 $\pm$ 0.2 \\
\enddata
\tablecomments{Column 1: Name of galaxy; 
Column 2: The optical radius, $\rm{r_{25}}$, in units of kpc, adopting 
the distance in \citet{walter08}; 
Column 3: The radial scale length, $\rm{r_{D}}$, in units of kpc derived by \citet{HunterElmegreen04} and \citet{leroy08}; 
Column 4: Velocity dispersion scale lengths from the single Gaussian fit; 
Column 5: Velocity dispersion scale lengths of the narrow component; 
Column 6:Velocity dispersion scale lengths of the broad component. 
 Bold font indicates spiral galaxies, whereas normal font represents 
 dwarf galaxies \citep[adopting the 
definition of][]{leroy08}.}
\end{deluxetable}
\capstarttrue

\begin{figure}
	\centering
    \includegraphics[scale=.27]{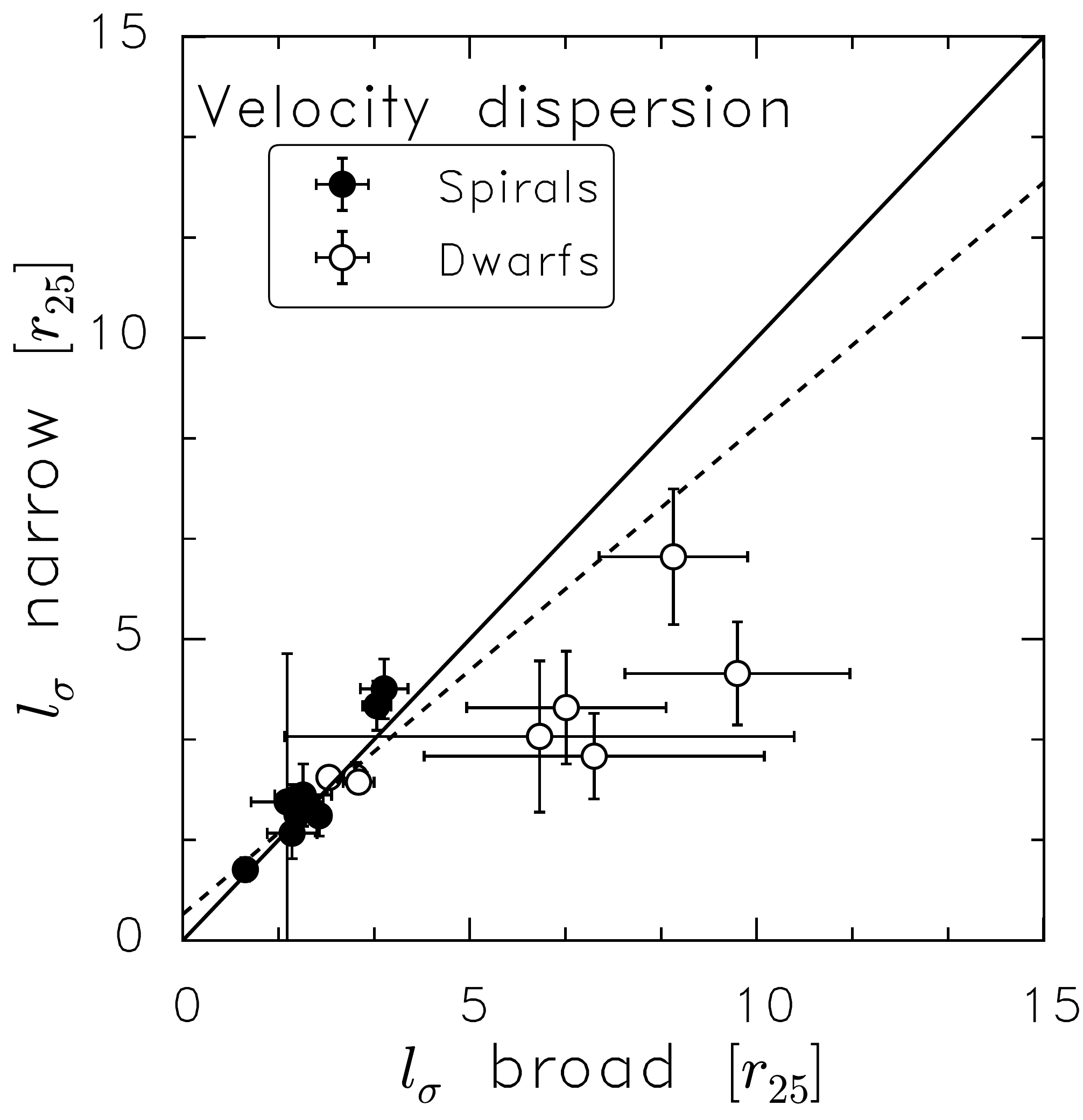} 
    \caption{Comparison of the scale length of the broad and the narrow components. 
    The dashed line is a linear fit to the data points. The solid line represents the line of equality. Note that 
    Ho I, M81 dw A, DDO 154 and DDO 53 are not shown in this plot as their radial velocity dispersion profiles are not exponential.}
    \label{fig:cv_slnb}
\end{figure}  

We compare the velocity dispersion scale lengths of 
the broad and narrow components in Figure~\ref{fig:cv_slnb}. We fit a linear function 
to this relation and we find a slope of $\sim$ 0.8, which indicates 
that the ratio between the velocity dispersion of the broad and narrow components is roughly constant 
with radius. The average velocity dispersion ratio between the narrow and broad component is  
$\langle \sigma_n/\sigma_{b} \rangle = 0.46\pm0.08$.  

\begin{figure*}[hbt]
    \begin{tabular}{l l l}
    \includegraphics[scale=.245]{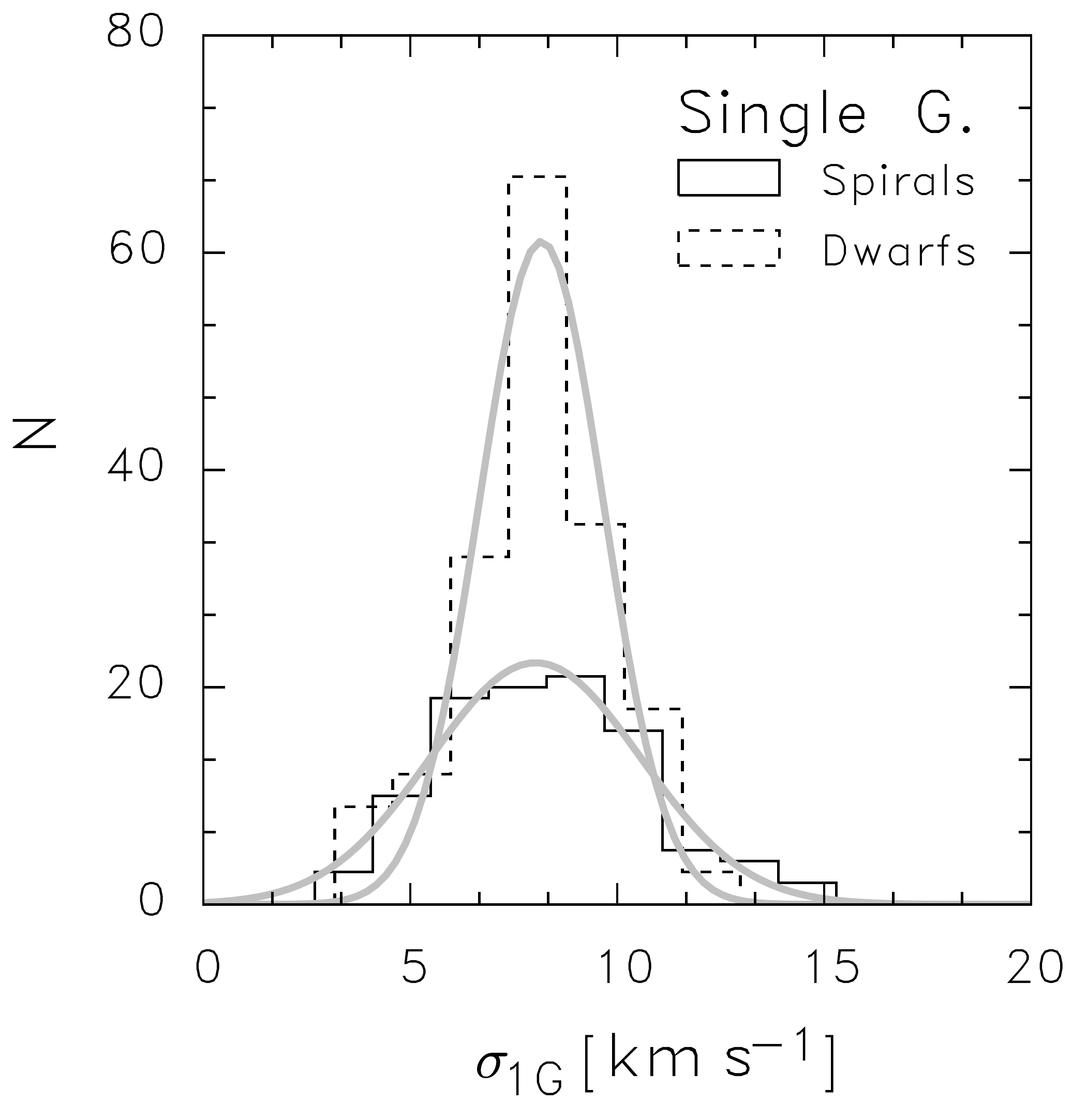} &
    \includegraphics[scale=.245]{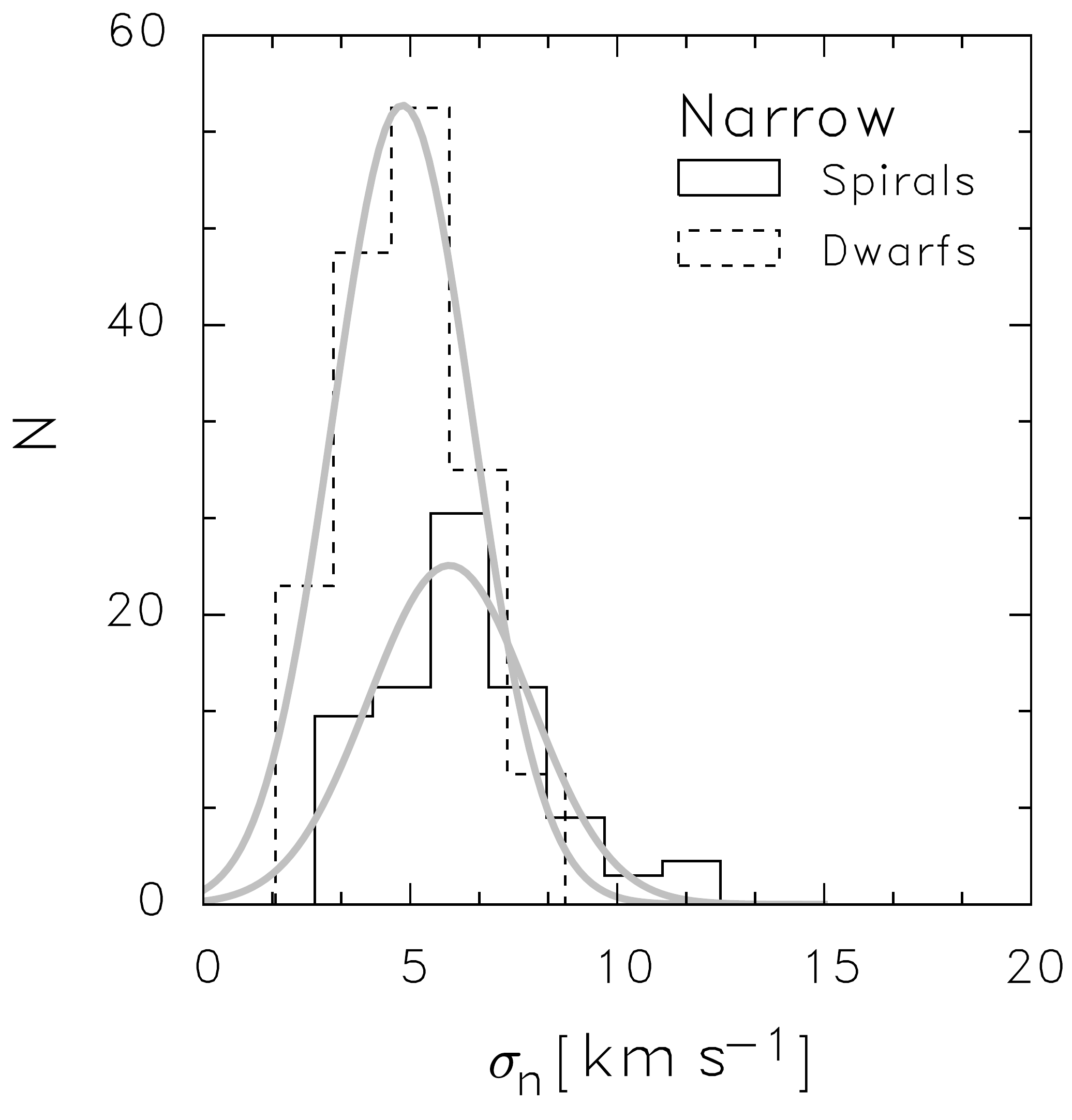}
    \includegraphics[scale=.245]{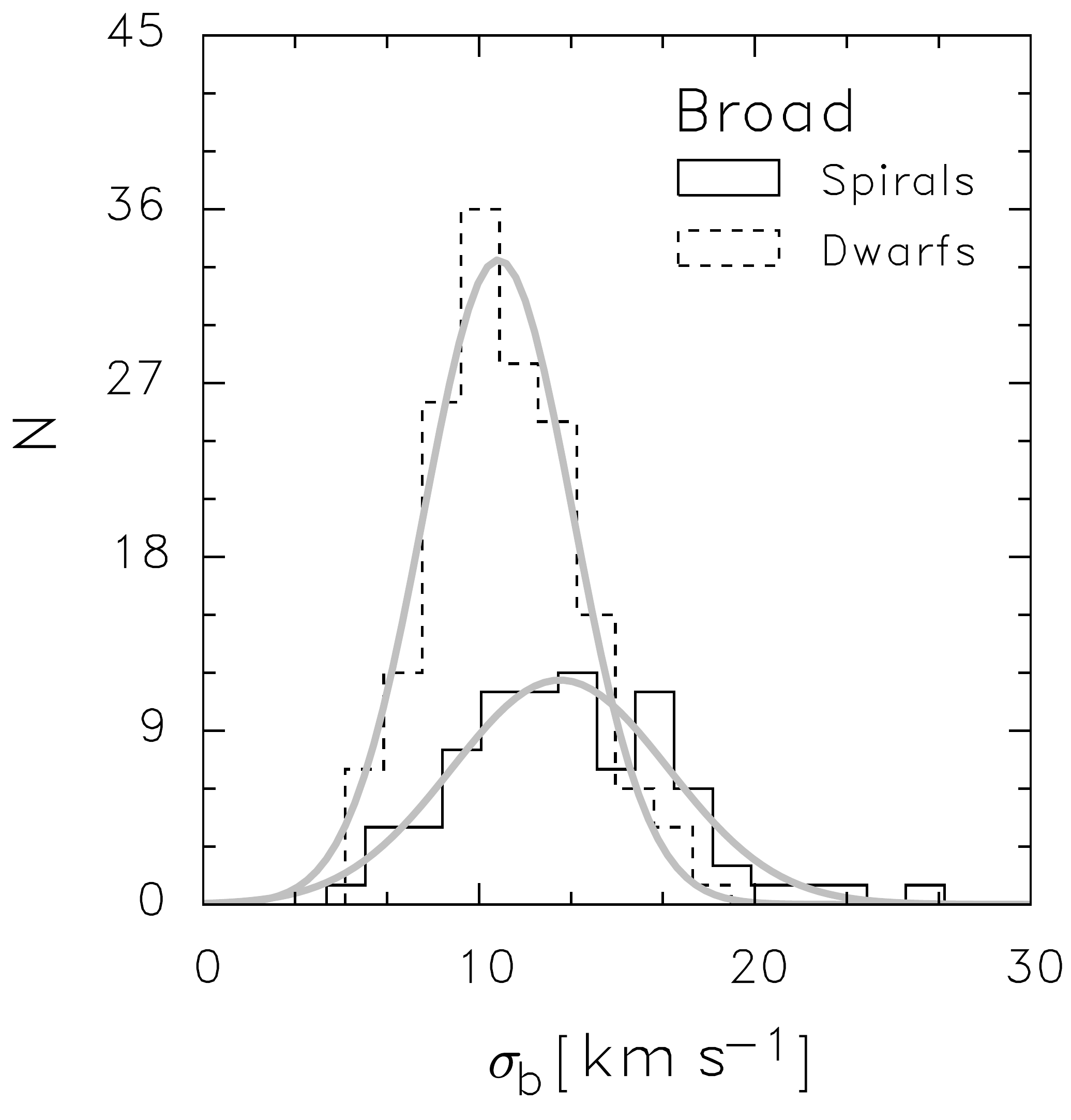}
    \end{tabular}
    \caption{Histograms of individual points in the radial velocity dispersion profiles of dwarf and spiral galaxies. Points 
    in the inner 0.2 $r_{25}$ are excluded as they are mostly affected by beam smearing and streaming motions.}
    \label{fig:hist_rad}
\end{figure*}

Figure~\ref{fig:hist_rad} shows histograms of the velocity dispersion values measured in radial bins for spiral 
and dwarf galaxies in our sample. Note that here we exclude values within 0.2 $\rm{r_{25}}$ 
where beam smearing and streaming motions are expected to play a role as reported by 
\citet{calduetal13}. We fit the histograms with a Gaussian function; the results of the fit are shown in Table~\ref{tab:hist}. 
As shown in the histograms and in Table~\ref{tab:hist}, spirals and dwarfs have roughly the same mean velocity dispersion values, 
with $\langle \sigma_{1G} \rangle \simeq 8~\rm{km~s^{-1}}$. This is somewhat smaller than previous estimates of 
the single Gaussian velocity dispersion of the THINGS galaxies. \citet {leroy08} found $\langle \sigma_{1G} \rangle $ $\simeq$ 11 $\pm$ 3 $\rm{km~s^{-1}}$, whereas \citet{ianjamasimananaetal12} and \citet{calduetal13} found $\langle \sigma_{1G} \rangle $ $\simeq$ 13 $\pm$ 4 $\rm{km~s^{-1}}$ and 12 $\pm$ 3 $\rm{km~s^{-1}}$, respectively. The difference can be explained by our use of natural residual scaled cubes as opposed to non-residual scaled cubes adopted by \citet {leroy08} and \citet{ianjamasimananaetal12}. 
E.g., for NGC 3184, the $\langle \sigma_{HI} \rangle$ from residual-scaled cubes and non-residual scaled cubes differ by $\sim$ 3 
$\rm{km~s^{-1}}$. The effects of residual scaling on H\,{\sc i} velocity dispersion are 
analysed in detail in a subsequent paper; but see also \citet{stilpetal13}. \citet{calduetal13} used residual-scaled cubes but considered a somewhat different radial range than is adopted in our analysis. They measured velocity dispersions of 12 spiral galaxies in regions where CO is present, mostly in the inner 60\% of the optical radius. The radial range being considered here is larger than that of \citet{calduetal13}, explaining why our mean dispersion value is smaller than theirs since we include more points from the outer disks where velocity dispersions tend to be smaller. We find, for spirals, $\langle \sigma_{1G} \rangle $ $\simeq$ 11 $\pm$ 2 $\rm{km~s^{-1}}$ and $\langle \sigma_{1G} \rangle $ $\simeq$ 8 $\pm$ 2 $\rm{km~s^{-1}}$ 
inside and outside 0.6 $\rm{r_{25}}$, respectively. This is consistent with the result of \citet{calduetal13}.

Although spirals and dwarfs have roughly the same $\langle \sigma_{1G} \rangle$ value, spirals  have broader $\sigma_{1G}$ 
distribution than dwarfs as shown in Figure~\ref{fig:hist_rad}. This may be caused by the effects of the presence of spiral arms or bars in spiral galaxies. Observational evidence for the effects of spiral arms on velocity dispersion 
has been presented by e.g., \citet{shostak84} for NGC 628. They found that the H\,{\sc i} velocity dispersion 
of NGC 628 was $\sim10~\rm{km~s^{-1}}$ in the spiral arms and $\sim7~\rm{km~s^{-1}}$ in the interarm regions. 

\begin{figure*}[hbt]
    \begin{tabular}{l l l}
    \includegraphics[scale=.245]{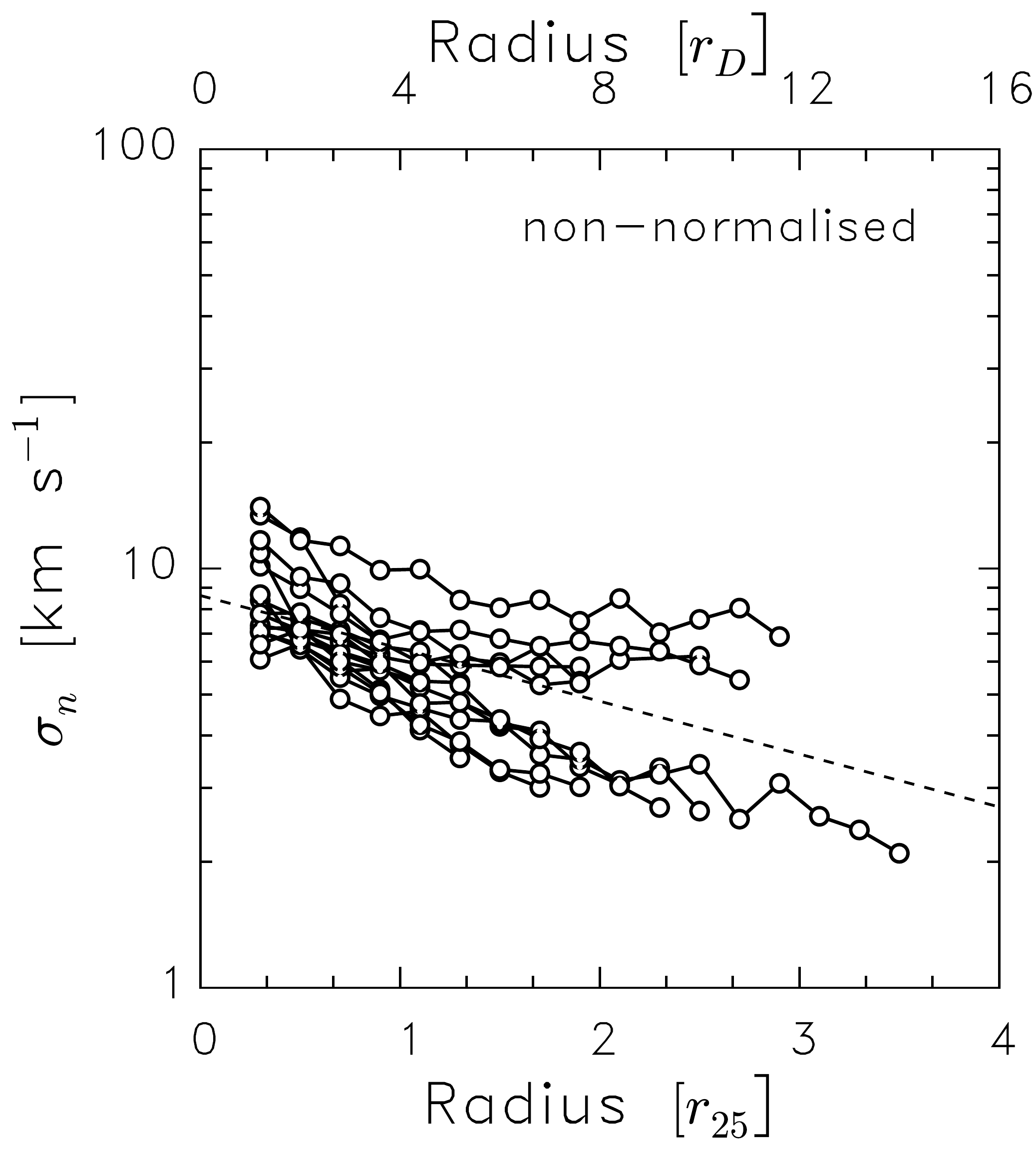} &
    \includegraphics[scale=.245]{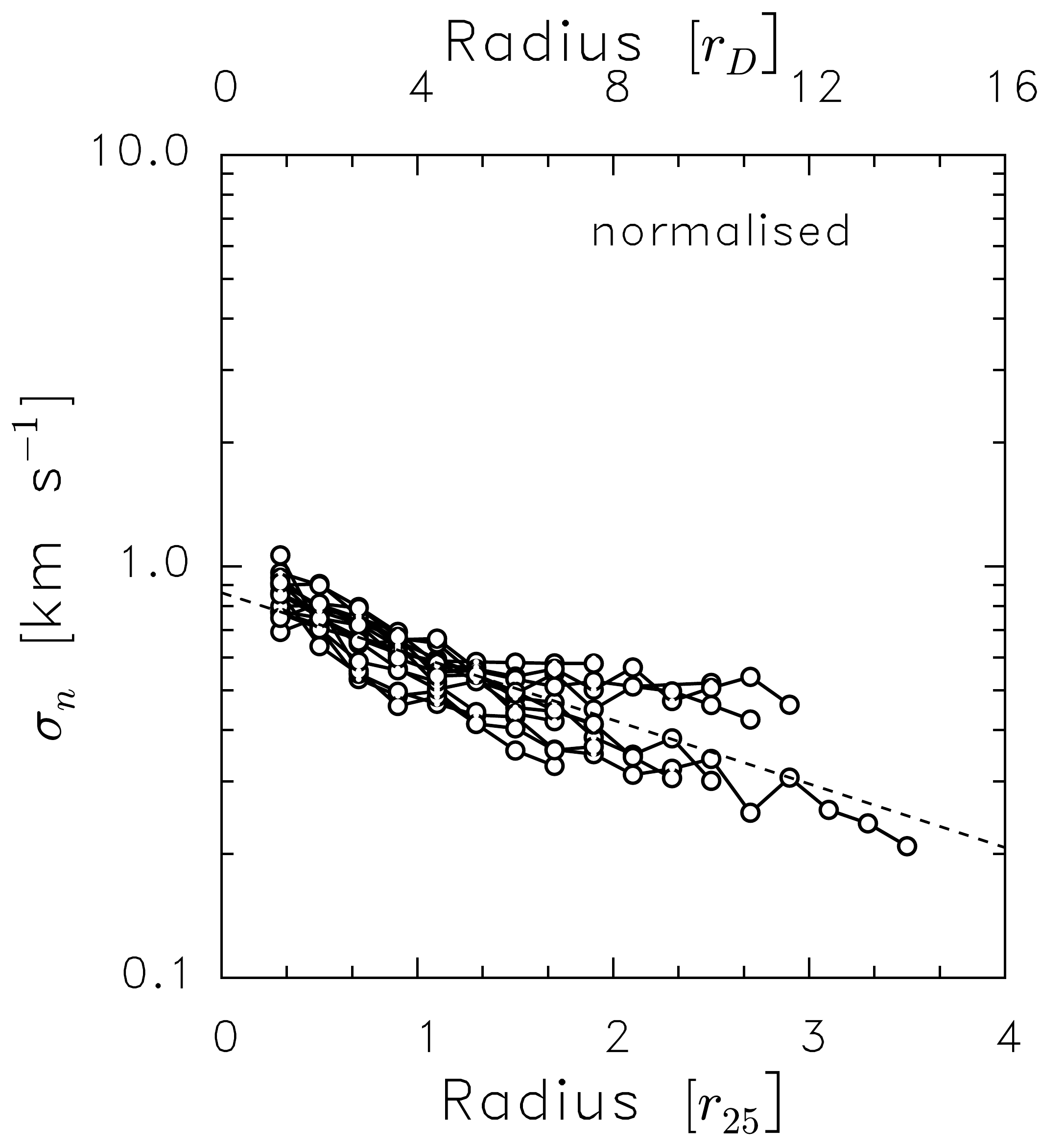}
    \end{tabular}
    \caption{Combined (narrow component) radial velocity dispersion profiles before (left panel) and after (right panel) the normalisation described in Section \ref{sub:rad_disp_prof}.}
    \label{fig:norm}
\end{figure*}

To have an overall representation of the radial velocity dispersion trend of the sample, 
we fit the complete set of radial velocity dispersion profiles with a single exponential function. 
Note that we exclude M81 dwA, DDO 154, Ho I, DDO 53 from this analysis due to their flat profiles. 
We further exclude IC 2574 and NGC 2366 as their single Gaussian velocity dispersion profiles are also flat.   
We try to remove the scatter between individual radial profiles by applying a normalisation technique 
similar to the one presented in \citet{schrubaetal11}. That is, we normalise the velocity dispersions so that 
an exponential fit to the profiles has a value of 
unity at 0.6 $\rm{r_{25}}$. Note that the choice of this value is arbitrary. 
This normalisation technique removes galaxy-to-galaxy variation.
We also fit the normalised radial velocity dispersion profiles with a single exponential function. 
We illustrate, for the narrow component, the combined radial profiles before and after normalisation in Figure 
\ref{fig:norm}. 
We show the fitted scale lengths of the  combined radial velocity dispersion profiles before and after normalisation 
in Table \ref{tab:l_comb}.  As expected, the (combined) normalised profiles show smaller scatter than the non-normalised ones.      
After the normalisation, the scatter has decreased by more than 90\%, whereas the fitted scale length values before and after the normalisation agree within 20\%. Thus, the scatter in the combined radial profiles is caused by galaxy-to-galaxy variation.  

\subsection{Comparison with literature values}\label{sub:comp}
\citet{Zhangetal2012} analysed the variations of the H\,{\sc i} power spectral index with channel width, 
for a sample of nearby dIrr galaxies, and suggested that the turbulent velocity dispersion of the 
coolest H\,{\sc i} ($\lesssim$ 600 K) should not exceed  $\sim5~\rm{km~s^{-1}}$.  For dwarf galaxies in our sample, we also find an average value of $\sim5~\rm{km~s^{-1}}$. For spirals, this value is somewhat higher, but this can be explained by the higher energy input into the ISM in these galaxies. 

A radial dependency of H\,{\sc i} velocity dispersions has previously been reported in the literature. 
\citet{dickeyetal90} and \citet{petricrupen07} studied the H\,{\sc i} kinematics of the face-on 
spiral galaxy NGC 1058 and found a radial decline of H\,{\sc i} velocity dispersion from 
$\sim12~\rm{km~s^{-1}}$ in the inner part to $\sim6~\rm{km~s^{-1}}$ in the outer part.
\citet{boulangerviallefond92} and \citet{boomsmaetal08} also found a radial fall-off of 
H\,{\sc i} velocity dispersion for the face-on spiral galaxy NGC 6946. 
As already mentioned, \citet{tamburroetal09} analysed the radial H\,{\sc i} second moment profiles of 11 THINGS galaxies 
(all but one galaxy are also in our sample), and found that all exhibited a clear radial decline. 
Our single Gaussian velocity dispersion profiles can be directly compared to those presented in \citet{boomsmaetal08} and 
\citet{tamburroetal09}. For NGC 6946, the radial profile from \citet{boomsmaetal08} shows a steep 
drop in velocity dispersion, followed by a linear decrease from about the optical radius to their 
outermost observed radius. Our velocity dispersion profile and that from \citet{tamburroetal09}, however, show a 
continual decline from $\sim$ 0.7 $\rm{r_{25}}$ to 
$\sim$ 1.8 $\rm{r_{25}}$ and then start to level off.

\begin{figure*}[hb]
\centering
\includegraphics[scale=0.53]{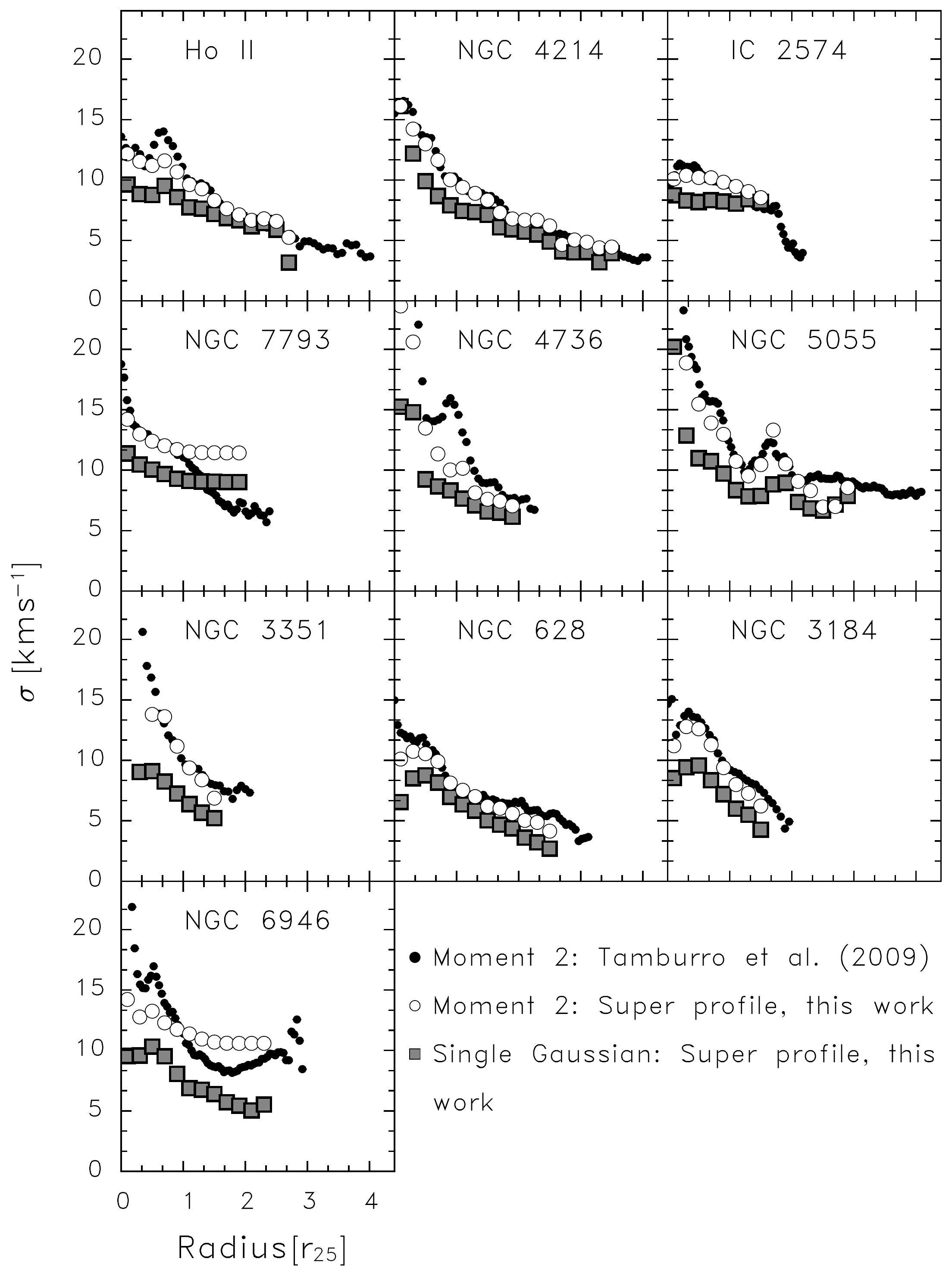}
\caption{Comparison of our velocity dispersions with those derived in \citet{tamburroetal09}. 
The solid black symbols represent the \citet{tamburroetal09} radial velocity dispersion profiles, derived 
from second moment maps. The square symbols are single Gaussian dispersions from our super profiles. The 
open circle symbols are second moment values calculated from the observed, total super profiles (i.e. no separation into broad and narrow components). \\\\}
\label{fig:grid}
\hspace*{-7cm}
\includegraphics[scale=0.32]{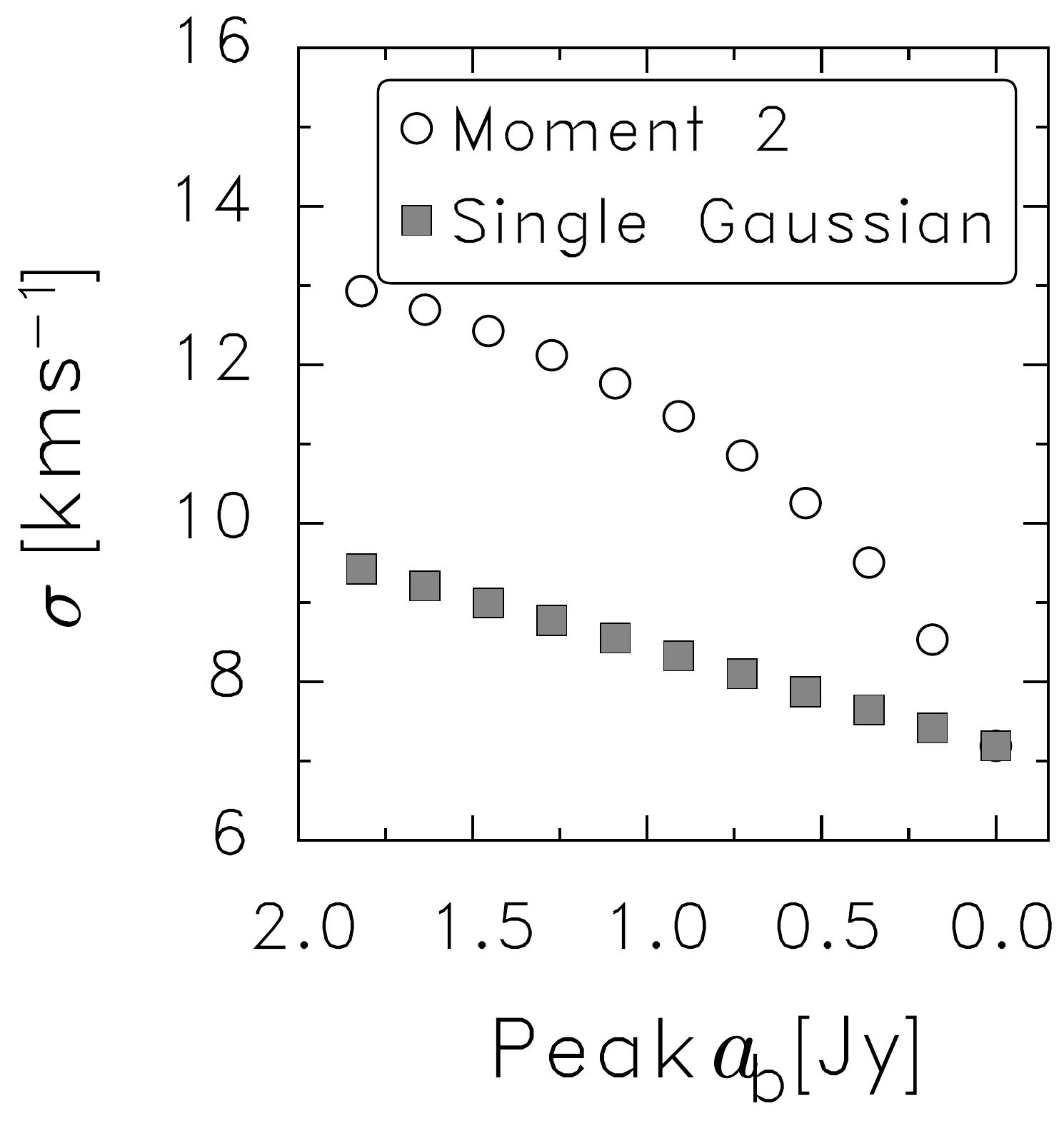}
\caption{Testing the dependence of single Gaussian dispersion and second moment values on the strength of profile wings. 
\textit{Open circle symbols}: second moment values. \textit{Square symbols}: single Gaussian velocity dispersions. $a_{\rm{b}}$: 
Broad component amplitude.}
\label{fig:wings}
\end{figure*}

We overplot our single Gaussian velocity dispersion profiles with the second moment velocity values of 
\citet{tamburroetal09} in Figure \ref{fig:grid}. Note that, in some cases, the radial velocity dispersion profiles of \citet{tamburroetal09} go further in radius than ours. 
This is because we have been more restrictive in fitting only high S/N profiles. 
We do not always find good agreement between our measured velocity 
dispersions and those from \citet{tamburroetal09}, especially in the central parts of galaxies where our velocity dispersions 
tend to be smaller. This can be explained by the fact that the shapes of the velocity profiles in the inner disks of galaxies 
are usually strongly non-Gaussian. For a perfectly Gaussian profile, the second moment value 
and the single Gaussian dispersion would be equal. However, for a velocity profile with 
prominent wings, the second moment value would be higher than that of the single Gaussian dispersion. 
\capstartfalse
\begin{deluxetable}{c c c c }[hm]
\centering
\tablecolumns{4}
\tablecaption{Results of Gaussian fits to the histograms of velocity dispersion values measured in radial bins\label{tab:hist}}
\tablehead{
   \multicolumn{1}{c}{}&\multicolumn{1}{c}{Single Gaussian}&\multicolumn{1}{c}{Narrow} & 
     \multicolumn{1}{c}{Broad}\\
   \multicolumn{1}{c}{}& $\langle \sigma_{1G} \rangle $ & $\langle \sigma_{n} \rangle $ &  $\langle \sigma_{b} \rangle $ \\
    &\multicolumn{1}{c}{$(\rm{km~s^{-1}})$}& \multicolumn{1}{c}{$(\rm{km~s^{-1}})$}& \multicolumn{1}{c}{$(\rm{km~s^{-1}})$} \\
}
\startdata
\multicolumn{1}{l}{Spirals:} & 8.0 $\pm$ 2.5 & 5.9 $\pm$ 1.9&  12.9 $\pm$ 3.9\\
\multicolumn{1}{l}{Dwarfs:} & 8.2 $\pm$ 1.5 & 4.8 $\pm$ 1.7& 10.7 $\pm$ 2.7                  
\enddata
\end{deluxetable}
\capstarttrue
In Figure \ref{fig:grid}, we also show second moment values derived from the observed, total super profiles (i.e. no separation into broad and narrow components was made). The offset between the single Gaussian dispersions and the second moment values become 
smaller with increasing radius and, for most galaxies, the two dispersions start to agree at 
a certain radius. This behaviour is due to the shapes of the profiles becoming more and more Gaussian toward larger radius. 
The radius where the values start to agree may 
indicate the point beyond which only the WNM is dominant. 
\capstartfalse
\begin{deluxetable}{c c c c c c c}[m]
\centering
\tablecolumns{8}
\tablecaption{Fitted scale lengths of combined radial velocity dispersion profiles.\label{tab:l_comb}}
\tablehead{
   \multicolumn{3}{c}{non-normalised}& \multicolumn{1}{c}{}& \multicolumn{3}{c}{normalised} \\
   \cline{1-3} \cline{5-7} \\
$l_{1G}$ & $l_{n}$ & $l_{b}$ & & $l_{1G}$ & $l_{n}$ & $l_{b}$ \\
$(\rm{r_{25}})$ & $(\rm{r_{25}})$& $(\rm{r_{25}})$& & $(\rm{r_{25}})$ & $(\rm{r_{25}})$ & $(\rm{r_{25}})$
}
\startdata
4.1$\pm$2.1&3.5$\pm$1.8&3.6$\pm$3.8& &3.7$\pm$0.1&2.8$\pm$0.1&3.0$\pm$0.2
\enddata
\tablecomments{The errors shown in the table are root mean square errors (i.e. scatter around best fit values).}
\end{deluxetable}
\capstarttrue

To better understand the behaviour of the second moment values as a function of 
profile shape, we proceed as follows. We take the fitted super profile of NGC 3184 
at 0.3 $\rm{r_{25}}$ (this is a typical non-Gaussian super profile with broad and narrow components of a face-on galaxy). 
We derive a single Gaussian dispersion and a second moment value from it. 
We then gradually suppress the wings of the profile by decreasing the broad component amplitude 
while keeping the narrow component parameters fixed. We re-measure the single Gaussian and the 
second moment values. We do this until the broad component is completely suppressed and we 
are left with a single Gaussian profile. We plot the derived second moment velocity values 
and single Gaussian velocity dispersions as a function of the broad component 
amplitude in Figure~\ref{fig:wings}. As expected, the disagreement between the two values becomes smaller 
with decreasing wing's strength. They are identical when the profiles are perfect Gaussians. It is also worth 
noting that the single Gaussian dispersion is less affected by the profile wings than the second moment value.        
\subsection{Surface density as a function of radius}\label{sub:flux}
In addition to measuring velocity dispersions, we can also use the super profiles to investigate the surface densities of 
the various components.
Here, we present radial surface density profiles of the single Gaussian component, $\rm{\Sigma_{1G}(R)}$, the narrow 
component, $\rm{\Sigma_{n}(R)}$, the broad component, $\rm{\Sigma_{b}(R)}$, and the $\rm{\Sigma_{n}(R)}$/$\rm{\Sigma_{b}(R)}$ 
ratio. If the narrow and broad components represent the CNM and the WNM, then, the $\rm{\Sigma_{n}}$/$\rm{\Sigma_{b}}$ 
ratio represents the surface density ratio between the CNM and the WNM. Theoretical models \citep{wolfireetal03} predict the 
CNM/WNM ratio to decrease as a function of radius due to a radially declining thermal pressure of the H\,{\sc i} gas. 
This has also been confirmed by observations of 11 nearby spiral galaxies by \citet{braun97}. 

\begin{figure*}
\begin{tabular}{l l l}
\includegraphics[scale=.22]{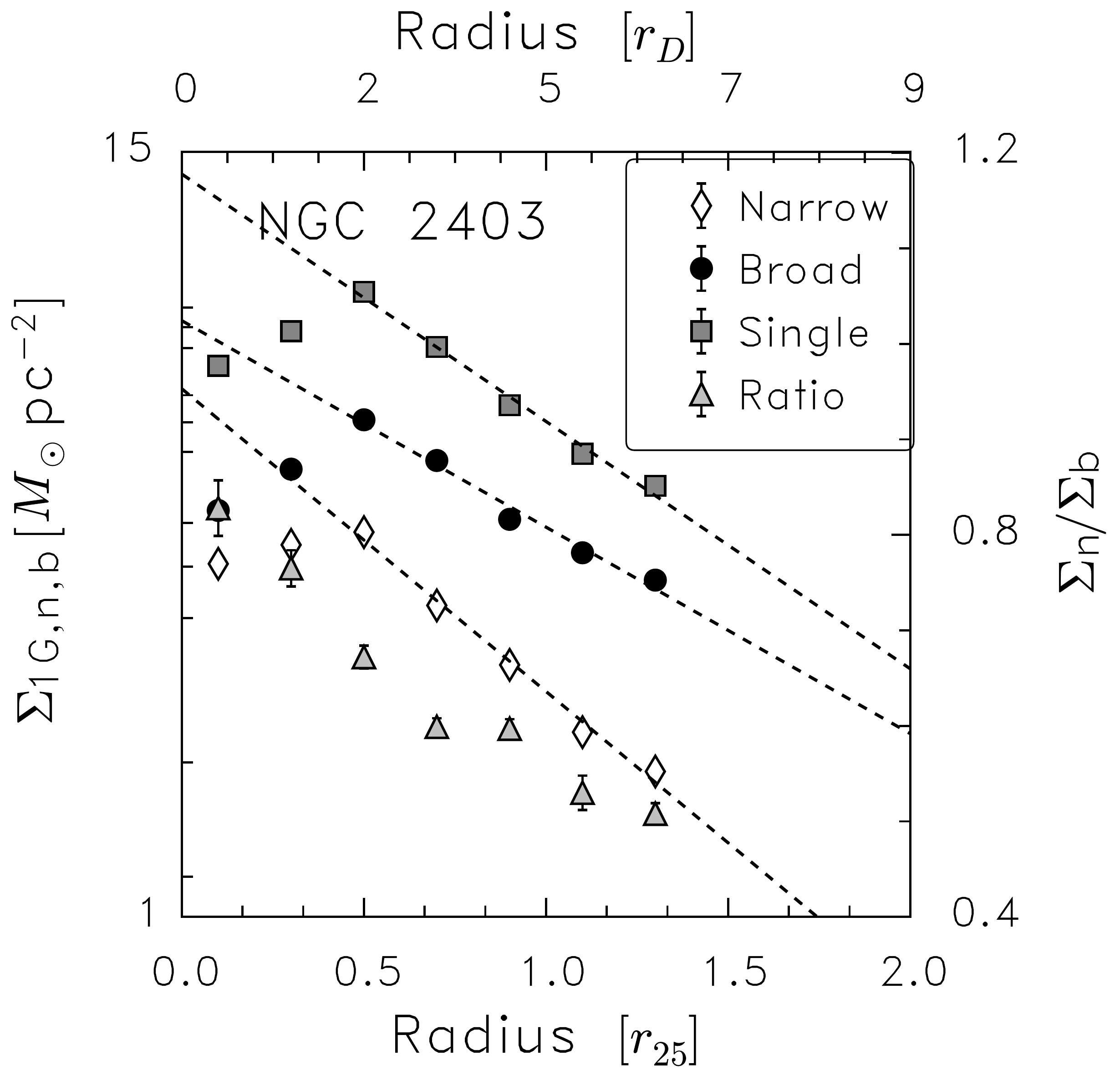}&
\includegraphics[scale=.22]{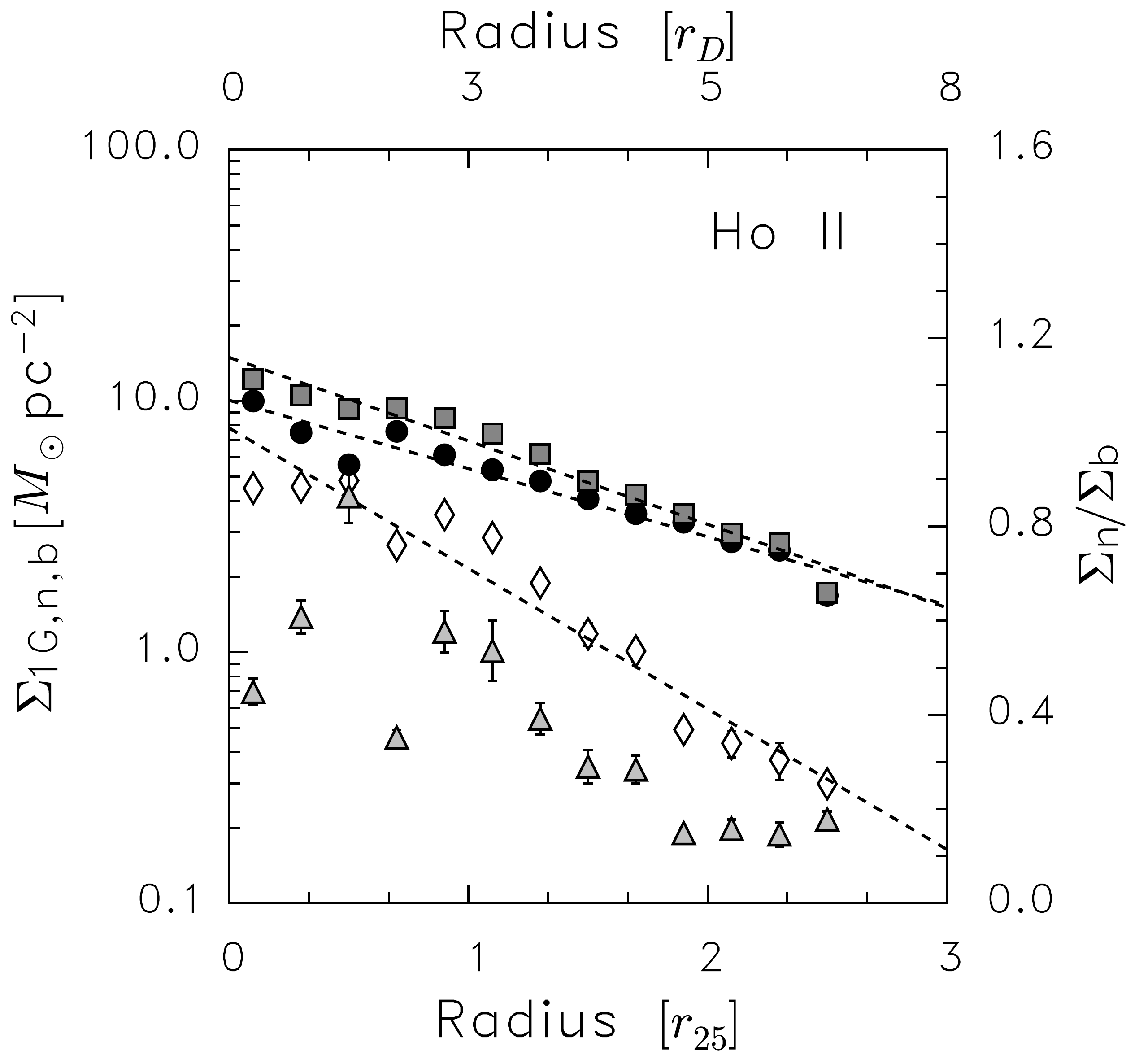}&
\includegraphics[scale=.22]{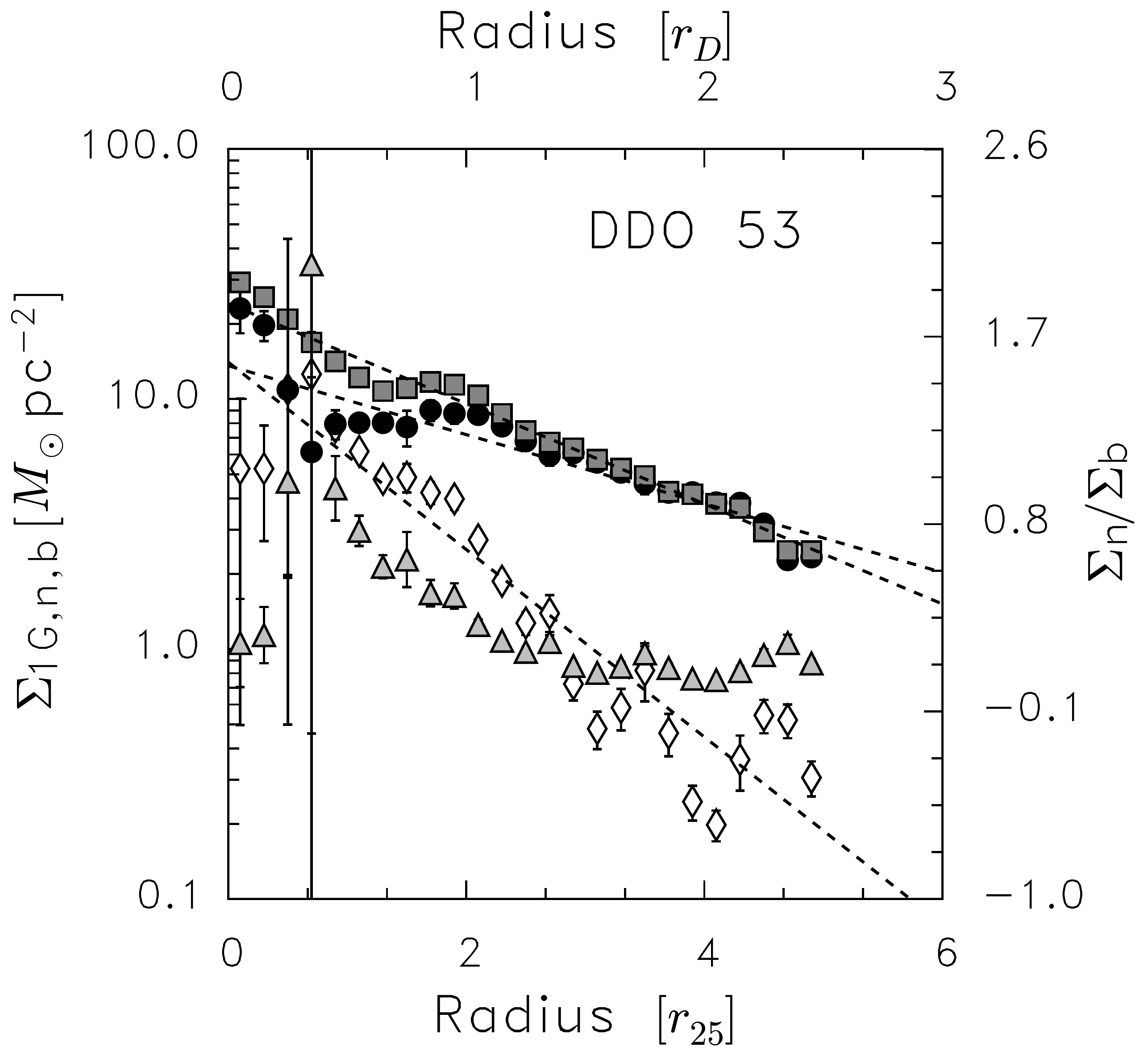}\\
\includegraphics[scale=.22]{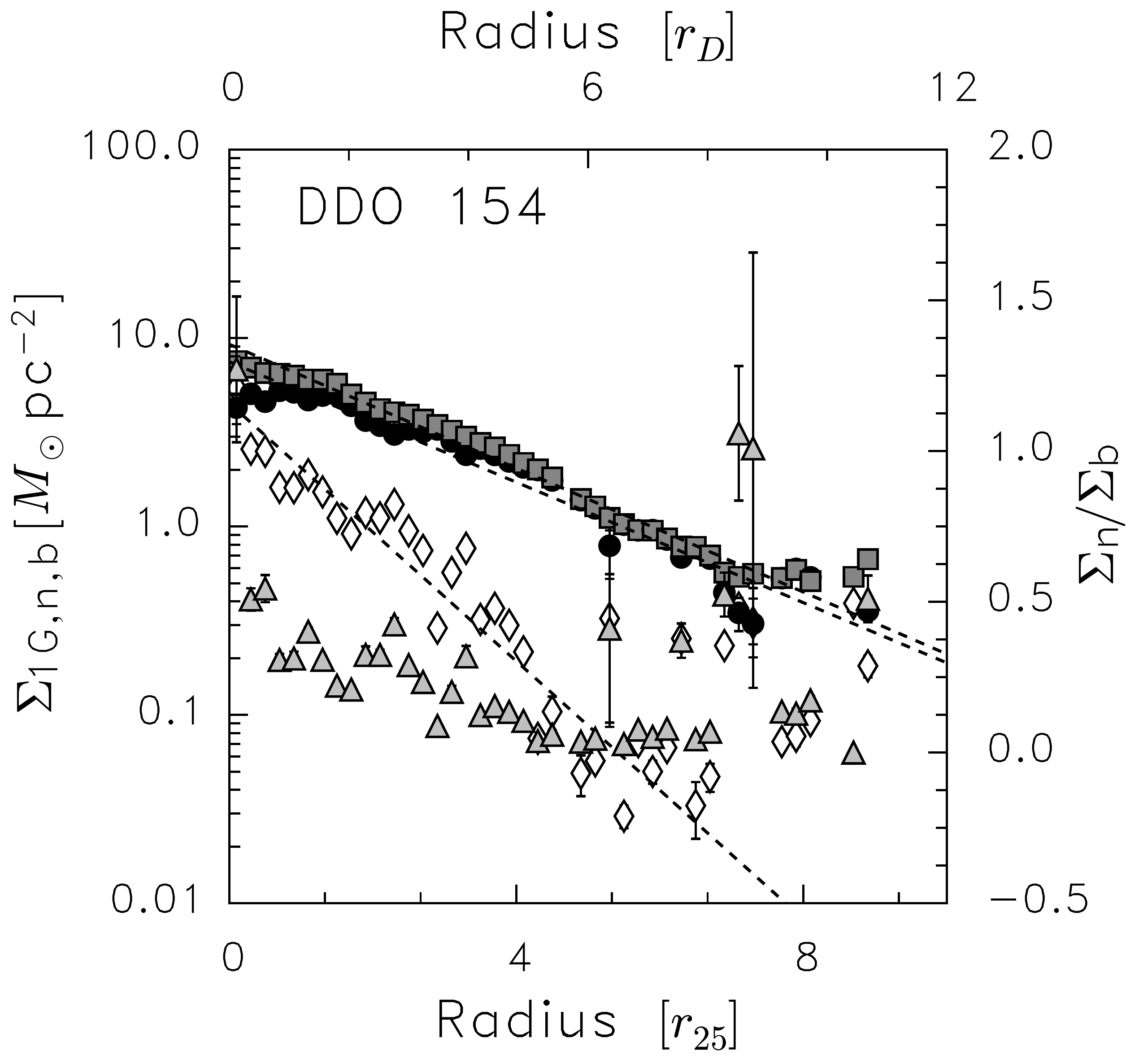}&
\includegraphics[scale=.22]{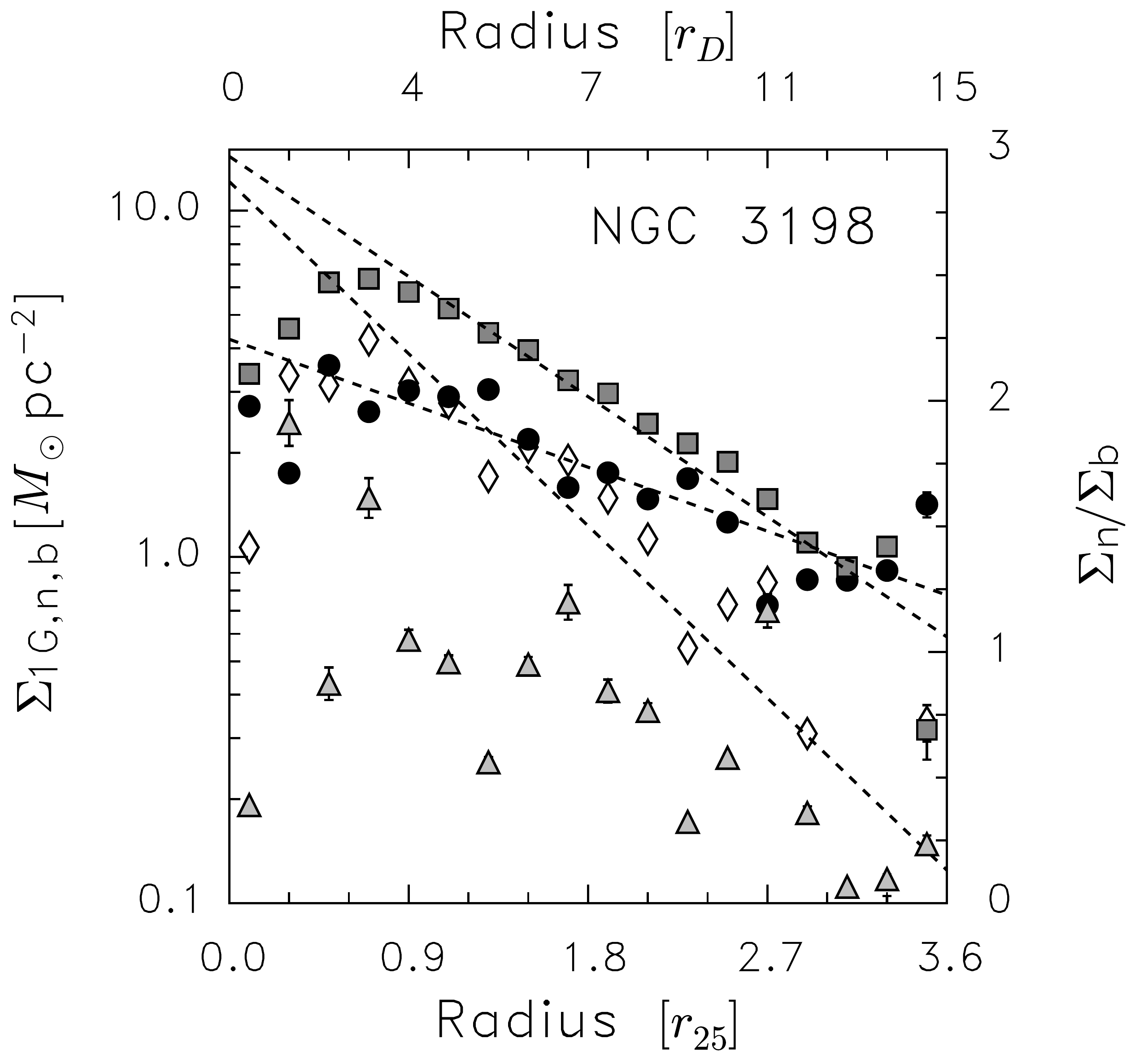}&
\includegraphics[scale=.22]{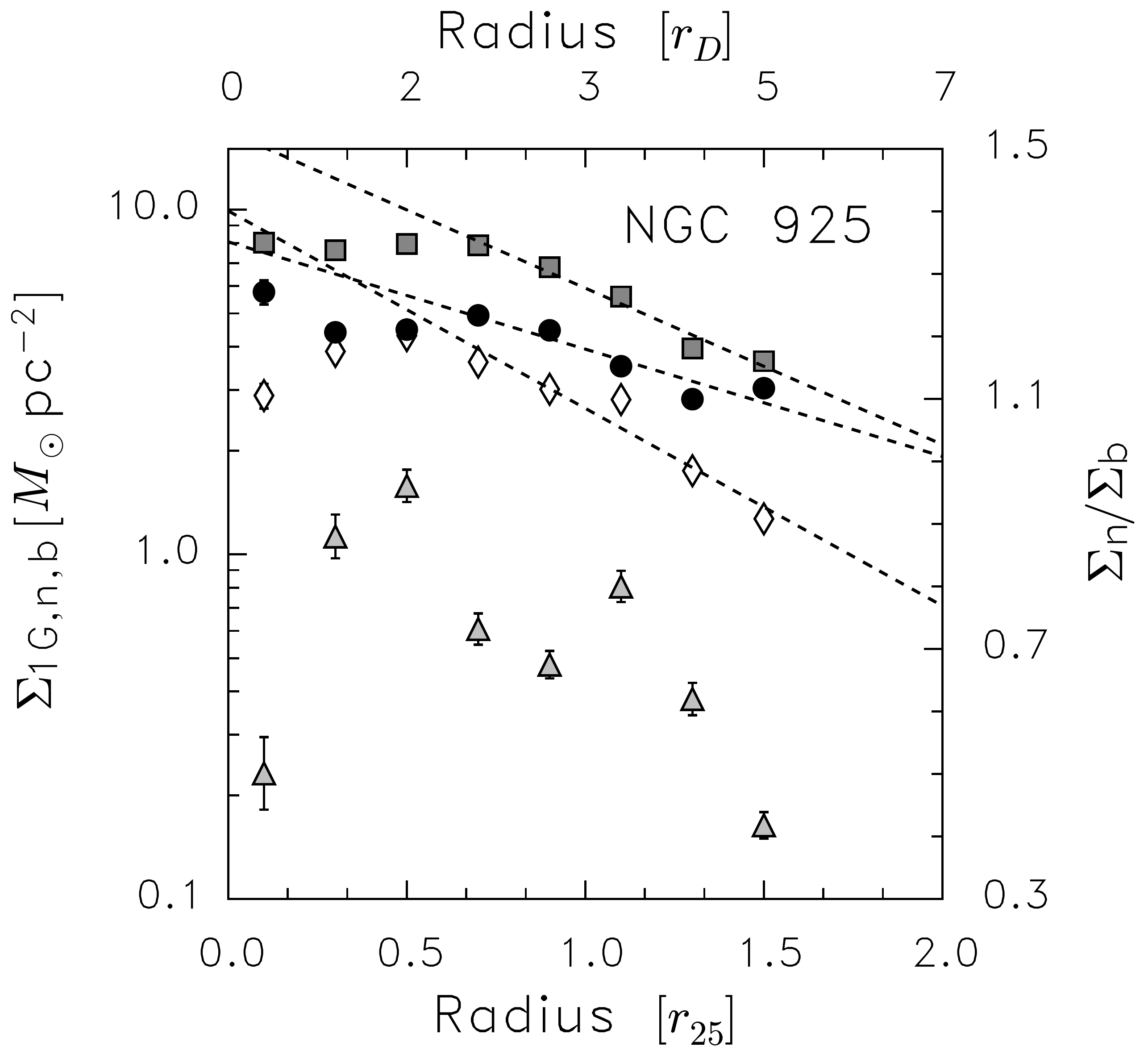}\\
\includegraphics[scale=.22]{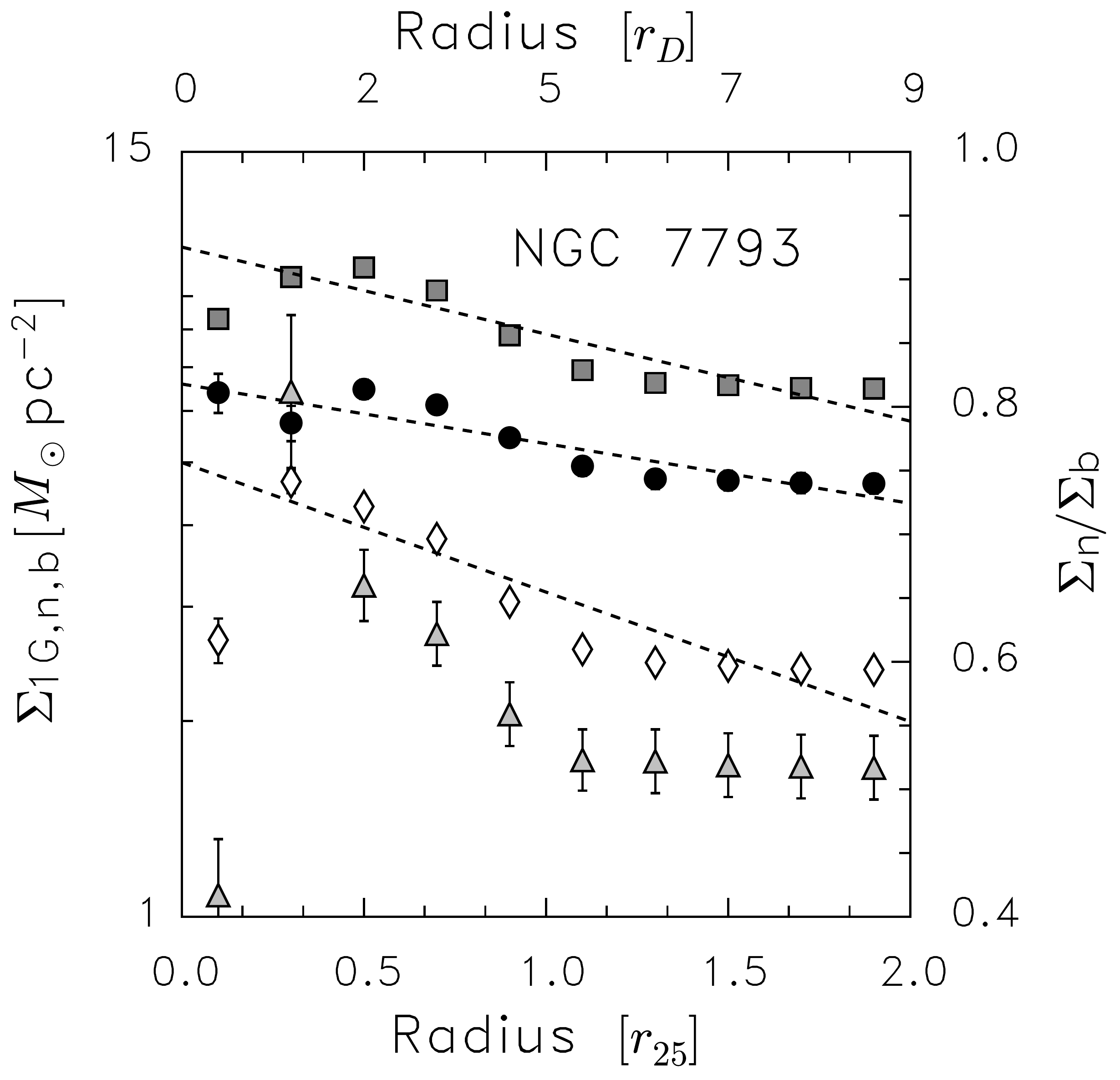}&
\includegraphics[scale=.22]{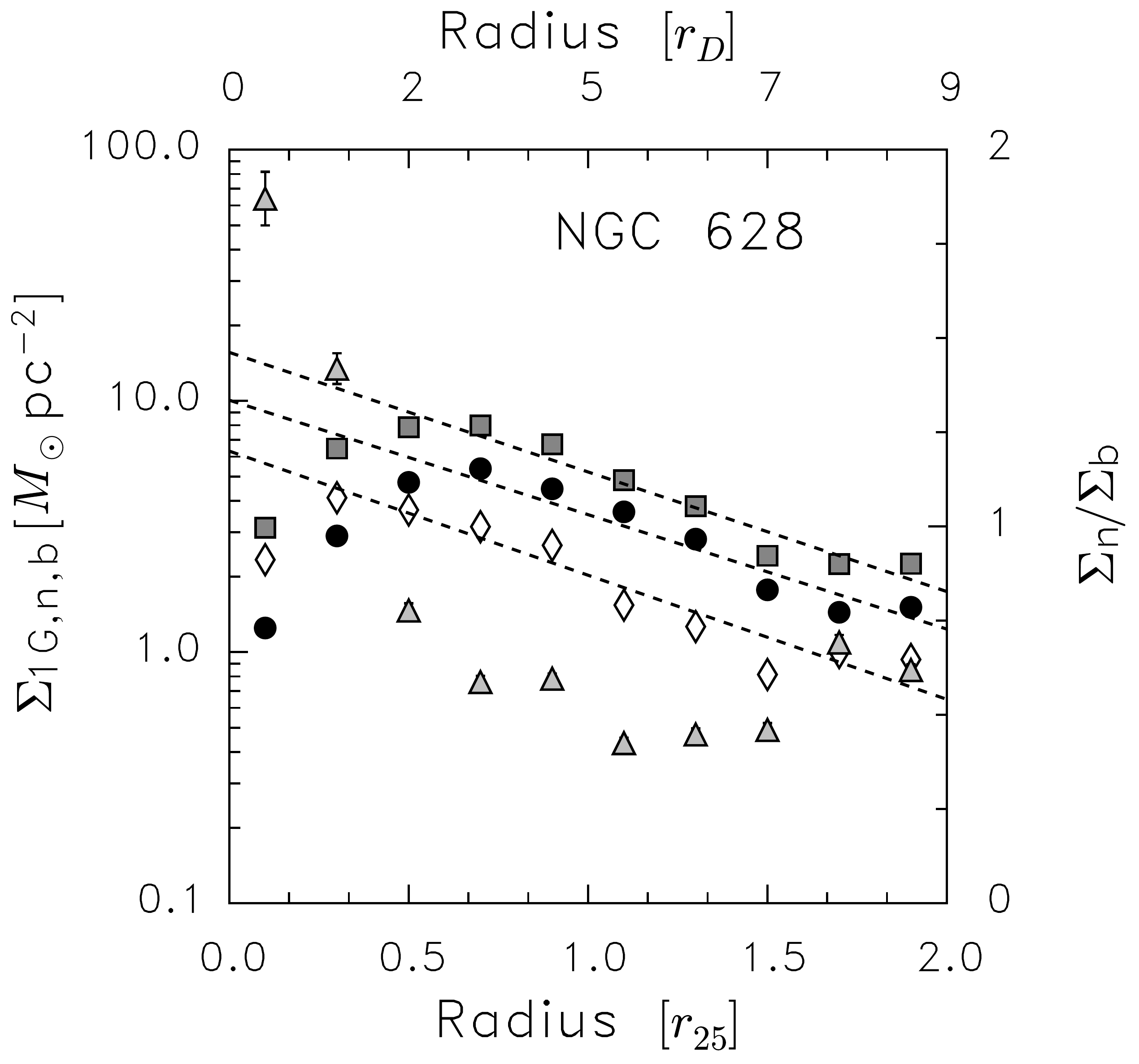}&
\includegraphics[scale=.22]{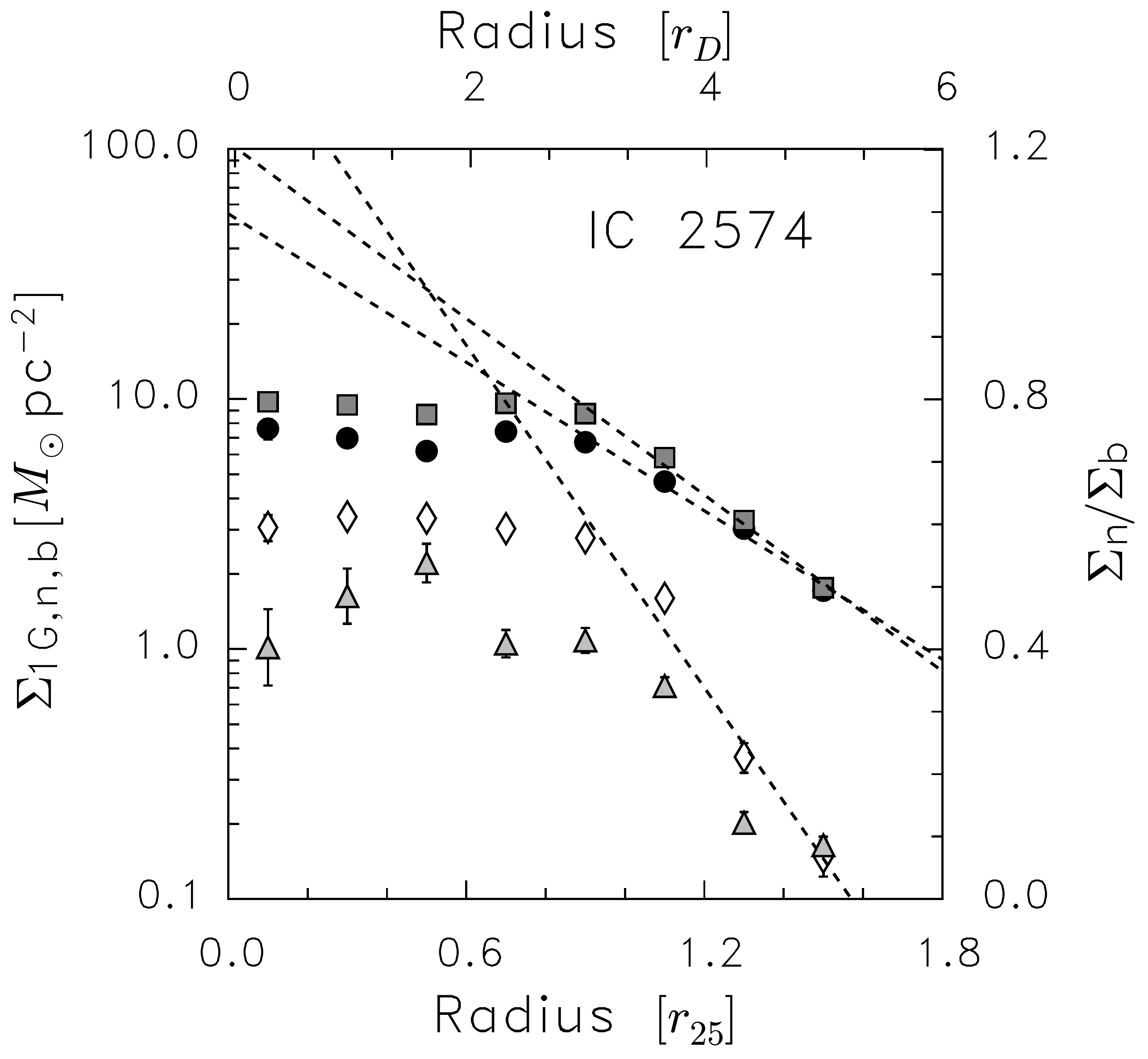}\\
\hspace*{-0.3cm}
\includegraphics[scale =.22]{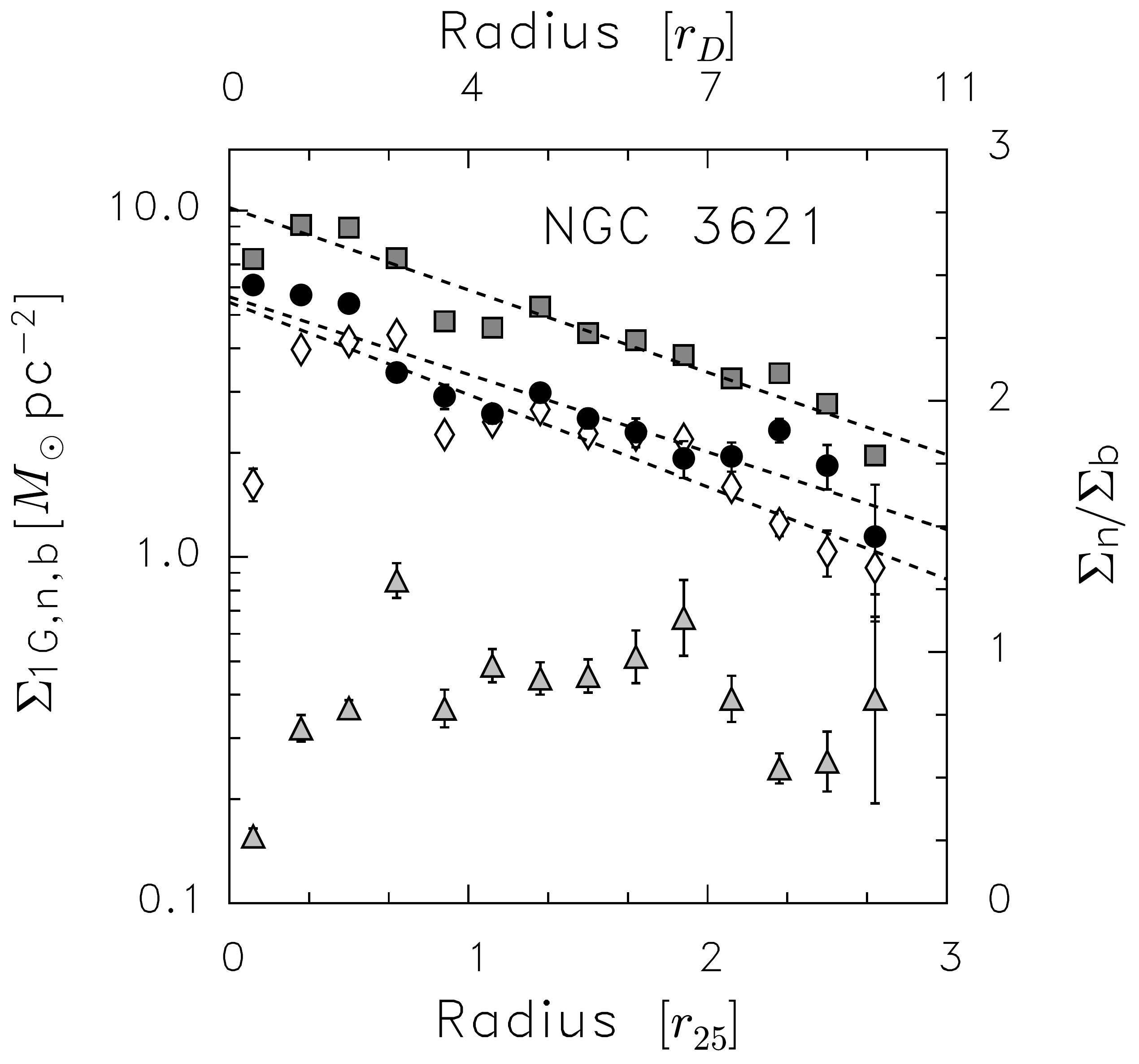}&
  \includegraphics[scale=.22]{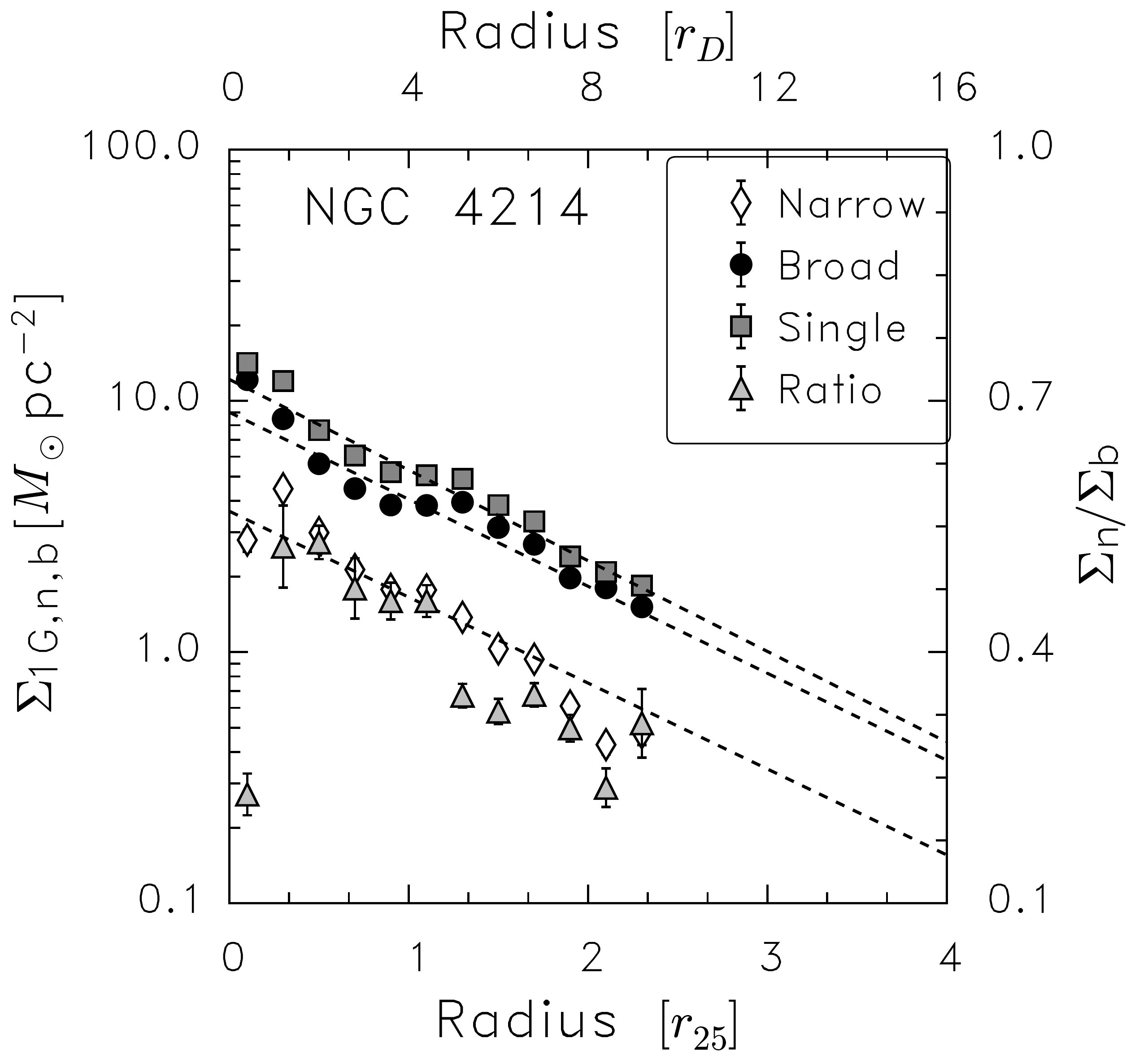}& 
  \includegraphics[scale=.22]{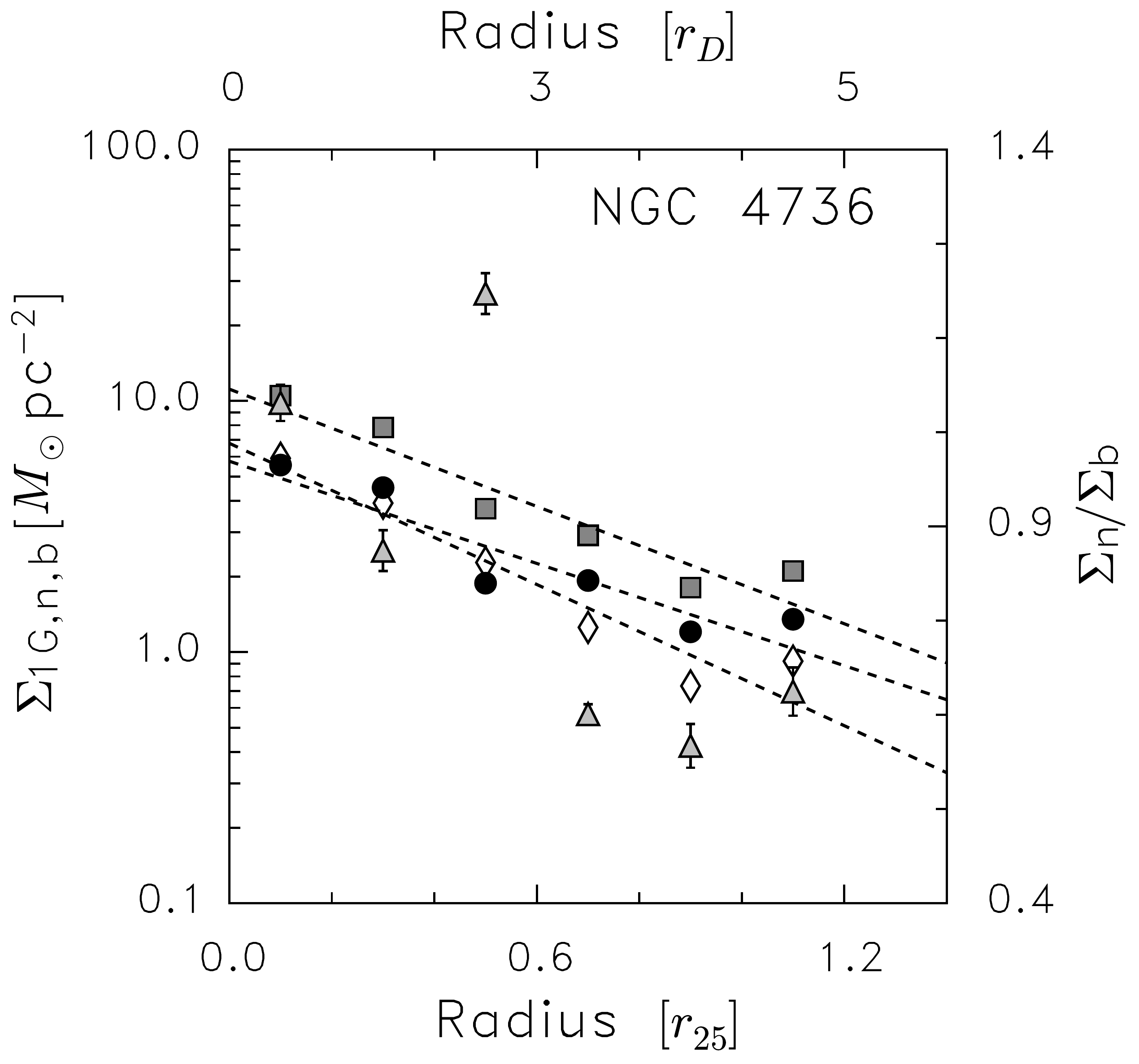}  
   \end{tabular}
\caption{\textit{Open diamond symbols}: surface density of the narrow component; \textit{solid black circle symbols}: 
surface density of the broad component; \textit{square gray symbols}: surface density of the single Gaussian 
component; \textit{triangle gray symbols}: ratio between the surface densities of 
the narrow and broad components; \textit{Dashed lines:} exponential fits.}
\label{fig:rad_area_main}
\end{figure*}

\begin{figure*}
\begin{minipage}[b]{0.45\textwidth}
\includegraphics[scale = 0.27]{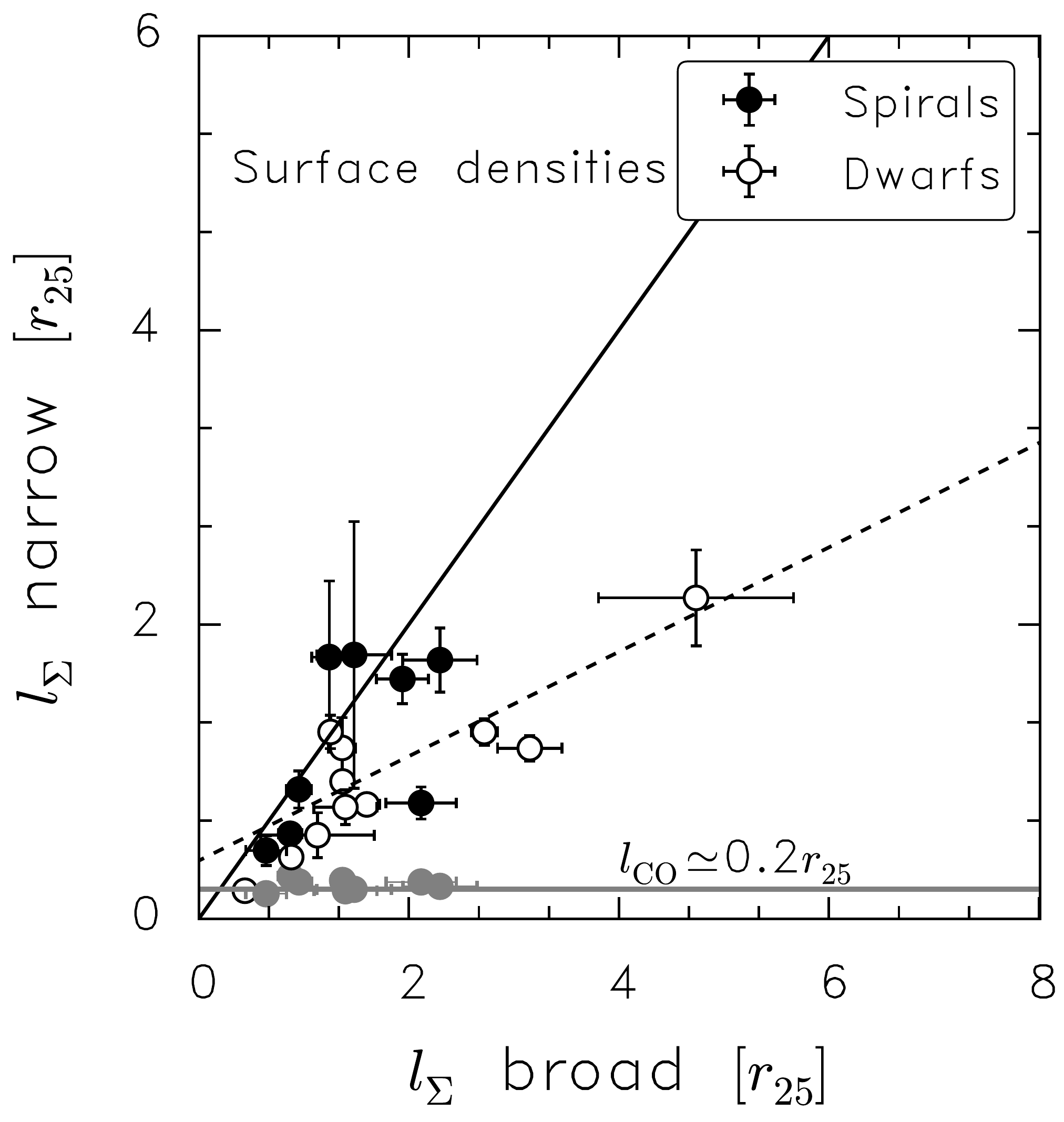}
\caption{Comparison of the scale lengths derived from the narrow and broad component surface density profiles. 
The solid line is a line of equality and the dashed line is a linear fit with a slope of 0.6. Gray points are 
CO surface density scale length by \citet{schrubaetal11}; the mean CO scale length is shown as a horizontal solid line.}
\label{fig:rad_surfarea_compsl}
\end{minipage}
\hspace{0.7cm}
\begin{minipage}[b]{0.5\textwidth}
\includegraphics[scale=0.27]{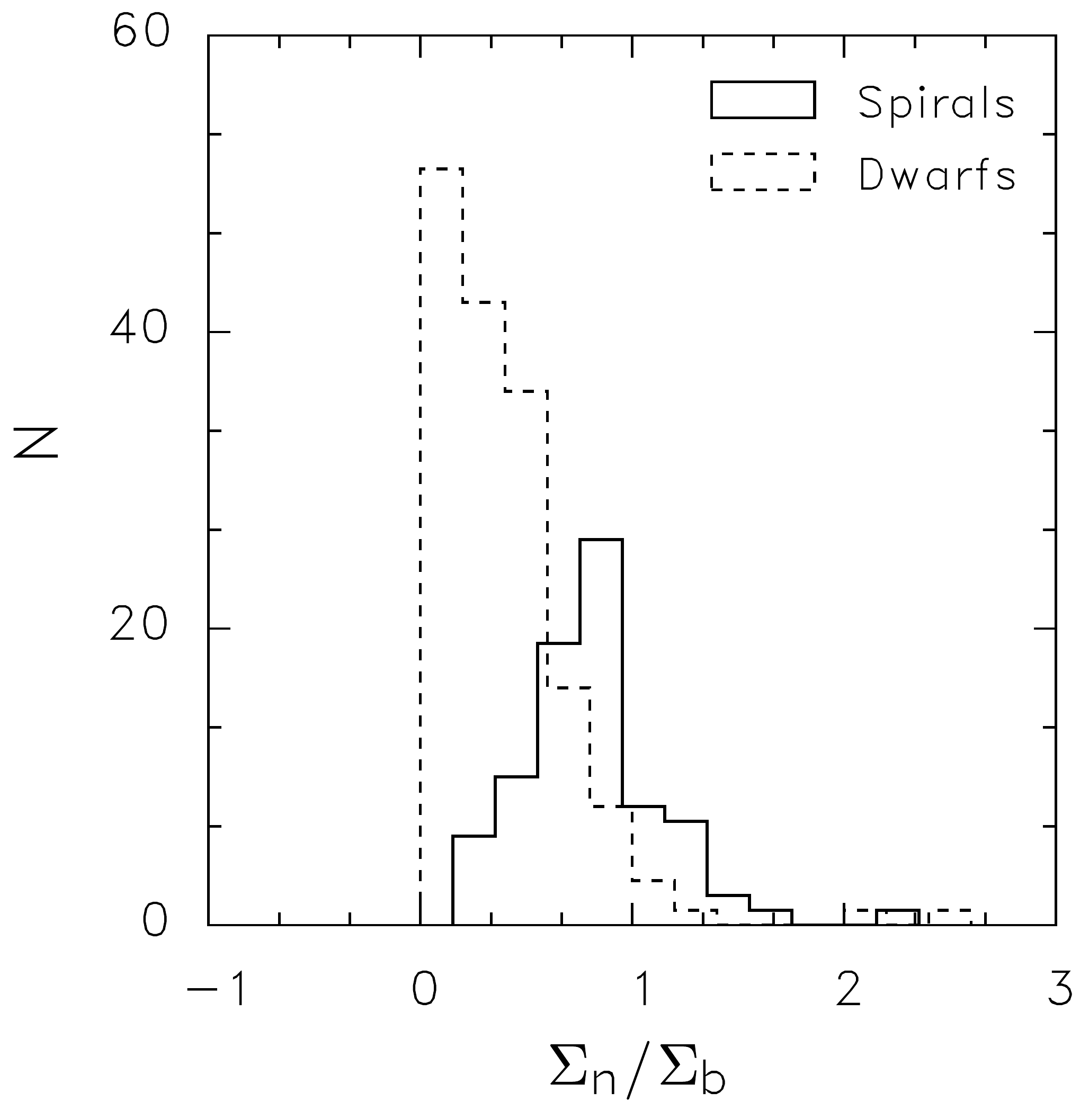}
\caption{Histograms of $\Sigma_{\rm{n}}/\Sigma_{\rm{b}}$ measured in radial bins. \vspace*{1.5cm} }
\label{fig:anabratio}
\end{minipage}
%
  \begin{tabular}{l l l}
  \hspace{-.3cm}
  \\
  \includegraphics[scale=.27]{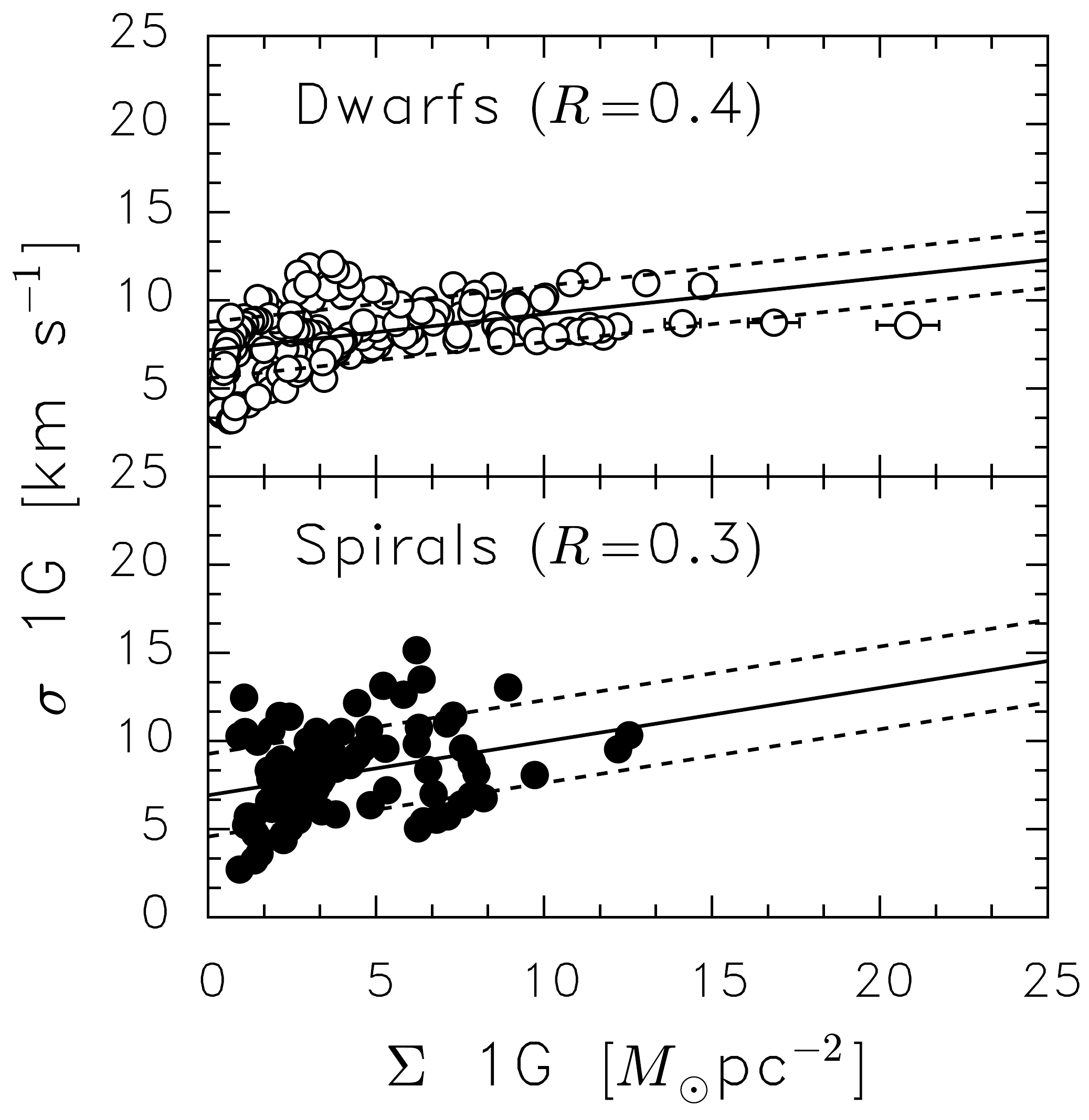}&
  \includegraphics[scale=.27]{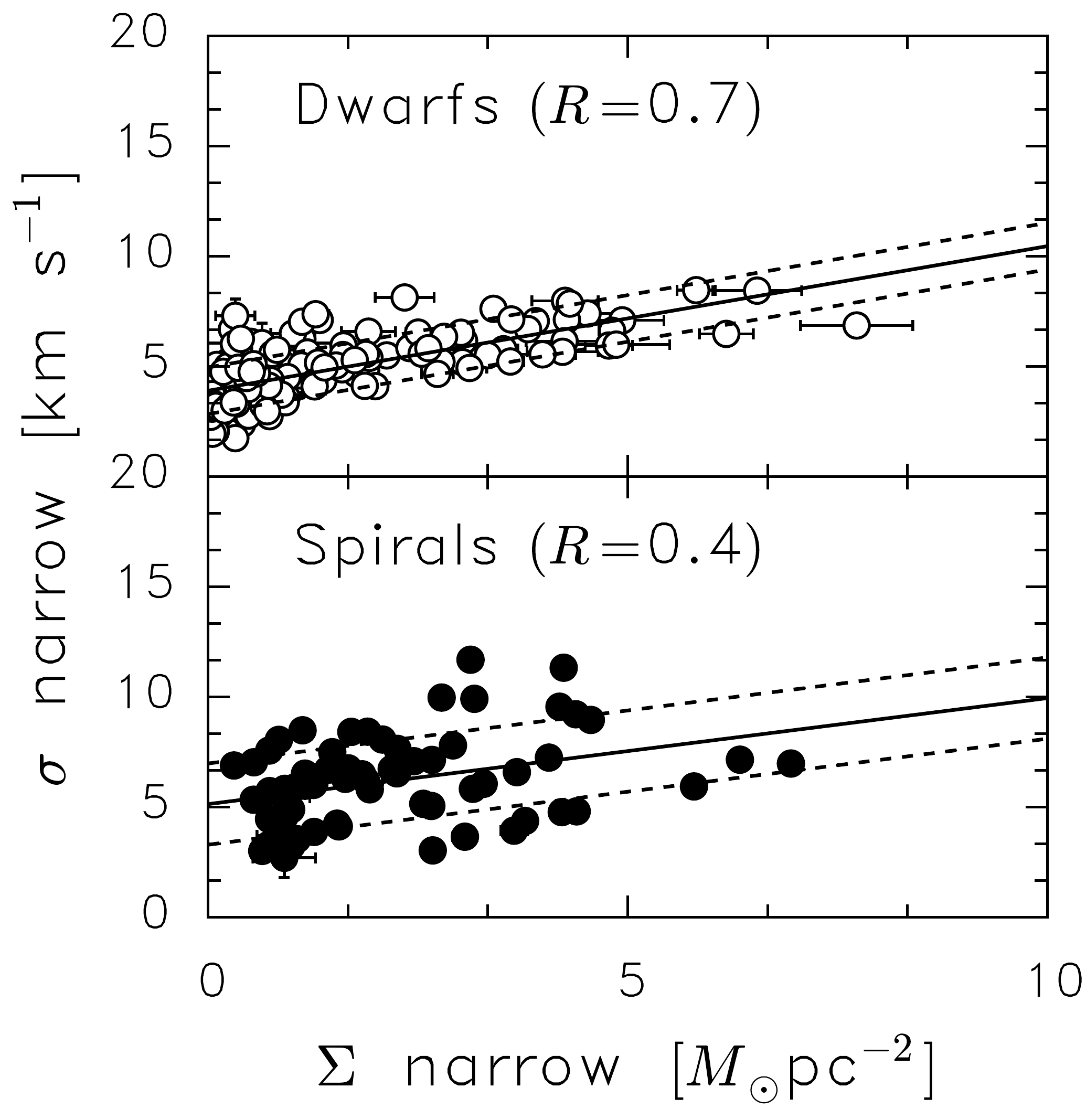}&
   \includegraphics[scale=.27]{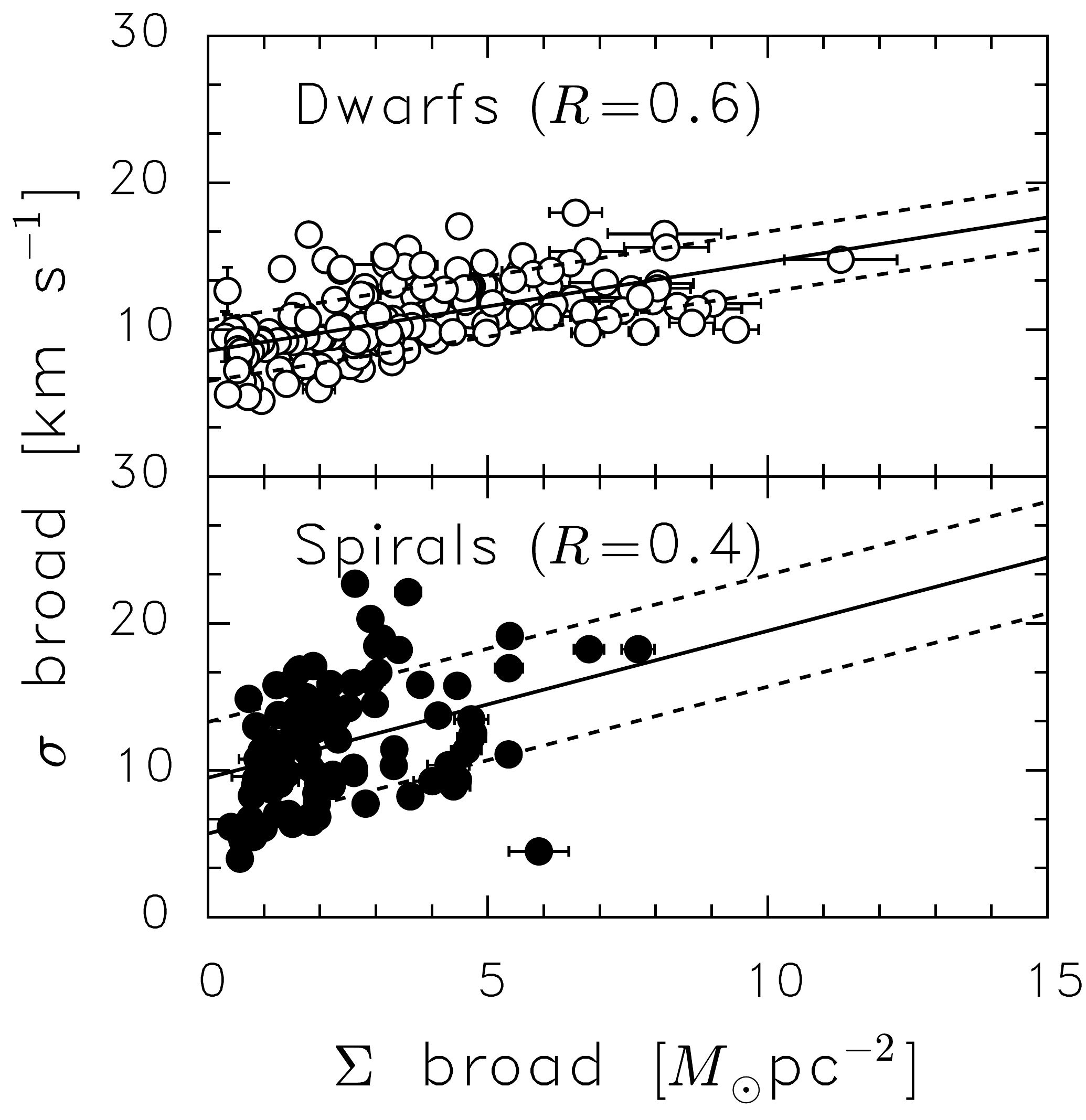}
  \end{tabular}
  \caption{Comparison of the velocity dispersions and the surface densities of the single Gaussian (left panel), narrow (middle panel) and broad (right panel) components. The solid lines are linear fits to the data; the dashed lines are one-sigma rms scatters. $R$ denotes the Pearson's rank correlation coefficients.\\} 
  \label{fig:comp_dispsl_surfsl}
\end{figure*}

Figure~\ref{fig:rad_area_main} shows $\rm{\Sigma_{1G}}$, $\rm{\Sigma_{n}}$, $\rm{\Sigma_{b}}$, and 
$\rm{\Sigma_{n}}$/$\rm{\Sigma_{b}}$ as a function of radius. 
We fit a single exponential function to the surface density profiles to facilitate comparisons with CO surface 
density profiles from \citet{schrubaetal11}. These authors fit a single exponential function to the CO surface 
density profiles of galaxies from the HERA CO-Line Extragalactic Survey \citep[HERACLES,][]{heracles}, a 
CO $J$ = 2$\rightarrow$ 1 survey which covers most of the THINGS sample. 
Exponential H\,{\sc i} surface density profiles were already found for 
late-type dwarf galaxies \citep{swaters02}. Here, we find that the radial decline in surface densities can be reasonably 
described by the exponential function. The narrow component surface densities tend to have smaller scale lengths than 
those of the broad component. The single Gaussian surface densities decline radially with a scale length somewhere between 
those of the narrow and broad components. 
We show a comparison of the scale lengths of the narrow and broad component's surface density profiles in Figure 
\ref{fig:rad_surfarea_compsl} and Table~\ref{tab:scale_lengths}. 
\capstartfalse

\begin{deluxetable}{l c c c c c}
\centering
\tabletypesize{\scriptsize}
\tablecaption{Fitted surface density scale lengths \label{tab:scale_lengths}}
\tablewidth{0pt}
\tablehead{
	\multicolumn{1}{c}{Galaxy}& $\rm{r_{25}}$ & $\rm{r_{D}}$ &\multicolumn{1}{c}{$l_{1g}$}&
	\multicolumn{1}{c}{$l_{n}$}&\multicolumn{1}{c}{$l_{b}$}\\
	&(kpc)&(kpc)&\multicolumn{1}{c}{$(\rm{r_{25}})$}&$(\rm{r_{25}})$&$(\rm{r_{25}})$\\
	\multicolumn{1}{c}{1} & \multicolumn{1}{c}{2} &\multicolumn{1}{c}{3} &\multicolumn{1}{c}{4}&\multicolumn{1}{c}{5}
}
\startdata 
NGC 2976 & 3.8 & 0.9 & 0.5 $\pm$ 0.1 & 0.4 $\pm$ 0.1 & 0.6 $\pm$ 0.1 \\ 
NGC 2366 & 2.2 & 1.3 &1.2 $\pm$ 0.0 & 1.2 $\pm$ 0.2 & 1.4 $\pm$ 0.1 \\ 
HOII & 3.3 & 1.2&1.1 $\pm$ 0.1 & 0.6 $\pm$ 0.0 & 1.4 $\pm$ 0.1 \\ 
NGC 4214 & 2.9 & 0.7&1.2 $\pm$ 0.1 & 0.9 $\pm$ 0.1 & 1.3 $\pm$ 0.1 \\ 
\textbf{NGC 3184} & 12.0 & 2.4 & 0.7 $\pm$ 0.1 & 0.6 $\pm$ 0.1 & 0.9 $\pm$ 0.1 \\ 
NGC 2403 & 7.4 & 1.6 & 1.1 $\pm$ 0.1 & 0.9 $\pm$ 0.1 & 1.4 $\pm$ 0.1 \\ 
NGC 7793 & 5.9 & 1.3& 3.2 $\pm$ 0.5 & 2.2 $\pm$ 0.3 & 4.7 $\pm$ 0.9 \\ 
IC2574 & 7.5 & 2.1 & 0.4 $\pm$ 0.0 & 0.2 $\pm$ 0.0 & 0.4 $\pm$ 0.0 \\ 
DDO154 & 1.2 & 0.8&2.6 $\pm$ 0.1 & 1.3 $\pm$ 0.1 & 2.7 $\pm$ 0.1 \\ 
\textbf{NGC 628} & 10.4 & 2.3 & 0.9 $\pm$ 0.1 & 0.9 $\pm$ 0.1 & 1.0 $\pm$ 0.1 \\ 
HOI & 1.8 & 0.8 &0.7 $\pm$ 0.0 & 0.5 $\pm$ 0.1 & 0.8 $\pm$ 0.1 \\ 
NGC 925 & 14.3 & 4.1& 1.0 $\pm$ 0.1 & 0.8 $\pm$ 0.1 & 1.4 $\pm$ 0.3 \\ 
\textbf{NGC 2903} & 15.2 & 2.9 & 0.6 $\pm$ 0.0 & 0.6 $\pm$ 0.0 & 0.8 $\pm$ 0.1 \\ 
\textbf{NGC 5055} & 17.3 & 3.2&1.8 $\pm$ 0.6 & 2.0 $\pm$ 1.1 & 1.5 $\pm$ 0.4 \\ 
DDO53 & 0.4 & 0.7& 2.2 $\pm$ 0.1 & 1.2 $\pm$ 0.1 & 3.2 $\pm$ 0.3 \\ 
\textbf{NGC 4736} & 5.3 & 1.2& 0.6 $\pm$ 0.1 & 0.5 $\pm$ 0.1 & 0.6 $\pm$ 0.1 \\ 
\textbf{NGC 3198} & 13.0 & 3.2 & 1.3 $\pm$ 0.0 & 1.0 $\pm$ 0.1 & 1.6 $\pm$ 0.2 \\ 
\textbf{NGC 3621} & 9.4 & 2.6 & 1.8 $\pm$ 0.2 & 1.6 $\pm$ 0.2 & 2.0 $\pm$ 0.3 \\ 
\textbf{NGC 6946} & 9.9 & 2.5& 2.4 $\pm$ 0.4 & 1.6 $\pm$ 0.2 & 3.5 $\pm$ 0.9 \\
\textbf{NGC 5236} & 10.1 & 2.5&1.2 $\pm$ 0.2 & 2.2 $\pm$ 1.0 & 1.3 $\pm$ 0.2 \\
\enddata
\tablecomments{Column 1: name of galaxy; 
Column 2: the optical radius, $\rm{r_{25}}$, in units of kpc, adopting 
the distance in \citet{walter08}; 
Column 3: The radial scale length, $\rm{r_{D}}$, in units of kpc derived by \citet{HunterElmegreen04} and \citet{leroy08}; 
Column 4: surface density scale lengths from the single Gaussian fit; 
Column 5: surface density scale lengths of the narrow component; 
Column 6: surface density scale lengths of the broad component. 
 Bold font indicates spiral galaxies, whereas normal font represents 
 dwarf galaxies \citep[adopting the 
definition of][]{leroy08}.}
\end{deluxetable}
\capstarttrue

To test whether the surface density 
of the narrow component traces the surface density of molecular gas, we compare the narrow component's surface density scale 
length with that of a tracer of the molecular gas component. \citet{schrubaetal11} 
found that the integrated CO intensities in their sample (a sample of 33 nearby spiral galaxies) decrease 
exponentially with a scale length of $\sim 0.2~\rm{r_{25}}$. This is much steeper than the scale length of the narrow 
component's surface density that we find in this analysis ($l\sim 1.1~\rm{r_{25}}$). This result suggests that the formation 
of cold H\,{\sc i} is not a sufficient condition for the formation of molecular gas \citep{schaye04}. As put forward by 
\citet{youngetal03}, it is also possible that there is 
a delay between the formation of cold H\,{\sc i} and $\rm{H_{2}}$. This is because factors such as pressure, 
column density, metallicity, play a role in the transition from H\,{\sc i} to $\rm{H_{2}}$ \citep{elmegreen93, honmaetal95}.  
Another possibility is that, in some regions, even if $\rm{H_{2}}$ 
formation proceeds after the formation of cold H\,{\sc i}, the $\rm{H_{2}}$ molecule is subsequently destroyed by FUV radiation 
\citep{elmegreen93}.  
    
The $\rm{\Sigma_{n}/\Sigma_{b}}$ ratio tends to decrease with radius. In 
a similar analysis, \citet{braun97} found that the fractional line flux of the 
HBN mentioned previously (attributed to the CNM) decreases abruptly near 
the optical radius, $\rm{r_{25}}$. However, only about 
20\% of our sample galaxies show an abrupt decrease in the $\rm{\Sigma_{n}/\Sigma_{b}}$ near 
the optical radius. For most galaxies, the $\rm{\Sigma_{n}/\Sigma_{b}}$ ratio continually 
declines without a clear break. We attempted to fit the super profiles with the 
sum of a Gaussian and a Lorentzian function, similar to \citet{braun97} but we still did not find a clear break 
in the $\rm{\Sigma_{n}/\Sigma_{b}}$ profiles. 
This difference could be due to the fact that we fitted the super profiles with the 
sum of a Gaussian plus Lorentzian function with all parameters free rather than with the physical model 
assumed by \citet{braun97}. Investigation of this will be left as future work.  

We compare the $\rm{\Sigma_{n}/\Sigma_{b}}$ ratio of dwarfs and spirals in Figure~\ref{fig:anabratio}. On average, 
spirals have higher $\rm{\Sigma_{n}/\Sigma_{b}}$ ratio (0.8 $\pm$ 0.3) than dwarfs (0.4 $\pm$ 0.3), consistent 
with the results presented in \citet{ianjamasimananaetal12}. 
We attribute the difference in $\rm{\Sigma_{n}/\Sigma_{b}}$ as an indication that the fraction of 
cold H\,{\sc i} component is higher in spirals than in dwarfs.       

Figure~\ref{fig:comp_dispsl_surfsl} shows a comparison of $\rm{H\,{\textsc i}}$ velocity dispersions and $\rm{H\,{\textsc i}}$ 
surface densities. 
Dwarf galaxies show stronger correlation, and less scatter, in the $\sigma_{\rm{H\,{\textsc i}}}$-$\Sigma_{\rm{H\,{\textsc i}}}$ 
relation than spirals. Possibly this is due to the presence of extra morphological and kinematical features such as bars and spiral arms which are not present in dwarf galaxies. Confirmation of this needs the stacking of super profiles in the arms and inter-arm regions, 
which is beyond the scope of the current analysis. We use the Pearson's rank correlation coefficient, $R$, to quantify the strength of the 
correlation between $\sigma_{\rm{H\,{\textsc i}}}$ and $\Sigma_{\rm{H\,{\textsc i}}}$. In general, $\sigma_{\rm{H\,{\textsc i}}}$ and $\Sigma_{\rm{H\,{\textsc i}}}$ correlate better to each other for the narrow component than for the broad and the single Gaussian components. If the narrow component is associated with star formation as shown before by \citet{younglo96} and 
\citet{deblokwalter06}, then the correlation between $\sigma_{\rm{H\,{\textsc i}}}$ and $\Sigma_{\rm{H\,{\textsc i}}}$ may be 
driven by star formation. Correlation between $\sigma_{\rm{H\,{\textsc i}}}$ and $\Sigma_{\rm{H\,{\textsc i}}}$ 
has previously been reported in the literature 
\citep{shostak84, stilpetal13}. \citet{shostak84} found a positive correlation between 
$\sigma_{\rm{H\,{\textsc i}}}$ and $\Sigma_{\rm{H\,{\textsc i}}}$ for NGC 628. Since regions 
of higher $\Sigma_{\rm{H\,{\textsc i}}}$ usually correspond to regions of active star formation, \citet{shostak84} attributed 
the increase in $\sigma_{\rm{H\,{\textsc i}}}$, in regions of high $\Sigma_{\rm{H\,{\textsc i}}}$, as a result of energy input by 
stellar winds from OB stars, supernovae and expanding HII regions.

\capstartfalse
\begin{figure*}
\begin{tabular}{l l l}
  \includegraphics[scale=.22]{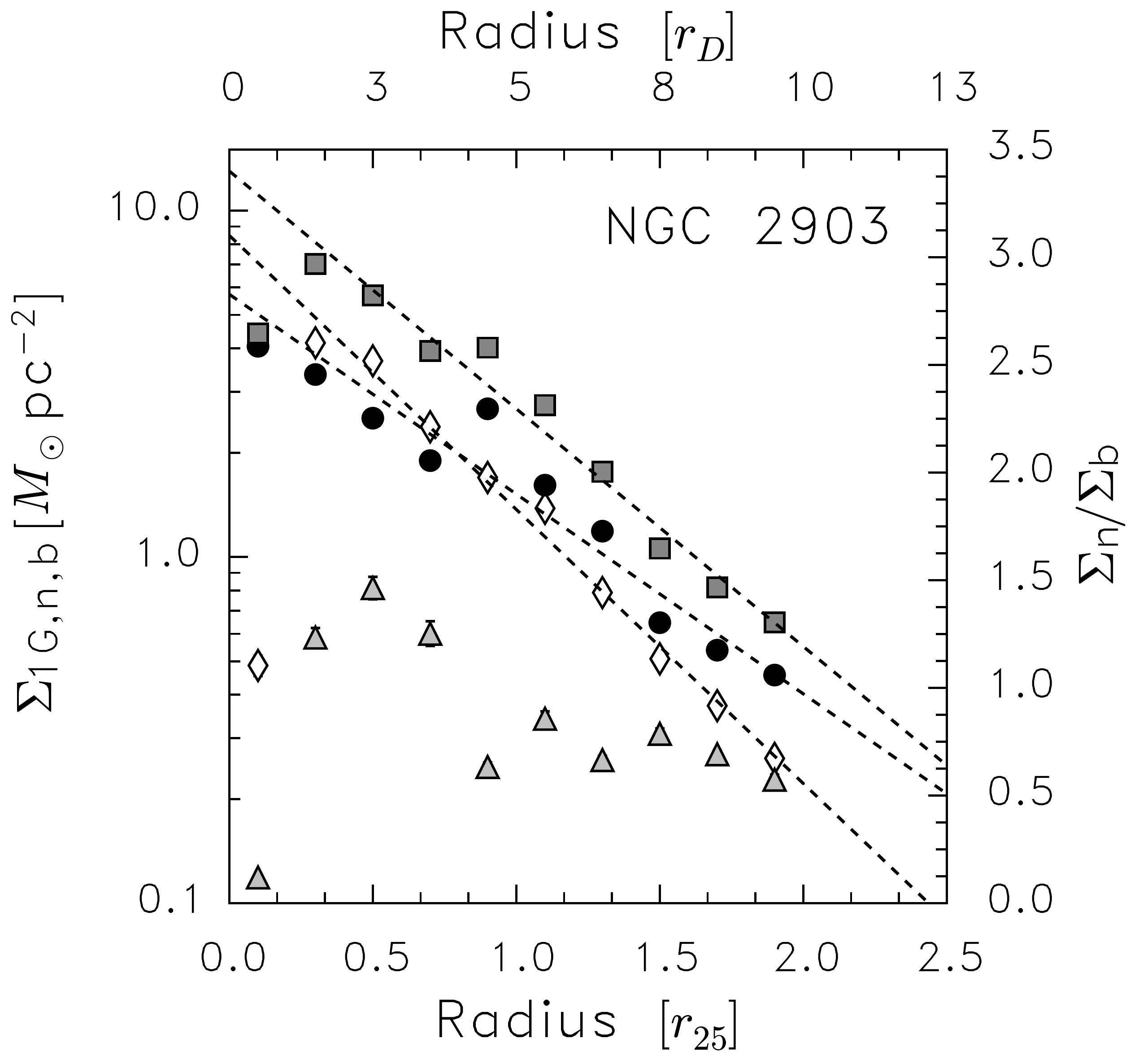}&
  \includegraphics[scale=.22]{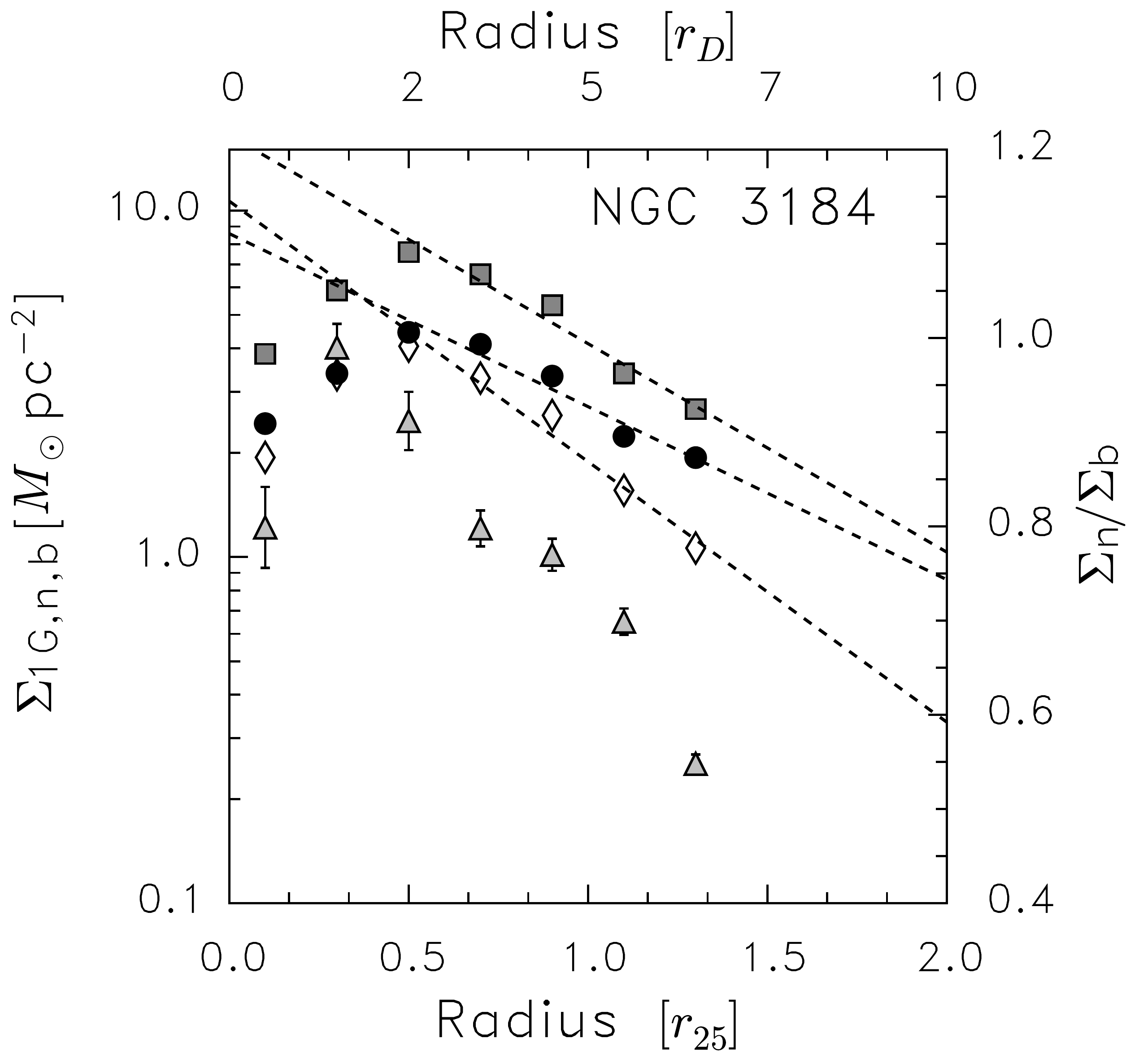}&
  \includegraphics[scale=.22]{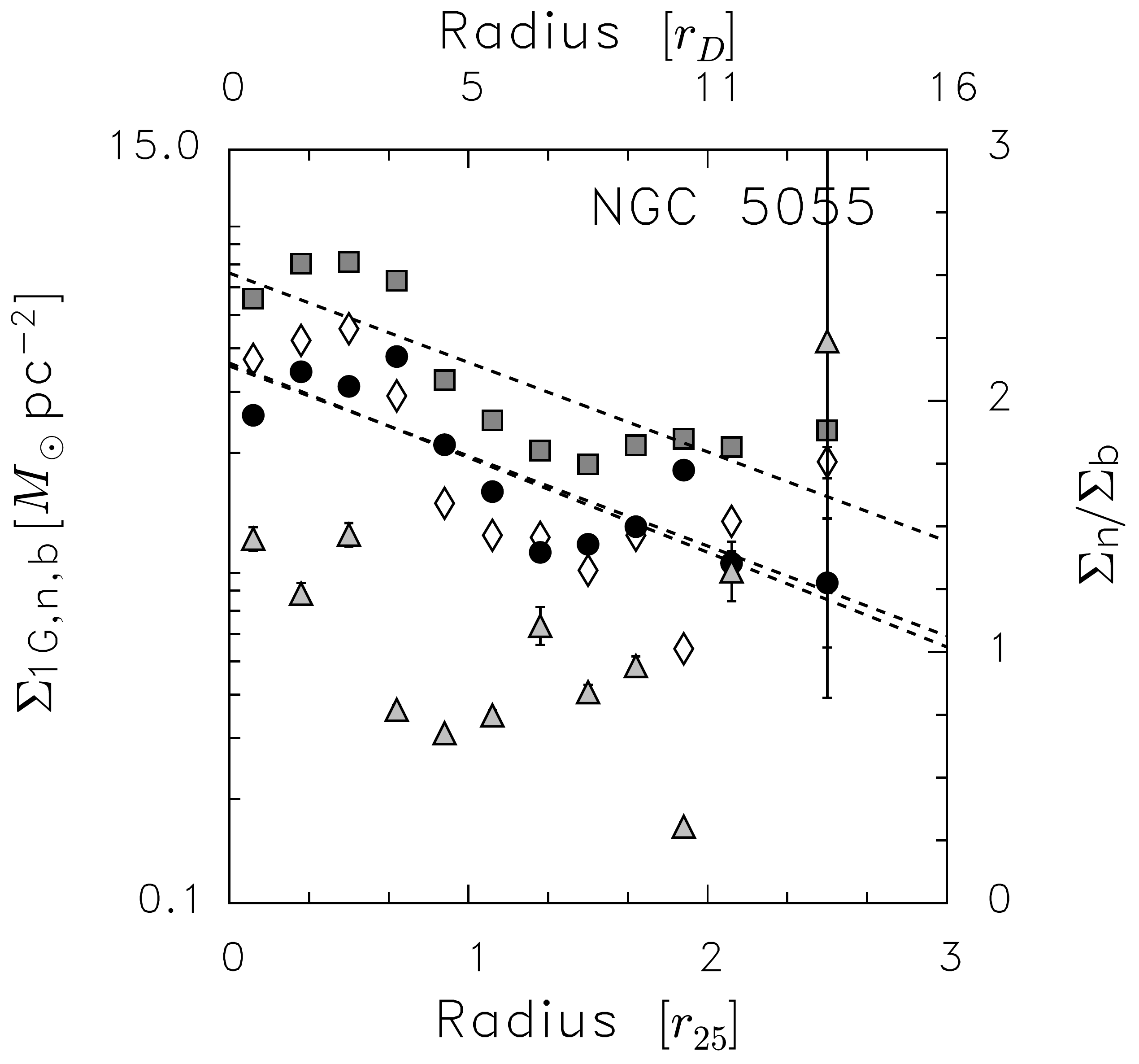}\\
  \includegraphics[scale=.22]{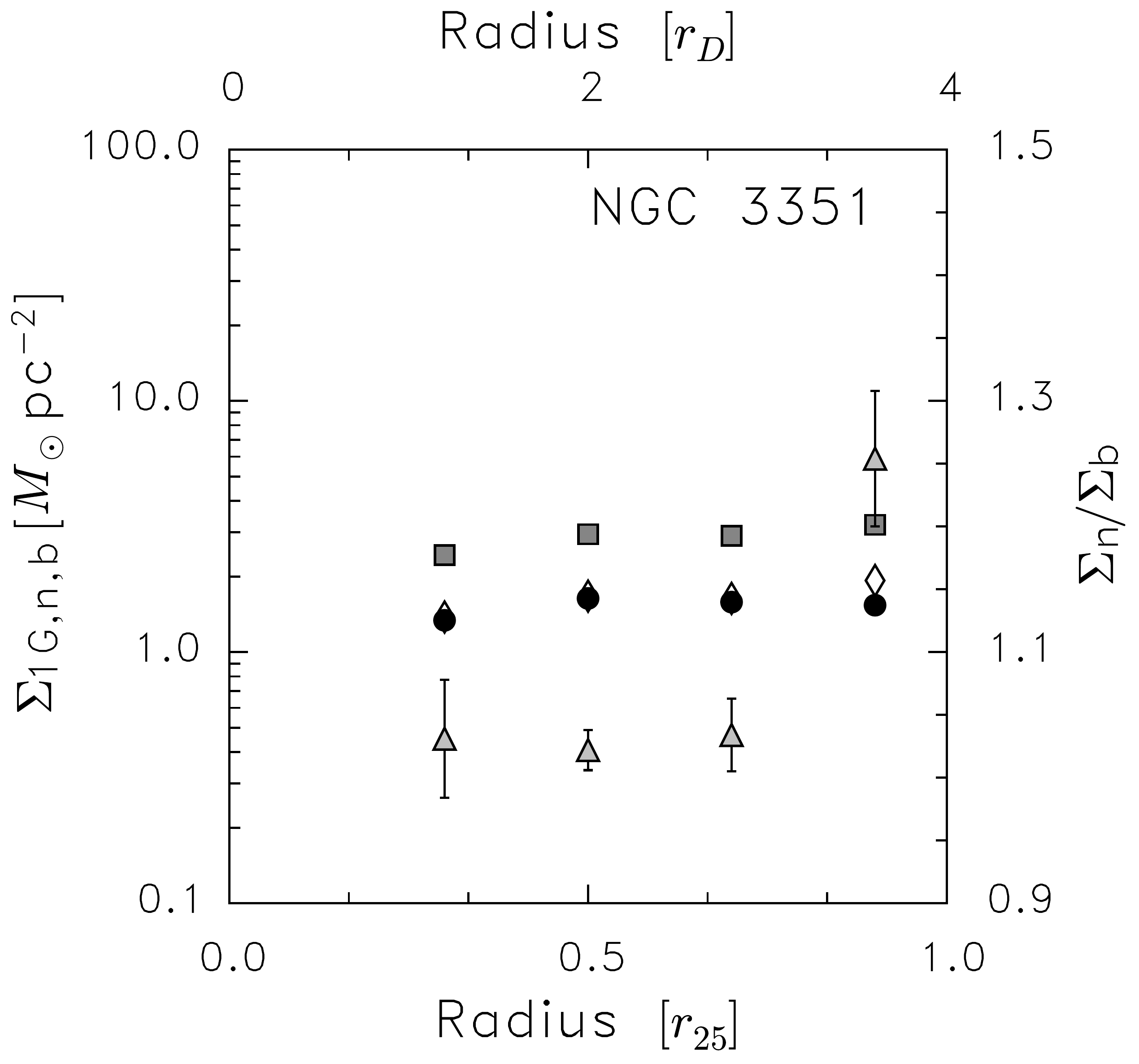}&
  \includegraphics[scale=.22]{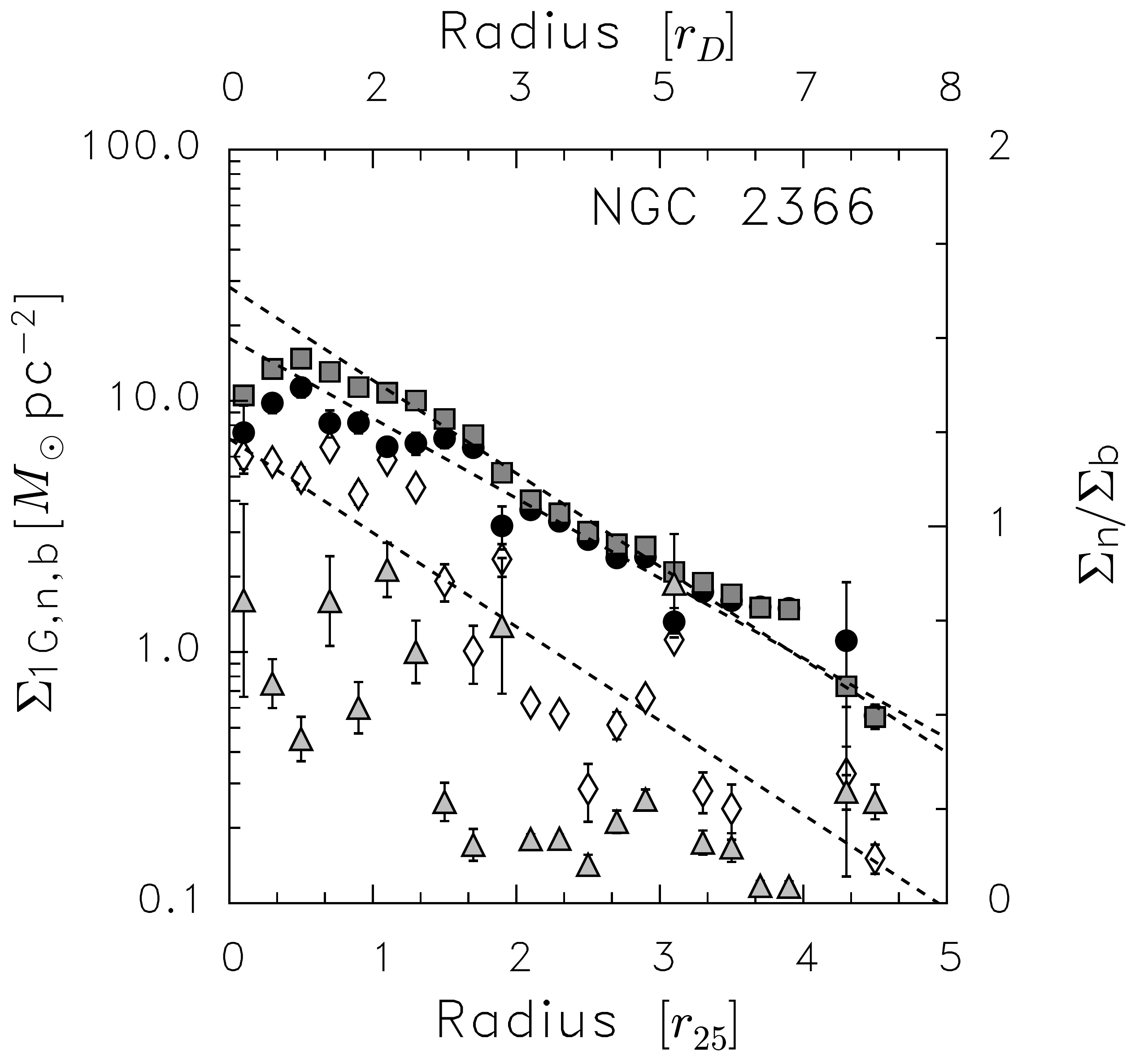}&
  \includegraphics[scale=.22]{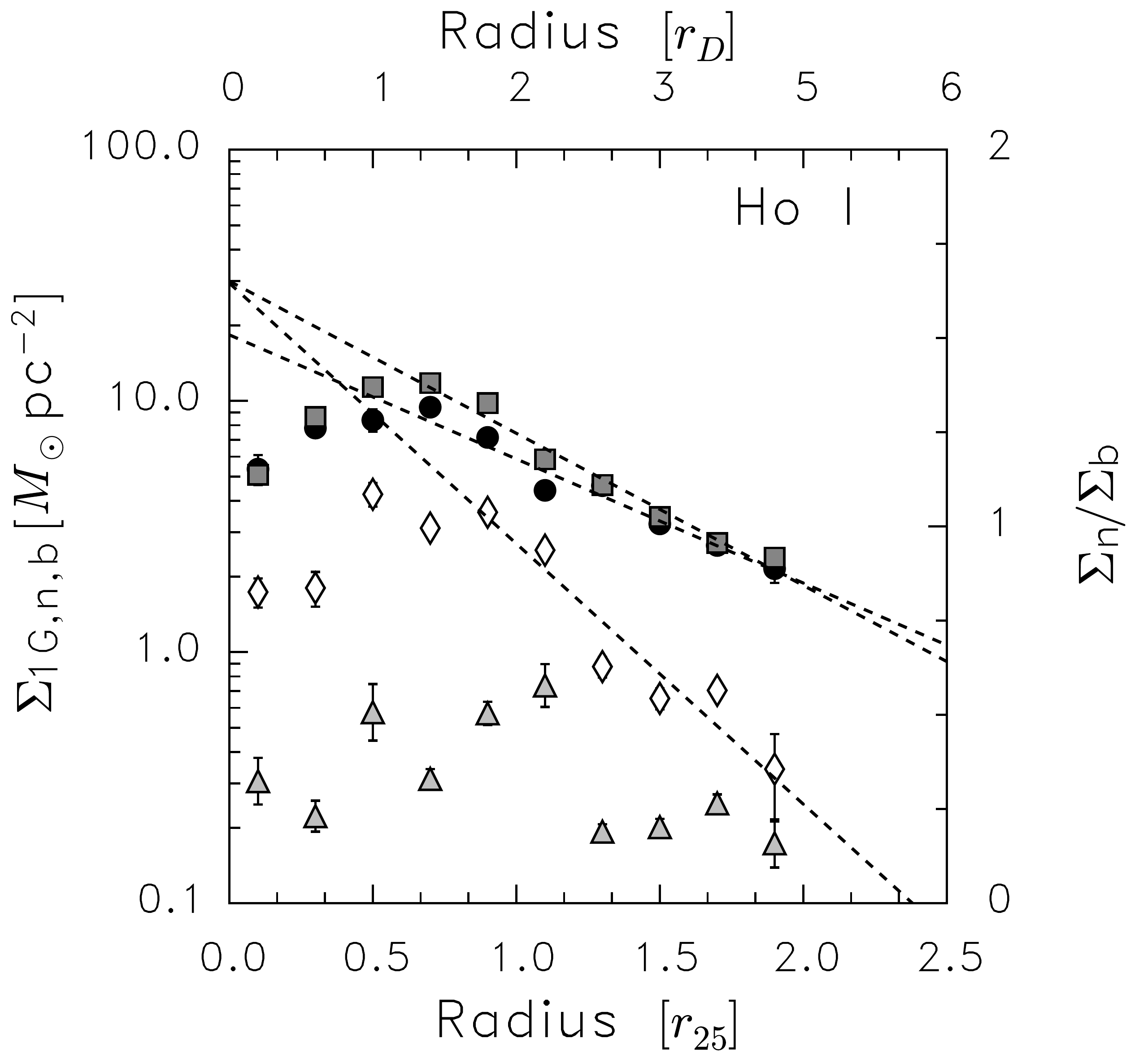}\\
  \includegraphics[scale=.22]{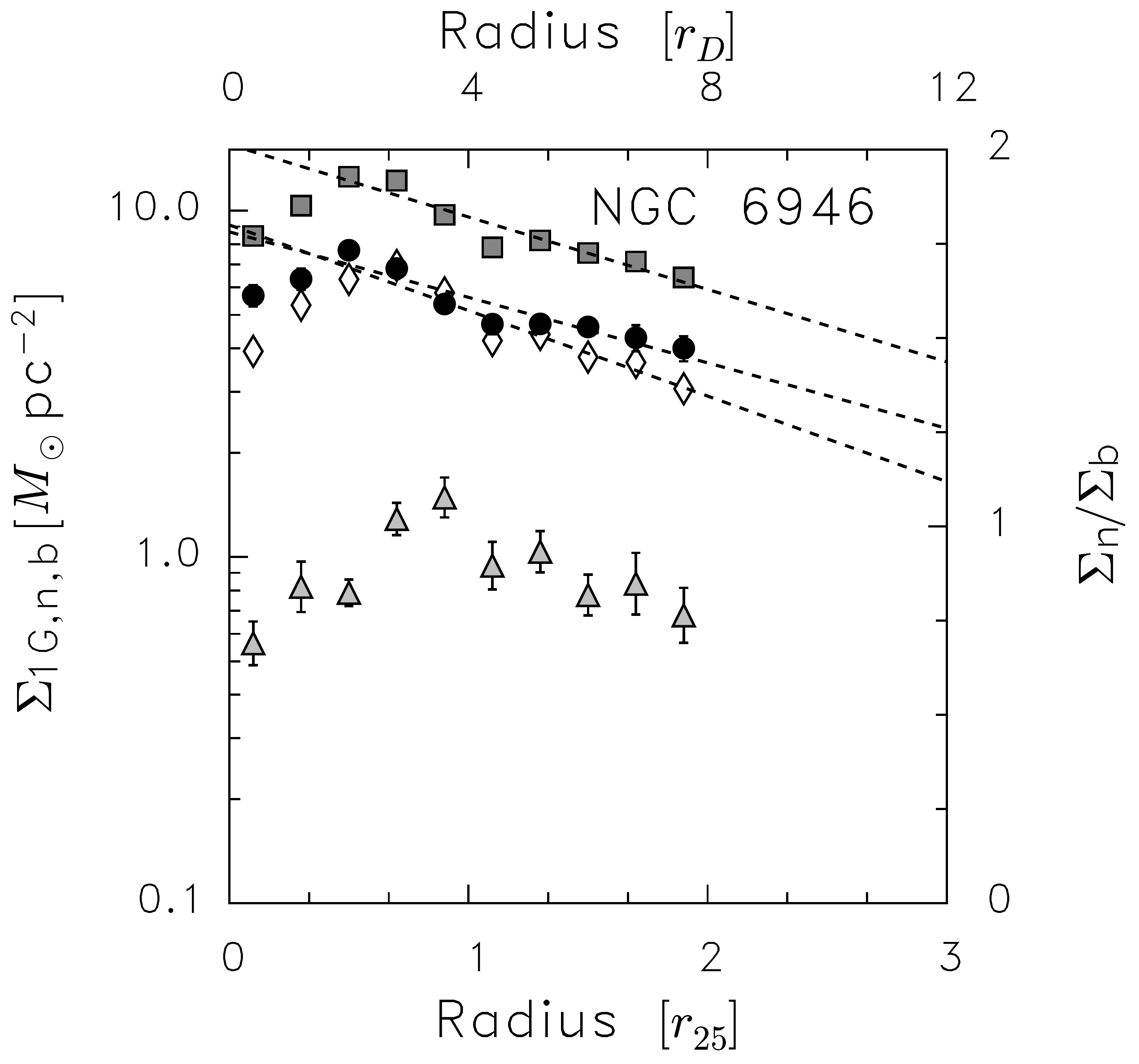} &
  \includegraphics[scale=.22]{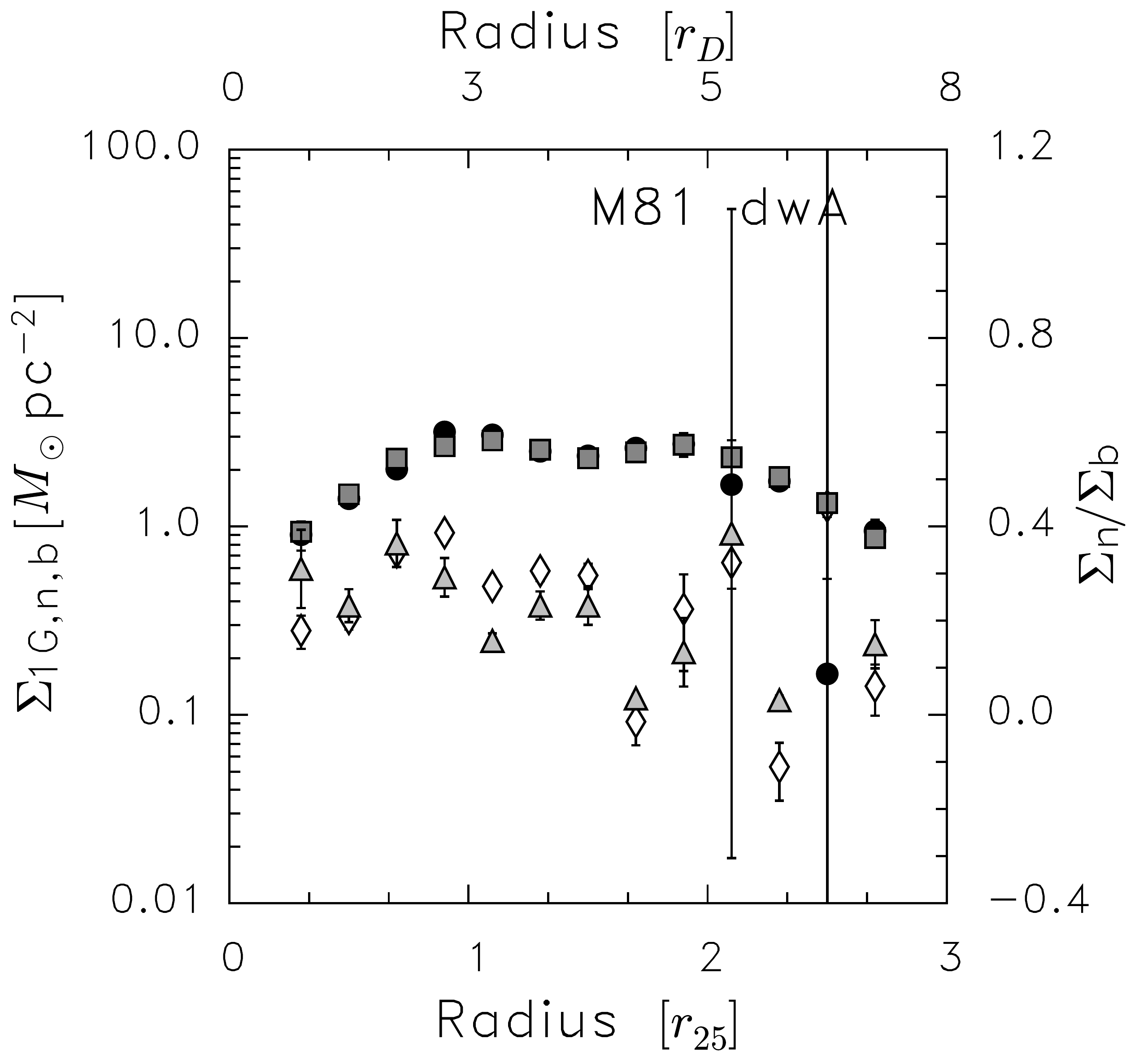}&
  \includegraphics[scale=.22]{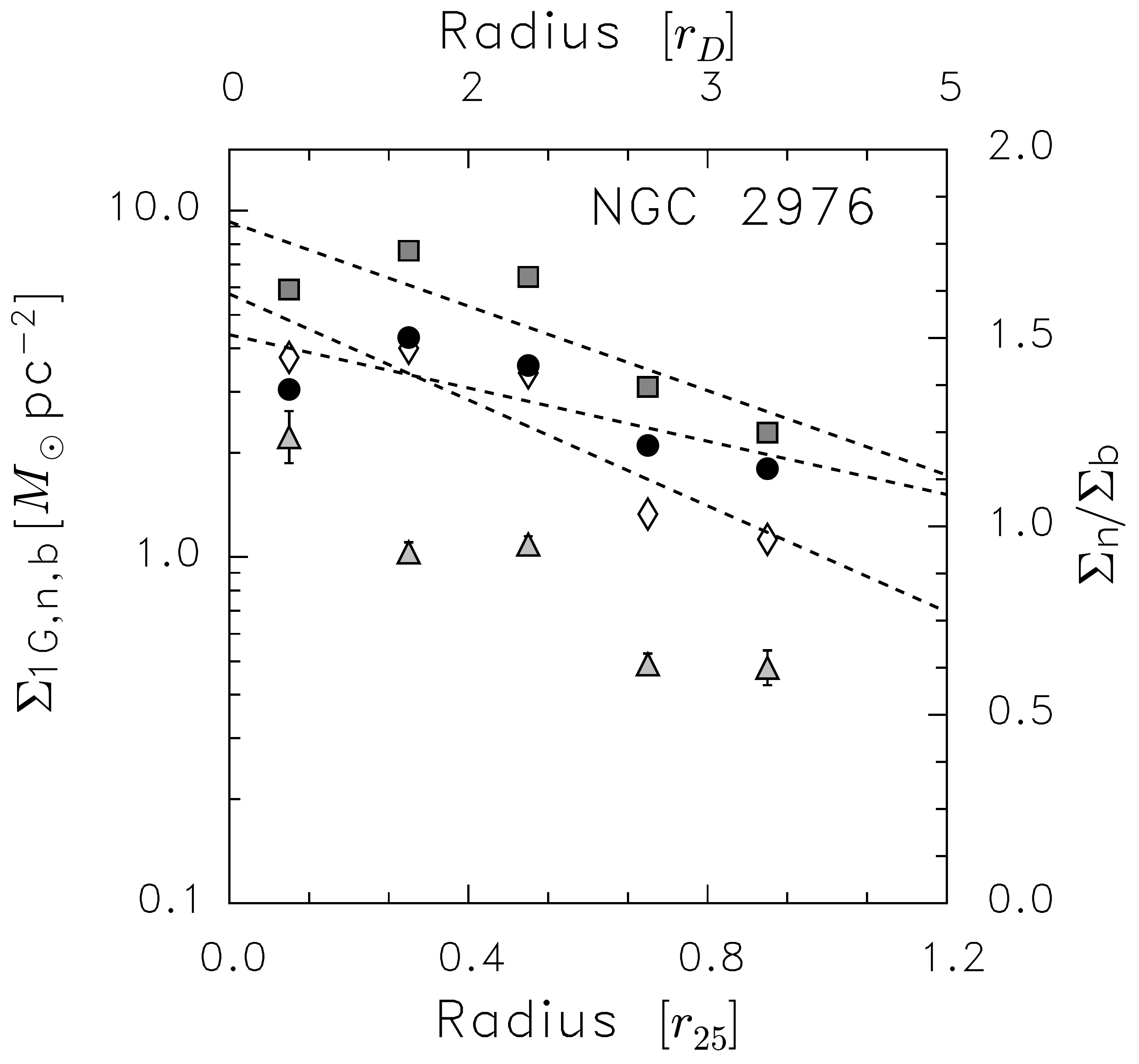}\\
  \includegraphics[scale=.22]{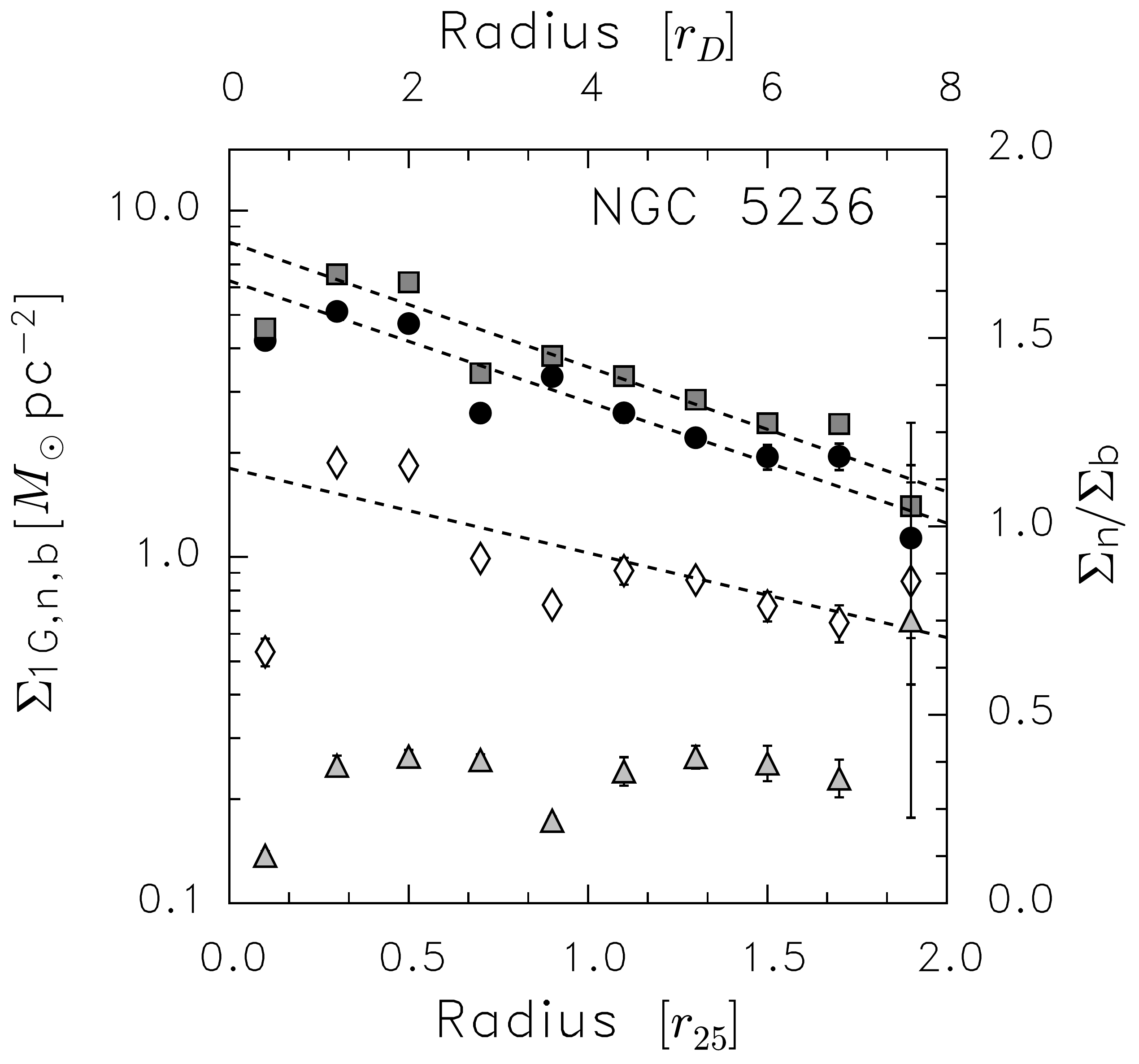}
\end{tabular}
\\\textbf{Figure \ref{fig:rad_area_main}}: (Continued).
\end{figure*}
\capstarttrue

\section{SUMMARY AND CONCLUSION}\label{sec:conc} 
This paper presents a comparison of the radial variation of H\,{\sc i} velocity dispersions and surface densities of 
the components of H\,{\sc i} in spiral and dwarf galaxies. 
Spirals exhibit a clear radial decline in velocity dispersion, which can be well 
described by an exponential function. Dwarfs show flatter or flat radial velocity dispersion profiles. 
On average, though, we find that spiral and dwarf galaxies have similar H\,{\sc i} velocity dispersion. 
  
The radial velocity dispersion profiles derived 
from single Gaussian fits to stacked H\,{\sc i} profiles (super profiles) tend to be flatter than those of the narrow and broad components, identified from a double Gaussian fit to the super profiles. In addition, 
the single Gaussian dispersion values approach those  of the broad component at large radius, 
suggesting the broad component is dominant in the outer disks.

In general, the velocity dispersion profiles in the outer parts do not drop at the same rate as the star formation rate profiles 
analysed by \citet{tamburroetal09}. We therefore confirm their conclusion that star formation is not the 
main drivers of the H\,{\sc i} dispersions in the outer disk of galaxies. Our measured H\,{\sc i} dispersions are, however, somewhat smaller than the second moment values of 
\citet{tamburroetal09}. This is because second moment values are sensitive to the shapes of the profiles 
and therefore do not always reflect the velocity dispersions.

In terms of surface densities, spiral and dwarf galaxies have similar radial behaviour. 
We find that, except in the innermost part, the surface densities exponentially decrease with radius.  
This decrease is steeper for the narrow component than for the single Gaussian and 
broad components. The surface density ratio between the narrow and broad components also 
tend to decrease with radius.  
Identifying the narrow component with the CNM and the broad component with the WNM, this indicates 
that the fraction of gas in the cold phase tends to be higher in the inner parts of the galaxy disks. 
On average, dwarfs have a lower surface density ratio ($\Sigma_{\rm{n}}/\Sigma_{\rm{b}}$ = 0.4 $\pm$ 0.3) 
than spirals ($\Sigma_{\rm{n}}/\Sigma_{\rm{b}}$ = 0.8 $\pm$ 0.3), indicating that dwarf galaxies have a 
relatively lower amount of cold H\,{\sc i} component than spiral galaxies.
Finally, we find that $\sigma_{\rm{H\,{\textsc i}}}$ is correlated with $\Sigma_{\rm{H\,{\textsc i}}}$, which we interpret as a correlation with 
star formation given the relationship between $\Sigma_{\rm{H\,{\textsc i}}}$ and $\Sigma_{\rm{SFR}}$.   
\acknowledgments
We thank the anonymous referee for the constructive comments that have improved this 
paper. R.I. acknowledges funding from the South African National Research Foundation (NRF grant number MWA1203150687), 
a Postdoctoral Grant of the University of South Africa, and the PPI grant of the University of Cape Town. 
R.I.'s working visit to the Netherlands was supported by NRF-NWO exchange programme in 
''Astronomy, and Enabling Technologies for Astronomy''. 
The work of W.J.G.d.B. was supported by the European Commission (grant FP7-PEOPLE-2012-CIG \#333939).
 
\end{document}